\documentclass[twocolumn,modern,astrosymb,tighten,trackchanges,article]{aastex631}
\usepackage{comment}
\usepackage{CJK}
\usepackage{amsmath}
\usepackage{appendix}
\usepackage{hyperref}

\newcommand{\eg}{e.g.\/}

\submitjournal{The Astrophysical Journal}

\shorttitle{Tracing Cosmic-Ray Ionization in NGC\,253}
\shortauthors{Behrens et al.}

\graphicspath{{./}{figures/}}

\begin{document}
\begin{CJK*}{UTF8}{gbsn}

\title{Neural Network Constraints on the Cosmic-Ray Ionization Rate and Other Physical Conditions in NGC\,253 with ALCHEMI Measurements of HCN and HNC}

\author[0000-0002-2333-5474]{Erica Behrens}
\affiliation{Department of Astronomy, University of Virginia, P.~O.~Box 400325, 530 McCormick Road, Charlottesville, VA 22904-4325, USA}
\affiliation{National Radio Astronomy Observatory, 520 Edgemont Road, Charlottesville, VA  22903-2475, USA}
\author[0000-0003-1183-9293]{Jeffrey G.~Mangum}
\affiliation{National Radio Astronomy Observatory, 520 Edgemont Road,
  Charlottesville, VA  22903-2475, USA}
%
%
\author[0000-0001-8504-8844]{Serena Viti}
\affiliation{Leiden Observatory, Leiden University, P.O. Box 9513, NL-2300 RA Leiden, The Netherlands}
\affiliation{Argelander Institut f\"{u}r Astronomie der Universit\"{a}t Bonn, Auf dem H\"{u}gel 71, 53121 Bonn Germany}
\affiliation{Department of Physics and Astronomy, University College London, Gower Street, London WC1E 6BT}
\author[0000-0003-4025-1552]{Jonathan Holdship}
\affiliation{Leiden Observatory, Leiden University, P.O. Box 9513, NL-2300 RA Leiden, The Netherlands}
\affiliation{Department of Physics and Astronomy, University College London, Gower Street, London WC1E 6BT}
\author[0000-0002-1227-8435]{Ko-Yun Huang}
\affiliation{Leiden Observatory, Leiden University, P.O. Box 9513, NL-2300 RA Leiden, The Netherlands}
\author[0000-0002-0370-8034]{Mathilde Bouvier}
\affiliation{Leiden Observatory, Leiden University, P.O. Box 9513, NL-2300 RA Leiden, The Netherlands}
\author[0000-0002-5353-1775]{Joshua Butterworth}
\affiliation{Leiden Observatory, Leiden University, P.O. Box 9513, NL-2300 RA Leiden, The Netherlands}

\author[0000-0002-1185-2810]{Cosima Eibensteiner}
\affiliation{National Radio Astronomy Observatory, 520 Edgemont Road, Charlottesville, VA  22903-2475, USA}
\author[0000-0002-6824-6627]{Nanase Harada}
\affiliation{National Astronomical Observatory of Japan, 2-21-1 Osawa, Mitaka, Tokyo 181-8588, Japan}
\affiliation{Department of Astronomy, School of Science, The Graduate University for Advanced Studies (SOKENDAI), 2-21-1 Osawa, Mitaka, Tokyo, 181-1855 Japan}
\author[0000-0001-9281-2919]{Sergio Mart\'in}
\affiliation{European Southern Observatory, Alonso de C\'ordova, 3107, Vitacura, Santiago 763-0355, Chile}
\affiliation{Joint ALMA Observatory, Alonso de C\'ordova, 3107, Vitacura, Santiago 763-0355, Chile}
%
\author[0000-0001-5187-2288]{Kazushi Sakamoto}
\affiliation{Institute of Astronomy and Astrophysics, Academia Sinica, 11F of AS/NTU
Astronomy-Mathematics Building, No.1, Sec. 4, Roosevelt Rd, Taipei 10617, Taiwan}
%
\author[0000-0002-9931-1313]{Sebastien Muller}
\affiliation{Department of Space, Earth and Environment, Chalmers University of Technology, Onsala Space Observatory, SE-43992 Onsala, Sweden}
%
\author[0000-0001-8153-1986]{Kunihiko Tanaka}
\affil{Department of Physics, Faculty of Science and Technology, Keio University, 3-14-1 Hiyoshi, Yokohama, Kanagawa 223--8522 Japan}

\author[0000-0001-8064-6394]{Laura Colzi}
\affiliation{Centro de Astrobiolog\'ia (CAB), CSIC-INTA,  Ctra. de Ajalvir km 4, E–28850, Torrej´on de Ardoz, Spain}






\author[0000-0002-7495-4005]{Christian Henkel}
\affiliation{Max-Planck-Institut f\"ur Radioastronomie, Auf dem H\"ugel   69, 53121 Bonn, Germany}
\affiliation{Xinjiang Astronomical Observatory, Chinese Academy of Sciences, 830011, Urumqi, China}





\author[0000-0001-9436-9471]{David S.~Meier}
\affiliation{New Mexico Institute of Mining and Technology, 801 Leroy Place, Socorro, NM 87801, USA}
\affiliation{National Radio Astronomy Observatory, PO Box O, 1003 Lopezville Road, Socorro, NM 87801, USA}



\author[0000-0002-2887-5859]{V\'ictor M.~Rivilla}
\affiliation{Centro de Astrobiolog\'ia (CAB), CSIC-INTA,  Ctra. de Ajalvir km 4, E–28850, Torrej´on de Ardoz, Spain}



\author[0000-0001-5434-5942]{Paul P.~van der Werf}
\affiliation{Leiden Observatory, Leiden University,
P.O. Box 9513, NL-2300 RA Leiden, The Netherlands}
 

\collaboration{20}{(ALMA Comprehensive High-resolution Extragalactic Molecular Inventory (ALCHEMI) collaboration)} 

\correspondingauthor{Erica Behrens} \email{eb7he@virginia.edu}



\begin{abstract}
We use a neural network model and ALMA observations of HCN and HNC to constrain the physical conditions, most notably the cosmic-ray ionization rate (CRIR, $\zeta$), in the Central Molecular Zone (CMZ) of the starburst galaxy NGC\,253. Using output from the chemical code \texttt{UCLCHEM}, we train a neural network model to emulate \texttt{UCLCHEM} and derive HCN and HNC molecular abundances from a given set of physical conditions. We combine the neural network with radiative transfer modeling to generate modeled integrated intensities, which we compare to measurements of HCN and HNC from the ALMA Large Program ALCHEMI. Using a Bayesian nested sampling framework, we constrain the CRIR, molecular gas volume and column densities, kinetic temperature, and beam-filling factor across NGC\,253's CMZ. The neural network model successfully recovers \texttt{UCLCHEM} molecular abundances with $\sim3$\% error and, when used with our Bayesian inference algorithm, increases the parameter inference speed tenfold. We create images of these physical parameters across NGC\,253's CMZ at 50\,pc resolution and find that the CRIR, in addition to the other gas parameters, is spatially variable with $\zeta \sim$ a few $\times 10^{-14}$\,s$^{-1}$ at $r\gtrsim100$\,pc from the nucleus, increasing to $\zeta >10^{-13}$\,s$^{-1}$ at its center. These inferred CRIRs are consistent within 1 dex with theoretical predictions based on non-thermal emission. Additionally, the high CRIRs estimated in NGC\,253's CMZ can be explained by the large number of cosmic-ray-producing sources as well as a potential suppression of cosmic-ray diffusion near their injection sites.

\end{abstract}


\section{Introduction} \label{sec:intro}

Understanding the physical processes associated with star formation is pivotal for identifying how star formation proceeds in different environments. Star-forming regions vary in many ways, including, but not limited to, their temperatures, densities, turbulence, radiation strength, and molecular complexity \citep[e.g.][]{Downes1998,Heyer2004,Padoan2002,Gong2020}. All of these factors combine to determine the nature and lifetime of star-forming regions and the galaxies that host them \citep[see, for example,][]{Corbelli2017,Chevance2020,Semenov2021,Kim2021}. Starburst galaxies feature some of the most extreme examples of star-forming environments, demonstrated by the high surface brightness of their emission as well as enhanced star formation rates \citep[e.g.][]{Schinnerer2007,Leroy2013,Murphy2015,Brunetti2021,Callanan2021,Eibensteiner2022}. Naturally, we expect that the physical processes occurring in starburst galaxies must differ from those in more quiescent regions, such as in the Milky Way, to account for their increased star formation output.

One important avenue of study is the feedback processes that occur as a result of star formation activity while also affecting the gas and dust that will form future generations of stars. Recently-formed stars produce large quantities of energy in the form of ultraviolet (UV) and X-ray radiation \citep{Hollenbach1999,Lepp1996}, turbulence from stellar winds and gravitational collapse \citep[e.g.][]{Loenen2008}, and ionizing cosmic rays \citep{Acero2009,Papadopoulos2010,bayet11,Padovani2022A&A}. All of these mechanisms contribute energy and pressure to the surrounding medium that can have a marked effect on the molecular gas chemistry \citep{holdship_c2h,Holdship2022,Behrens2022,Huang2023}. However, the relative dominance of each of these feedback mechanisms can vary based on environment \citep[\eg][]{Schinnerer2024arXiv, Cenci2024MNRAS}, so it is important to understand the role of each process in different star-forming contexts.


NGC\,253 is one of the nearest \citep[$d \sim 3.5 \pm 0.2$\,Mpc,][]{Rekola2005} starburst galaxies; thus it has been the subject of many studies investigating its intense and complex star formation \citep[e.g.][]{Sakamoto2011,Rosenberg2013,leroy15,leroy18,mangum19,Krieger2020ApJ,Levy2022ApJ}. Though only featuring a modest total star-formation rate (SFR) of $\sim 5$\,M$_{\odot}$\,yr$^{-1}$, nearly half of the galaxy's star formation is occurring in its central kiloparsec, making NGC\,253's Central Molecular Zone (CMZ) a very active star-forming environment \citep{leroy15}. For comparison, the SFR in NGC\,253's CMZ is larger than the total SFR across the entire Milky Way galaxy \citep[1.65--1.9\,M$_{\odot}$\,yr.$^{-1}$,][]{Chomiuk2011,Licquia2015}. Recent observations and analyses from the ALMA Comprehensive High-resolution Extragalactic Molecular Inventory (ALCHEMI) Large Program \citep{ALCHEMI-ACA} have revealed a wealth of molecular complexity, making it possible to study shocks, heating processes, masers, and molecular distributions at $\sim 28$\,pc resolution \citep{holdship_c2h, harada21, Haasler2022, Holdship2022, Humire2022, Harada2022, Behrens2022, Huang2023,Butterworth2024, Tanaka2024,Bouvier2024,Harada2024ApJS,Bao2024arXiv}.

Previous studies by \cite{holdship_c2h,Holdship2022} and \cite{Behrens2022} have investigated the heating processes associated with the intense star formation occurring in NGC\,253's nucleus. Their results point to high cosmic-ray ionization rates (CRIRs; $\sim 10^{-14}-10^{-12}$\,s$^{-1}$) across the NGC\,253 CMZ, as well as show a spatially-varying CRIR distribution with higher ionization rates in the center that decrease towards the edge of the CMZ. For comparison, \cite{Ravikularaman2024} found a CRIR of $2\times 10^{-14}$\,s$^{-1}$ in the Milky Way CMZ, which is 2--3 orders of magnitude larger than the CRIR in other parts of the Galaxy. \cite{holdship_c2h,Holdship2022} and \cite{Behrens2022} also report that these high CRIRs are the main contributors to the high kinetic temperatures ($\geq 300$\,K) measured by \cite{mangum19}, making them the dominant heating mechanism in the CMZ. Additionally, \cite{Behrens2022} found a positive correlation between the CRIR and the location and number of radio continuum sources (primarily H\,II regions and supernova remnants) from recent star formation. However, due to computational limitations, these studies limited their focus to 5 \citep{Holdship2022} and 10 \citep{Behrens2022} Giant Molecular Cloud (GMC) regions in the CMZ that were identified by \cite{leroy15} based on peaks in the dense gas emission. Thus these investigations were unable to fully spatially sample the CRIR and other physical conditions across the entire CMZ. 

Chemical modeling codes, such as \texttt{UCLCHEM}\footnote{\url{https://uclchem.github.io/}} \citep{Holdship2017}, which was used in \cite{Holdship2022} and \cite{Behrens2022}, are powerful tools for understanding the chemistry of star-forming regions. By providing chemical models with sets of physical conditions; such as kinetic temperature, volume density, and CRIR; we can predict the abundances with respect to total hydrogen of various species within a region of interest. 
For instance, \cite{Behrens2022} found that the abundances of HCN and HNC are greatly influenced by cosmic-ray chemistry in the UV-shielded environments of NGC\,253. While previous studies suggested that HCN and HNC were sensitive to changes in kinetic temperature for $T_\text{K} \lesssim 50$\,K \citep{hacar20,Colzi2018}, \cite{Behrens2022} found that at high cosmic-ray ionization rates, the formation of HCN and HNC is dominated by reactions of HCNH$^+$ with electrons.
Since both HCN and HNC are destroyed at relatively equal rates via reactions with ions, their cosmic-ray-influenced formation is primarily responsible for their relative abundances \citep[for more details on the HCN and HNC chemistry, see][]{Behrens2022}. 

When we combine information derived from these chemical pathways with radiative transfer calculations, we can directly compare model output to measured molecular intensities, thereby allowing us to constrain the gas conditions in specific star-forming regions and make inferences about the physical processes occurring there. However, chemical models are computationally expensive and time-consuming. Previous parameter-inference studies in the NGC\,253 CMZ were able to provide strong constraints on the conditions in only a handful of regions. In order to perform a comprehensive study of the gas conditions across NGC\,253's CMZ, we must construct a method that allows us to thoroughly sample the CMZ's spatially-varying chemistry in a reasonable time frame.

In this paper, we present a neural network that can emulate and adopt the role of a chemical model in our physical parameter-inference algorithm, allowing us to ascertain the gas conditions in a given region ten times faster than with the chemical model alone. This new implementation to our algorithm allows us to infer the physical properties across the entire NGC\,253 CMZ at 50\,pc resolution and will pave the way toward extending this analysis to other galaxies, as well as images with higher resolution or larger fields of view. Ultimately, the ability to constrain molecular gas conditions and make inferences about the physical processes associated with starburst activity will help us better understand the relationship between star formation and galaxy evolution in starburst galaxies.

We present our data and identification of regions of interest in Section \ref{sec:data}. In Section \ref{sec:NN}, we describe our neural network architecture and its implementation into our Bayesian nested sampling algorithm. We summarize our results in Section \ref{sec:results} and discuss the implications of our work on star formation and galaxy evolution in Section \ref{sec:disc}. 

\vspace{-10pt}
\section{OBSERVATIONAL DATA} \label{sec:data}


\begin{deluxetable*}{cccccccccc}
\centering
\tablecolumns{9}
\tablecaption{Example average integrated intensity and RMS values for regions in the outer (region 26) and inner (region 53) CMZ.\tablenotemark{a} \label{tab:obs}}
\tablehead{
 & &  &  & \multicolumn{2}{c}{Region 26} & & \multicolumn{2}{c}{Region 53} \\ 
 \cline{5-6} \cline{8-9}
 & \colhead{Transition} & \colhead{Rest Frequency} & \colhead{$E_{\text{u}}$} &\colhead{$\langle\int S_\nu d\nu \rangle$} &\colhead{RMS} & &\colhead{$\langle\int S_\nu d\nu \rangle$} & \colhead{RMS} \\ 
 &  & (GHz) & (K) & (Jy\,km\,s$^{-1}$ & (Jy\,km\,s$^{-1}$ & & (Jy\,km\,s$^{-1}$ & (Jy\,km\,s$^{-1}$ \\ [-4pt]
 & & &  & beam$^{-1}$) & beam$^{-1}$) & & beam$^{-1}$) & beam$^{-1}$)
}
\startdata
HCN     &  $1-0$  &  88.6316 & 4.25 & 4.7  &  0.7 & &  8.8 &  1.3 \\
        & $2-1$ &   177.2611 & 12.76 & 13.3  &  2.0 & &  33.5 &  5.0  \\
        & $3-2$ &   265.8864 & 25.52 & 13.9 &  2.1 & &  56.1 &   8.4  \\
        & $4-3$ &   354.5055 & 42.53 & 11.0 &  1.7 &  & 68.6 &   10.3  \\ \hline 
HNC     &  $1-0$ &  90.6636 & 4.35 &2.2 &  0.3 &  & 7.9 &  1.2  \\
        & $2-1$ &  181.3248 & 13.05 &2.2 &  0.7 & & 19.9 &   3.0  \\
        & $3-2$ &   271.9811 & 26.11 &4.3 &  0.7 & & 37.9 &  5.7  \\
        & $4-3$ &   362.6303 & 43.51 & 2.4 &  0.4 & & 40.9 &   6.1 \\
\enddata
\tablenotetext{a}{Average integrated intensity and RMS values for all 94 regions are provided in a machine-readable table.}
\end{deluxetable*}

\vspace{-8mm}
\subsection{ALCHEMI Observations} \label{sec:obs}

\begin{figure*}[!h]
    \includegraphics[trim = 0 0 7mm 0, clip=True, scale=0.8]{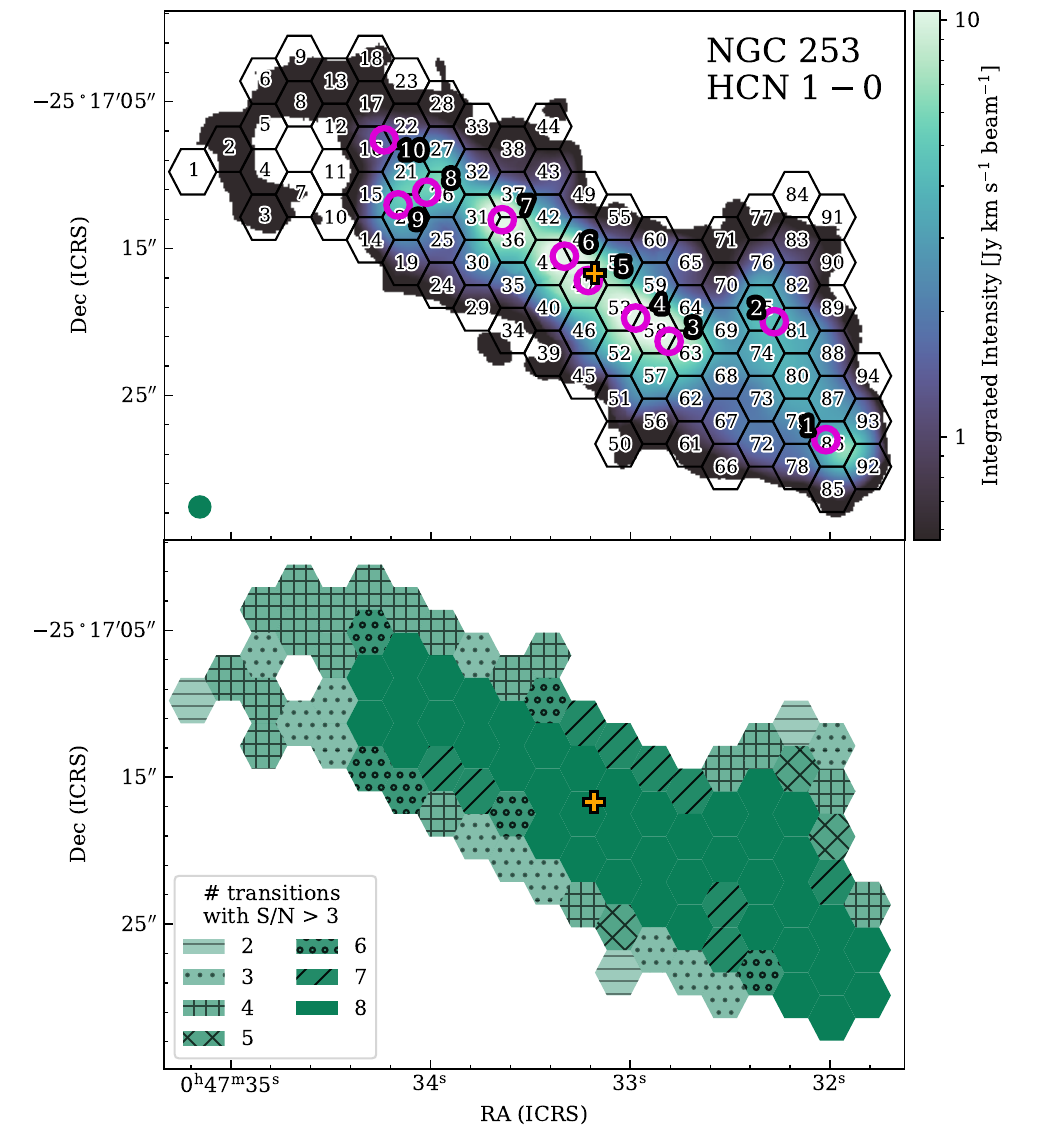}
    \caption{Top: Map of NGC\,253 HCN 1--0 integrated emission with S/N $> 3$ overlaid with the regions analyzed in this article. Purple circles indicate locations of GMCs studied in \cite{Holdship2022} and \cite{Behrens2022} (note \cite{Holdship2022} only investigated GMCs 3---7). The circle in the bottom left corner indicates the size of the 1.\!\!$^{\prime\prime}6$ ($\sim$28\,pc) ALCHEMI beam. Orange crosses in both panels indicate the location of the galaxy's kinematic center \citep{Turner1985}. Bottom: Map of the hexagonal regions used in this analysis with hatching to indicate how many total HCN and HNC transitions have been detected with S/N $>$ 3.}
    \label{fig:num_ref}
\end{figure*}

We use data from the ALMA Large Program ALCHEMI: the ALMA Comprehensive High-resolution Extragalactic Molecular Inventory \citep{ALCHEMI-ACA}, which imaged the NGC\,253 CMZ in ALMA Bands 3--7 (84--373\,GHz) at 1.\!\!$^{\prime\prime}6$ ($\sim28$\,pc) resolution during ALMA Cycles 5 (2017.1.00161.L) and 6 (2018.1.00162.S). Details of the observations and data reduction can be found in \cite{ALCHEMI-ACA} and \cite{Behrens2022}. As noted in \cite{Behrens2022}, we extract the HCN and HNC $J=1-0$, $2-1$, $3-2$, and $4-3$ transition integrated intensities using the \texttt{CubeLineMoment}\footnote{\url{https://github.com/keflavich/cube-line-extractor}} code. The frequencies and upper-state energies for our transitions of interest are shown in Table \ref{tab:obs}.

\subsection{Region Selection and Moment-0 Maps} \label{sec:regions}

In order to sample the entire nuclear region in NGC\,253, we divide the CMZ into adjoining hexagonal regions 50\,pc in height using the integrated emission averaging code \texttt{HERA}\footnote{\url{https://github.com/ebehrens97/HERA}}, the HExagonal Region Averager code. This code divides integrated intensity maps for each transition of interest into either adjacent circular or adjoining hexagonal regions of a user-specified size, averages the integrated emission over the region, and calculates noise statistics associated with the averaged emission. We choose a common region size of 50\,pc, which is the typical size of a GMC structure in the NGC\,253 CMZ \citep{leroy15}. The code-generated regions are pruned such that only regions that meet a specified signal-to-noise (S/N) threshold (in our case, S/N $\geq 3$) are retained. This information is written to a file from which we can extract and analyze it. Figure \ref{fig:num_ref} shows the final 94 regions we selected to study, where each of the hexagonal regions has S/N $\geq 3$ in at least one transition of interest. 

However, unlike the regions studied in \cite{Holdship2022} and \cite{Behrens2022}, not all hexagonal regions contain significant emission for every studied transition (see the bottom panel in Figure \ref{fig:num_ref}). In some cases, a portion of the image pixels within a given hexagon are masked due to low S/N. In other cases, entire hexagonal regions are masked.
Rather than removing regions which do not feature a full set of 8 detected transitions for HCN and HNC or, alternatively, constraining models of those regions with fewer than 8 transitions, we fill in low S/N image pixels with $3\sigma$ upper-limit integrated intensities. These limits allow us to constrain our models with an equal number of transitions per region while also testing the efficacy of this algorithm in regions with low levels of emission. We derive these $3\sigma$ limits using our single-channel RMS $\sigma_\text{noise}$, flux-calibration uncertainty $\sigma_\text{fluxcal}$, and an assumed linewidth $\Delta v$. Based on an inspection of the range of linewidths seen in the CMZ for HCN and HNC near the edge of our S/N threshold, we assume a random $\Delta v$ between 20 and 100\,km\,s$^{-1}$ for each low S/N image pixel.
We integrate over a Gaussian with width $\Delta v / \sqrt{8 \ln 2}$ and a peak value equal to 3 times the noise value, or $3 \times \sqrt{\sigma_\text{noise}^2 + \sigma_\text{fluxcal}^2}$, where $\sigma_\text{fluxcal}$ is equal to 15\% of the measured flux. Using these upper-limit values in place of masked pixels, we then obtain a unique average integrated intensity and RMS for each hexagonal region, which we use to constrain our physical and chemical models, as discussed in Section \ref{sec:NN}. Average integrated intensity measurements and RMS values for example outer and inner CMZ regions (regions 26 and 53, respectively) are shown in Table \ref{tab:obs}. Average integrated intensities and the associated RMS values for each of the 94 regions is provided as a machine-readable table.

\begin{figure*}
    \centering
    \includegraphics[scale=0.6, trim = 5mm 0mm 5mm 0mm]{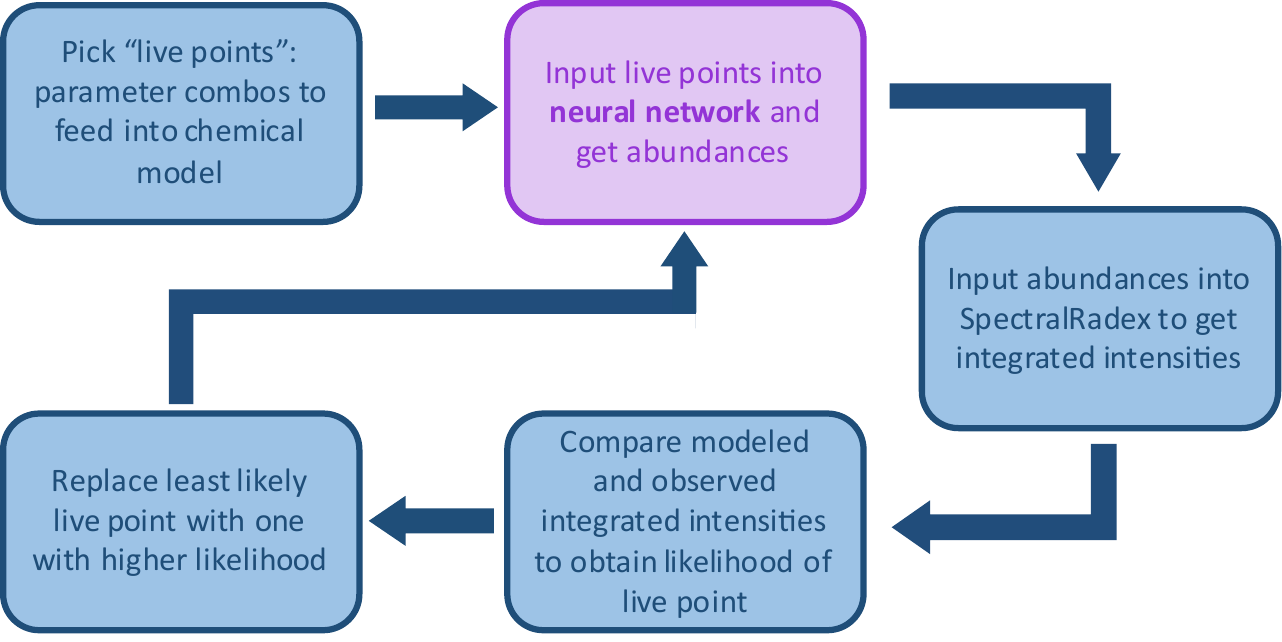}
    \caption{Flow chart describing our Bayesian nested sampling algorithm, where the highlighted purple rectangle indicates the neural network's position in our modeling process.}
    \label{fig:flow_chart}
\end{figure*}

\section{Bayesian Inference with Neural Networks} \label{sec:NN}

To constrain the gas properties in each of our designated regions, we consider the chemical pathways and molecular emission resulting from a given set of input gas conditions. Previous studies \citep{Holdship2022,Behrens2022,Huang2023} used the chemical modeling code  \texttt{UCLCHEM}\footnote{\url{https://uclchem.github.io/}} \citep{Holdship2017} to calculate molecular abundances given a set of input conditions. Using the radiative transfer code \texttt{SpectralRadex}\footnote{\url{https://spectralradex.readthedocs.io/en/latest/}}, these studies associated modeled abundances, in combination with the input gas conditions, with modeled integrated intensities for each transition of interest, which could be directly compared to observations. The aforementioned studies successfully demonstrated the validity of these methods by estimating certain gas properties, such as volume density and CRIR, for a small number of regions \citep[5--12, from][]{leroy15,Huang2023} in the NGC\,253 CMZ. However, the use of a chemical modeling code significantly prolongs this process, 
making this strategy untenable for modeling a larger number of regions. We address this issue by developing a neural network model that can replace and fulfill the role of a chemical model in our Bayesian inference algorithm (see Figure \ref{fig:flow_chart}). Further details regarding our Bayesian inference analysis are described in \cite{Holdship2022} and \cite{Behrens2022}.

\subsection{Neural Network Architecture} \label{sec:architecture}

We use the python package \texttt{TensorFlow}\footnote{\url{https://www.tensorflow.org/}} to build and train our neural network model. We find that a neural network with fully connected layers and 12,020,002 trainable parameters produces a model that best reproduces the \texttt{UCLCHEM} abundances (note that these trainable parameters---neural network weights and biases, see Appendix \ref{sec:NN_appendix}---are separate from the physical parameters listed in Table \ref{tab:priors}). We start with an input layer of four nodes, where each node corresponds to one of our chemical model parameters\footnote{Note that the beam-filling factor is not an input for the neural network, as it is applied after the integrated intensities are calculated with \texttt{SpectralRadex} (see Section \ref{sec:nested_samp}).} (kinetic temperature $T_K$, H$_2$ volume density $n$, CRIR $\zeta$, and H$_2$ column density $N$). After values from the input layer are passed through the hidden layers, the values calculated in the neural network's output layer represent the molecular abundance predictions of that epoch for each physical parameter combination in the training set. An epoch is defined as the time it takes for the neural network to be trained on the entire training set one time. See Appendix \ref{sec:NN_appendix} for more details on the neural network training process. 

Periodically throughout each training epoch, our training algorithm compares the node values for a given layer to the actual values provided by the training data using a loss function. We use a mean squared error loss function, and the calculated loss is used to tweak the neural network in order to improve its performance in future epochs. At the end of each epoch, the loss for a separate datatset, the validation data, is calculated (see Section \ref{sec:train_set} for a breakdown of the training and validation sets). The neural network does not ``see" the validation data until the end of an epoch, and thus the validation set is used to assess the neural network's efficacy so the model can be improved for the next round of training. This process continues over several epochs until the neural network reaches its specified stopping criterion. In this case, we cease neural network training once the calculated loss has not improved over 20 consecutive epochs, and the neural network then reverts back to using the weights from the epoch with the lowest validation loss. We employ this criterion to avoid overfitting, which occurs when the model simply memorizes the abundances associated with each parameter combination without learning how to predict abundances from new, unseen data.

\vspace{-1mm}
\subsection{Training Dataset}
\label{sec:train_set}

\begin{deluxetable*}{ccccc}
\centering
\tablecolumns{3}
\tablewidth{0pt}
\tablecaption{Prior Distributions and Training Data \label{tab:priors}}
\tablehead{
\colhead{} &  \colhead{Parameter} & \colhead{Range (Buffer Range)} & Distribution Type & \# Points (\# with Buffer)
}
\startdata
$T_\text{K}$  & Temperature (K) & 50--300 & Linear & 15\\ [5pt]
$n_{\text{H}_2}$  & Volume Density (cm$^{-3}$) & 10$^3$--10$^7$ $(10^{2.6}$--$10^{7.4})$ & Log & 20 (24)\\ [5pt]
$\zeta$ & Cosmic Ray Ionization Rate ($\zeta_{0}$\tablenotemark{a})& 10--10$^7$ & Log & 25\\[5pt]
$N_{\text{H}_2}$ & H$_{2}$ Column Density (cm$^{-2}$) & 10$^{22}$--$10^{25}$ $(10^{21.6}$--$10^{25.4})$ & Log & 15 (19)\\[5pt]
$\eta_{ff}$ & Beam-filling factor  & 0--1 & Linear & N/A\tablenotemark{b} \\ [5pt]
\enddata
\tablenotetext{a}{$\zeta_0$ = $1.36\times10^{-17}$\,s$^{-1}$}
\tablenotetext{b}{$\eta_{ff}$ is not a neural network parameter, as it is applied after the intensities are calculated by the neural network and \texttt{SpectralRadex}.}
\end{deluxetable*}

\vspace{-8mm}
We train one neural network model to predict the abundance of both HCN and HNC since previous ALCHEMI studies have shown that these molecules have similar spatial distributions \citep[see][]{Behrens2022}. We derive our training dataset from a grid of \texttt{UCLCHEM} models for HCN and HNC using input parameters that span our parameter space (see Table \ref{tab:priors}). We use a two-phase \texttt{UCLCHEM} model structure\footnote{Note that this two phase approach differs from that used in \cite{Behrens2022}, which only included the second phase of this modeling process. Further discussion on this subject can be found in Section \ref{sec:meas_compare}.} that takes only the assumed volume density as input in phase 1 and models the free-fall collapse of a gas cloud with an initial density of 100\,cm$^{-3}$ down to the specified density at a fixed temperature of 10\,K, transforming the gas from atomic/ionic to molecular form. This stage is followed by phase 2, which uses the calculated abundances inherited from phase 1 and four of our five input parameters ($T_\text{K}$, $n_{\text{H}_2}$, $\zeta$, and $N_{\text{H}_2}$) to chemically model the cloud itself over a timescale of 6\,Myr. The output from phase 2 makes up the training data for our neural network. We note that the abundances used at the beginning of stage 2 may be sensitive to the initial conditions (e.g. UV field, elemental abundances, mode of collapse)--varying these parameters would require reconstructing the training datset, which is not a case considered here. Here we assume that we are probing the inner, shielded part of the cloud (visual extinction $A_v$ of 10 mag) such that UV radiation (set to 1\,Habing in both \texttt{UCLCHEM phases}) is negligible. X-ray ionization is not included in \texttt{UCLCHEM}, but other studies have shown that the ionization contribution from X-rays in the regions we study is likely several orders of magnitude lower than the $\zeta$ values we infer \citep[$\zeta_{x-ray} \lesssim 10^{-17}$\,sec$^{-1}$ for N(H$_2$) $\gtrsim 10^{23}$\,cm$^{-2}$; ][]{Phan2024}, thus making X-ray ionization a negligible contributor to the heating budget within the NGC\,253 CMZ. See \cite{Holdship2017} for more details on \texttt{UCLCHEM} and its capabilities. Table \ref{tab:priors} shows the parameter space over which we run these models and also indicates the types of prior distributions we assume for each parameter, as well as the number of values we sample for the training set for each parameter within our desired range. Including buffer points, which will be discussed later, each initial dataset contains $\sim$170,000 parameter combinations. 

Any parameter combinations that result in very low abundances are likely to be dominated by numerical error rather than chemistry \citep{heyl2023}. Hence  
we retain only parameter combinations that result in abundances for both HCN and HNC above $10^{-12}$ (a conservative value corresponding to typically the lowest observed molecular abundances), leaving $\sim$115,000 data points. 

We divide the \texttt{UCLCHEM} output into a training set (60\% of data, $\sim$69,000 points), validation set (50\% of data, $\sim$17,000 points),  and a test set (25\% of data, $\sim$28,000 points). We use the parameter values in our training set as input for the neural network's first layer (using the log of the parameter value for $n_{\text{H}_2}$, $\zeta$, and $N_{\text{H}_2}$), and these values are then propagated through the neural network (see Appendix \ref{sec:NN_appendix} for more details), ultimately resulting in molecular abundance predictions for both HCN and HNC. We use the log of the \texttt{UCLCHEM} fractional abundance values when training the neural network in order to increase the magnitude of differences between abundances. We calculate the mean squared error loss between the neural network result and the training data continuously throughout each epoch, but we calculate the validation loss, or loss derived from the validation data, only once at the end of each epoch (see Figure \ref{fig:loss}). Once we reach our stopping criterion (see Section \ref{sec:architecture}), we test the newly-trained neural network model on the test set, a subset of the \texttt{UCLCHEM} grid that was not included in the training process. We then compare the model's predicted abundances to the actual abundances produced by \texttt{UCLCHEM} in order to assess its accuracy.

\subsection{Training Results}
\label{train_results}

\begin{figure*}
\centering
\gridline{\leftfig{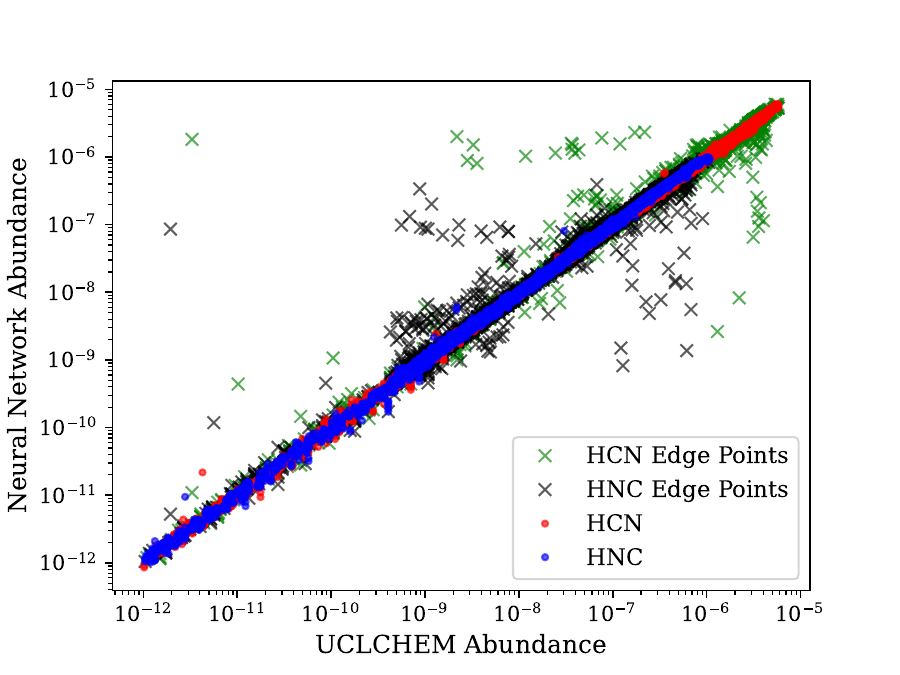}{0.5\textwidth}{(a)} \rightfig{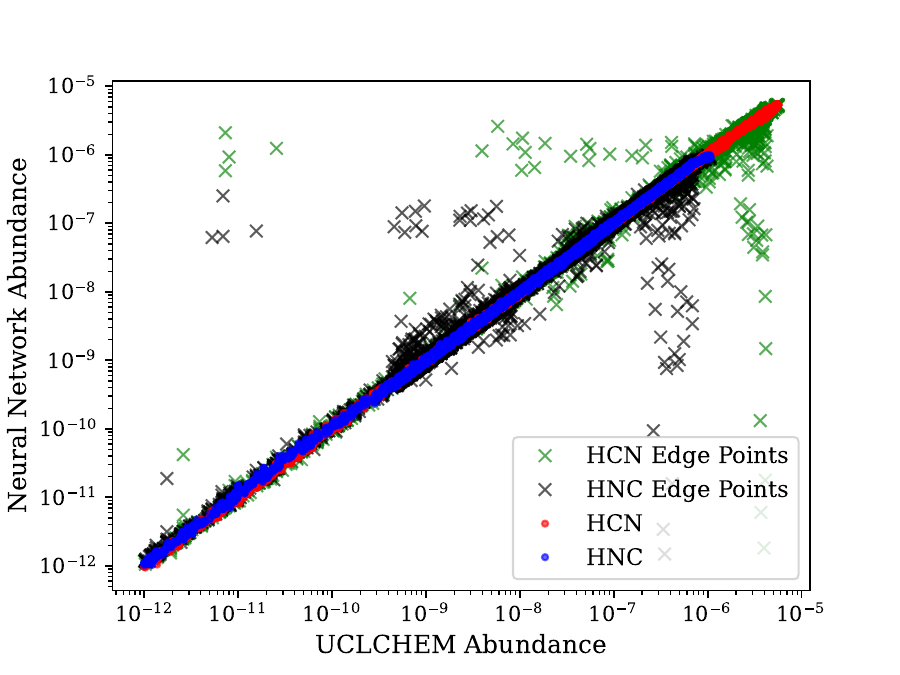}{0.5\textwidth}{(b)}}
\gridline{\fig{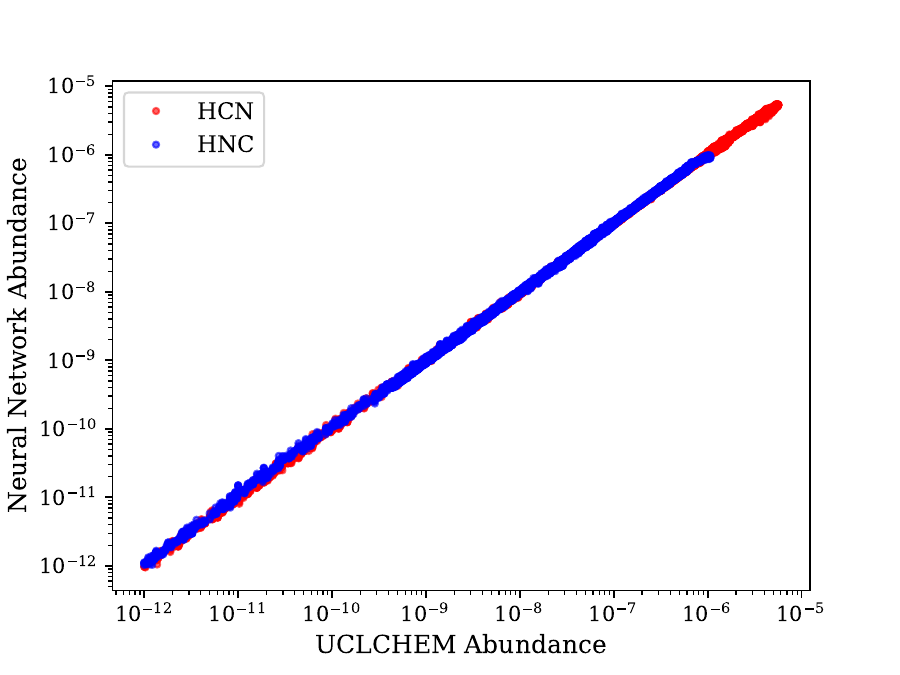}{0.5\textwidth}{(c)}}
\caption{(a) Neural network predictions of test set parameter combinations where the model was trained using the original parameter ranges in Table \ref{tab:priors} without buffer points. Predictions for parameter combinations that used maximum or minimum values for column or volume density are indicated with X's. (b) Same as (a), but neural network training set included additional buffer points for volume and column density.(c) Neural network predictions using the same model as in (b), but buffer points are not plotted.}
\label{fig:NN_tests}
\end{figure*}

Figure \ref{fig:NN_tests} shows the results from our neural network training process, including the results for the best model (panel C) which we later use in our Bayesian inference algorithm. We compare the neural network model-predicted values against those calculated with \texttt{UCLCHEM} and find that, on average, the neural network model is able to recover molecular abundances within 3\% of the chemical model's values. 

However, Figure \ref{fig:NN_tests}a shows there are outliers, making up $\sim1$\% of the test set where the neural network was unable to recover an abundance within one order of magnitude of the \texttt{UCLCHEM} value (green and black Xs in Figure \ref{fig:NN_tests}a). Initial testing revealed that the neural network was unable to predict abundances for parameter combinations that included the minimum and maximum column and volume density values that we probed ($10^{22}$\,cm$^{-2}$, $10^{25}$\,cm$^{-2}$ and $10^{3}$\,cm$^{-3}$, 10$^{7}$\,cm$^{-3}$, respectively). Excluding the edge points, all other column and volume density values within our parameter ranges are padded above and below by values of that parameter that the neural network can use to predict an abundance. However, values on the edge of our parameter space are not padded by additional parameter values on each side, resulting in poorer abundance predictions for parameter combinations that contain these edge points. To ensure that the neural network is equally trained and sensitive to all values that fall within our desired parameter space, we include additional column and volume density buffer points outside of our desired parameter ranges (see Table \ref{tab:priors}) when training our neural network model, though values outside of this original range will not be considered later in our Bayesian analysis. We found no evidence that the neural network was poorly predicting abundances for other parameter combinations including temperature and CRIR values on the edges of our parameter space, so we use the original parameter ranges for $T_K$ and $\zeta$ that are listed in Table \ref{tab:priors}.

Figure \ref{fig:NN_tests}b shows the predictions of our neural network model when including the buffer points shown in Table \ref{tab:priors} in our training set. These buffer points become the new edge points (green and black Xs in Figure \ref{fig:NN_tests}b) in this expanded parameter space and are thus subject to the same scatter as the edge points in Figure \ref{fig:NN_tests}a. However, when plotting the results of this same model but removing these buffer points (Figure \ref{fig:NN_tests}c), we see that the abundance predictions from parameters with values in our original parameter space have significantly less scatter. By including buffer points while training our neural network model, we ensure that our model can robustly predict abundances across our desired parameter space.

\subsection{Implementation with Bayesian Nested Sampling} \label{sec:nested_samp}

We implement our fully-trained neural network model for HCN and HNC as a replacement for the chemical model's role in our inference procedure. The original implementations of this algorithm \citep[see][]{Holdship2022,Behrens2022} used the nested sampling code \texttt{UltraNest}\footnote{\url{https://johannesbuchner.github.io/UltraNest/}} \citep{ultranest14,ultranest19,ultranest21} to feed parameter combinations into our chemical model, \texttt{UCLCHEM}, which would calculate and return abundance estimates for our molecules of interest. In our updated algorithm, our neural network model fills this role instead (see Figure \ref{fig:flow_chart}). The HCN and HNC model predicts abundances that we then feed into our radiative transfer code, \texttt{SpectralRadex}, in order to obtain integrated intensities for these molecular species. Note that \texttt{SpectralRadex} does a radiative transfer calculation which accounts for optical depth. We use collisional excitation rates from \cite{HernandezVera2017}, who calculated rates for HCN and HNC collisions with ortho- and para-H$_2$ \cite[see Appendix \ref{sec:coll_rates} for a detailed analysis of these new rates as compared to those used in][]{Behrens2022}. We adopt a  3:1 ortho:para hydrogen ratio. After the integrated intensities are calculated, we apply a beam-filling factor, which is a free parameter in our algorithm ranging from 0--1. We then compare these modeled integrated intensities $F_{t}$, which are calculated for some set of parameters $\theta$, to our observed intensities $F_{d}$ and their associated uncertainties $\sigma_{F}$ using the following log-likelihood function:
\begin{equation}
    P(\mathbf{F_d}|\boldsymbol{\theta}) = \exp\left(-\frac{\displaystyle 1}{\displaystyle 2} \sum\limits_{i}^{N} \frac{(F_{d,i} - F_{t,i})^2}{\sigma^2_{F,i}}\right)
    \label{eq:lik_fn_sum},
\end{equation}
where we sum the values for each transition $i$. 

\texttt{UltraNest} uses the results from each log-likelihood comparison to choose additional parameter combinations to sample within our designated parameter space (see Figure \ref{fig:flow_chart} and Table \ref{tab:priors}). This process concludes when \texttt{UltraNest} has sampled the majority of the probability density.

\begin{figure*}
    \gridline{\leftfig{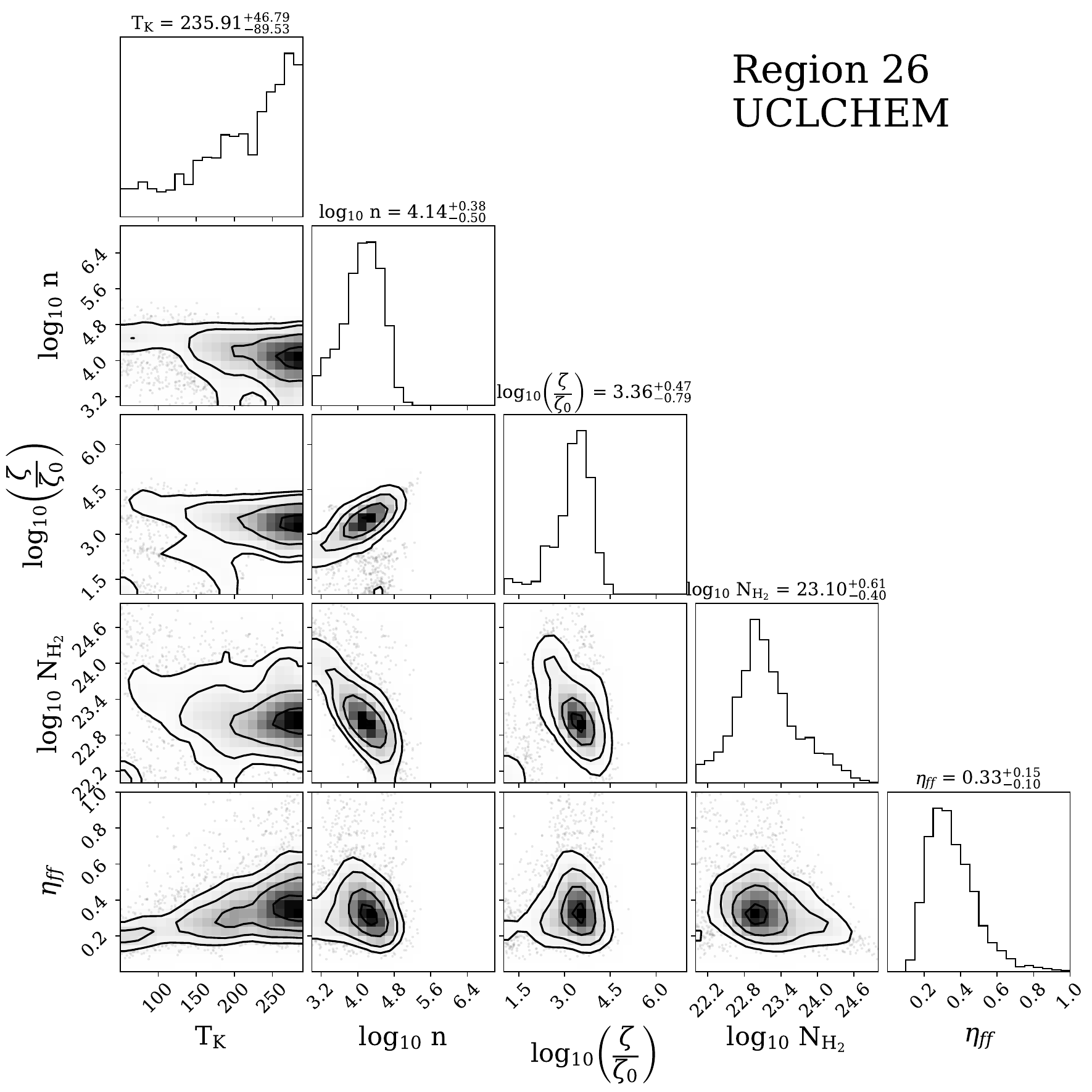}{0.5\textwidth}{(a)}
    \rightfig{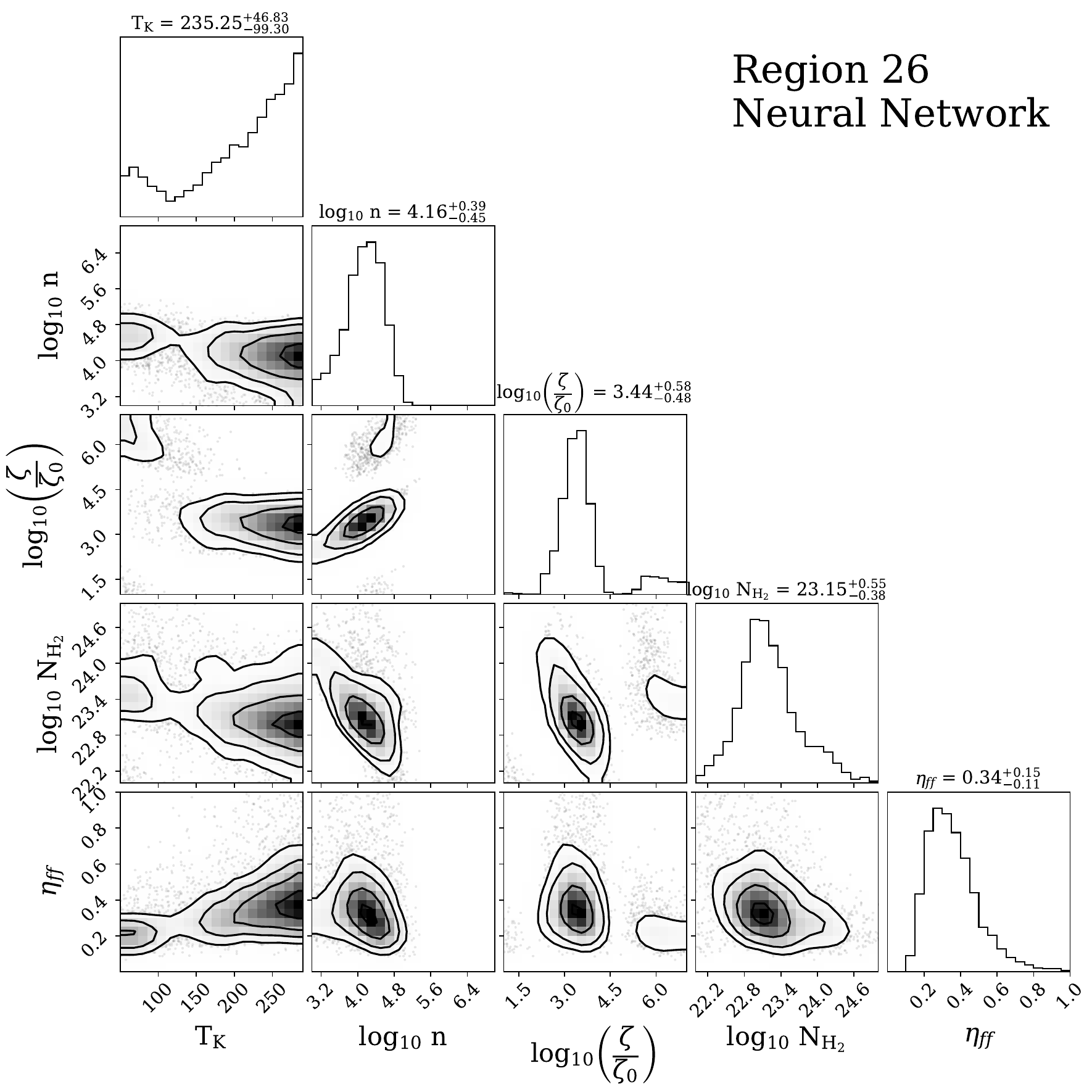}{0.5\textwidth}{(b)}}
    \gridline{\leftfig{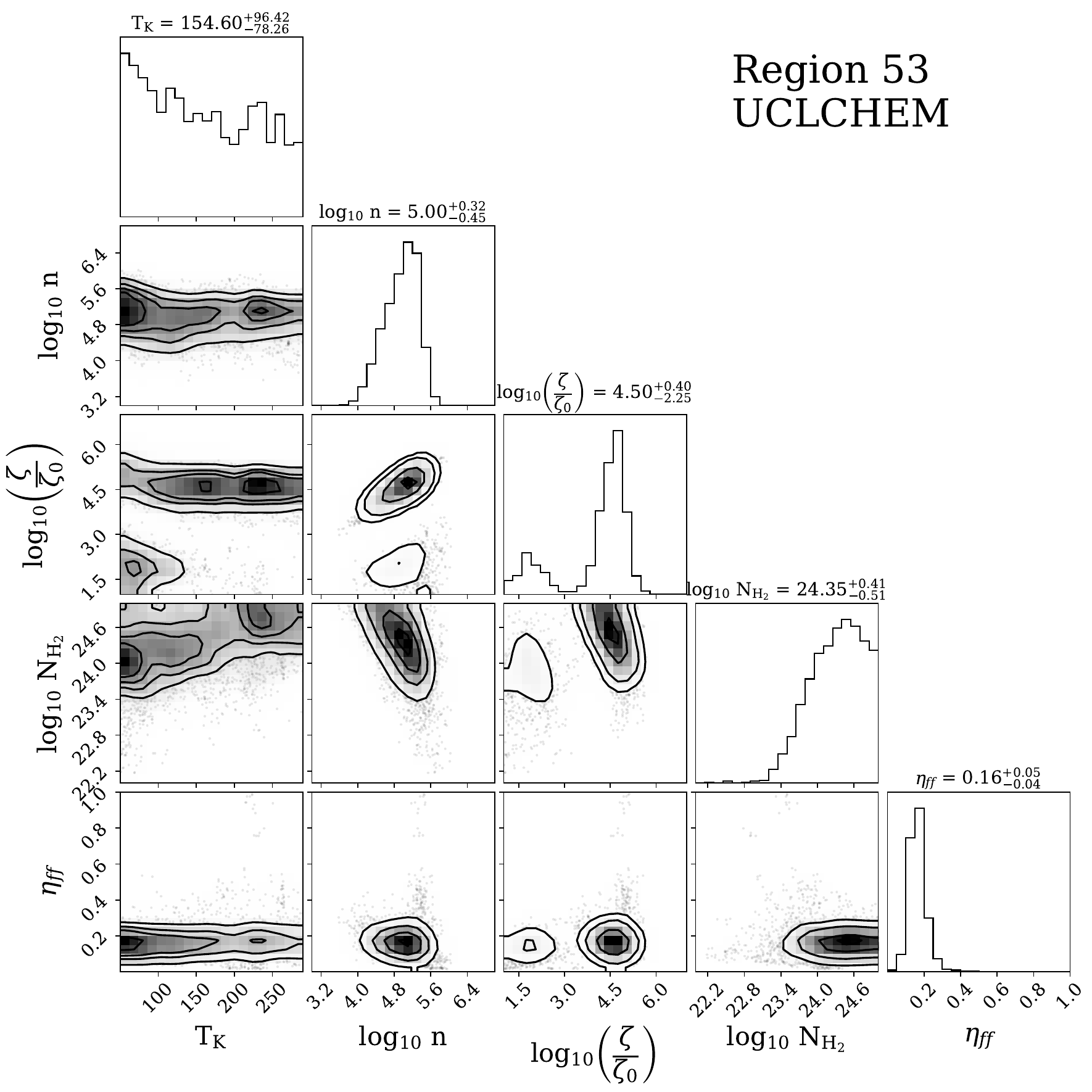}{0.5\textwidth}{(c)}
    \rightfig{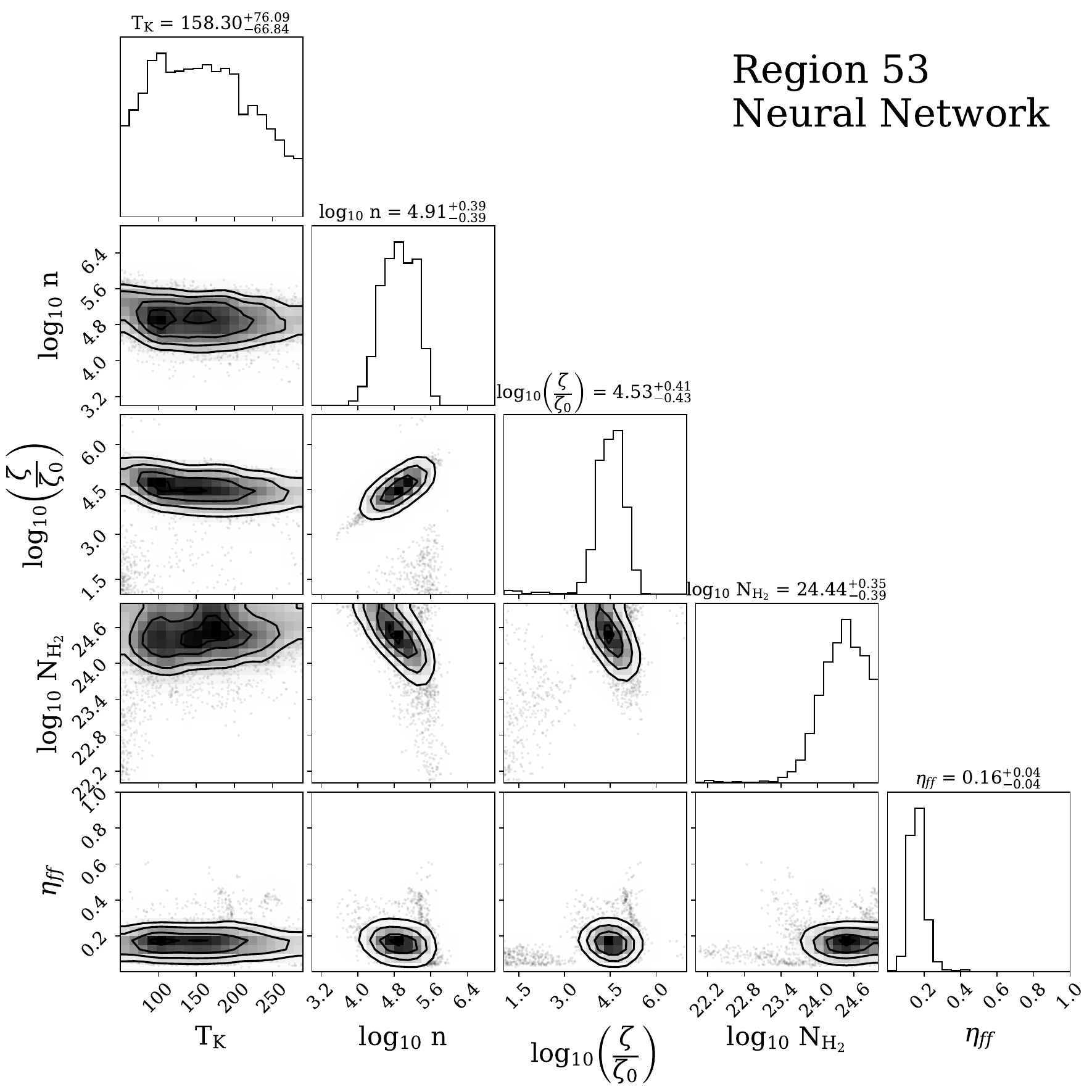}{0.5\textwidth}{(d)}}

    \caption{\textbf{(a)} Corner plot for region 26 using \texttt{UCLCHEM}. \textbf{(b)} Corner plot for region 26 using the HCN/HNC neural network. \textbf{(c)} Corner plot for region 53 using \texttt{UCLCHEM}. \textbf{(d)} Corner plot for region 53 using the HCN/HNC neural network.}
    \label{fig:hcnhnc_uclchem_comparison}
\end{figure*}

\begin{figure*}
    \includegraphics[scale=0.66, trim=4mm 6mm 0mm 0mm, clip=True]{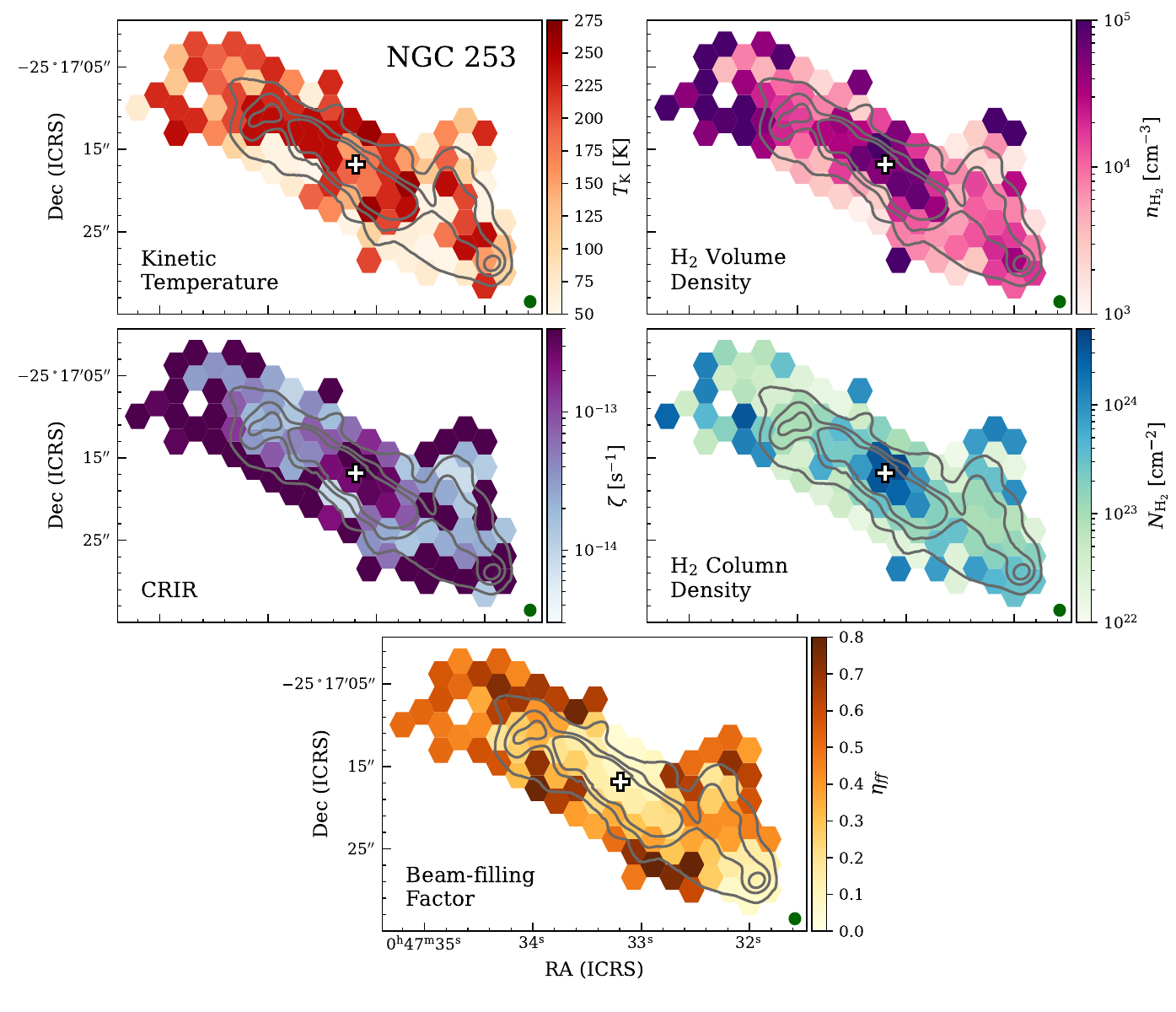}
    \caption{Parameter inference results from our Bayesian nested sampling plus neural network algorithm for kinetic temperature (upper left), H$_{\textrm{2}}$ volume density (upper right), CRIR (middle left), H$_{\textrm{2}}$ column density (middle right), and beam-filling factor (bottom). Gray contours correspond the the integrated HCN $1-0$ emission at levels of 1, 3, and 5 Jy\,km\,s$^{-1}$\,beam$^{-1}$. White and black crosses indicate the galaxy's kinematic center \citep{Turner1985}. Green circle in the bottom right corner of each map represents ALCHEMI's 1.\!\!$^{\prime\prime}6$ ($\sim$28\,pc) beam.}
    \label{fig:allParam_hexmap}
    
\end{figure*}

\begin{figure*}
    \includegraphics[scale=0.67]{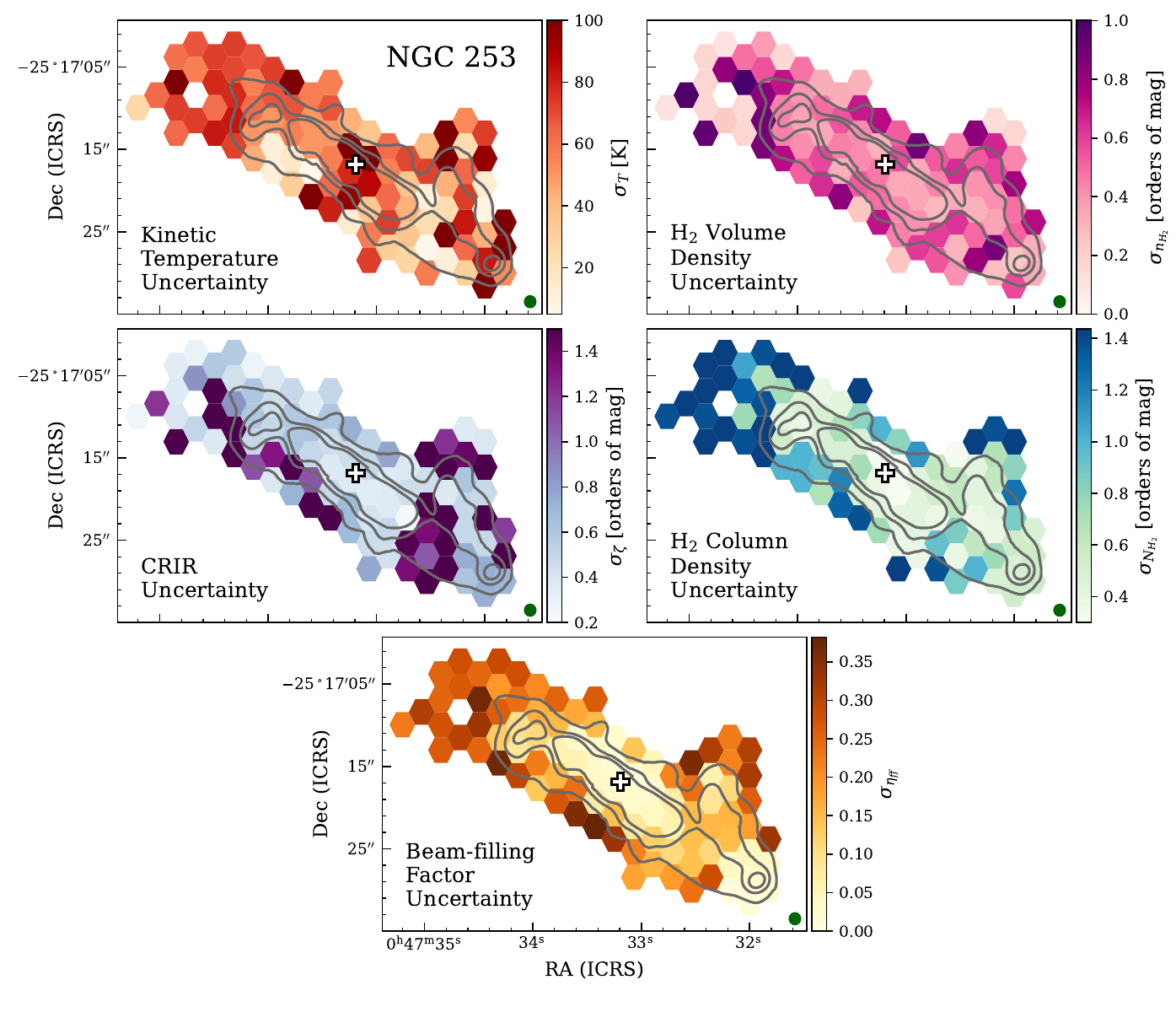}
    \caption{Uncertainties on our five inferred parameters. For the parameters over which we consider a log-uniform distribution ($n_{\text{H}_2}$, $\zeta$, and $N_{\text{H}_2}$), we show their uncertainties as orders of magnitude from the inferred value, as shown above the corner plot 1-D distributions in Figure \ref{fig:hcnhnc_uclchem_comparison}.}
    \label{fig:errMaps}
\end{figure*}

\vspace{5mm}
\section{Results} \label{sec:results}

\subsection{Neural Network vs. Chemical Model}
We compare the nested sampling results using our neural network model to those obtained with \texttt{UCLCHEM} to confirm that the neural network can reliably replace the role of a chemical model in our nested sampling algorithm. Figure \ref{fig:hcnhnc_uclchem_comparison} compares the inference algorithm results using \texttt{UCLCHEM} and the HCN/HNC neural network for regions 26 and 53 (see Figure \ref{fig:num_ref} for region numbers), which were chosen to represent regions in the outer and inner CMZ, respectively. The neural network clearly replicates the results we obtain using \texttt{UCLCHEM}. The final inferred parameter values and their uncertainties derived from \texttt{UCLCHEM} and our neural network are nearly identical. Additionally, both the 1D and 2D posterior distributions for the two methods feature very similar shapes and characteristics, demonstrating that the neural network can reliably reproduce \texttt{UCLCHEM} results in this context in a fraction of the time. We note that the 2D posteriors generated when including the neural network model are somewhat smoother than those produced when our algorithm includes \texttt{UCLCHEM}. This difference may result from the fact that the neural network models are simply an approximation of the \texttt{UCLCHEM} calculations. This approximation may smooth over the functional relationships between model parameters and output abundances and could be further related to the spacing between parameter points in our training grid. Further testing is required to determine whether a more finely-spaced grid would more closely resemble the structure seen in the 2D \texttt{UCLCHEM} posterior distributions. Nevertheless, the parameter values inferred by our neural network algorithm are well within the uncertainties obtained from the same analysis with \texttt{UCLCHEM}. The neural network$+$\texttt{UltraNest} algorithm completed its explorations of the parameter space in $\sim7$\% and $\sim15$\% of the time it took for \texttt{UCLCHEM}$+$\texttt{UltraNest} to do the same for regions 26 and 53, respectively.

\subsection{Parameter Maps} \label{sec:results_paramMaps}

We show maps of the inferred parameter values derived from our combined Bayesian nested sampling and neural network algorithm in Figure \ref{fig:allParam_hexmap} and their associated uncertainties in Figure \ref{fig:errMaps}. In general, we see notable differences in the inferred parameter values in the central 100\,pc versus the rest of the CMZ. H$_{2}$ volume and column density are enhanced by one to two orders of magnitude in the nucleus of the CMZ with values reaching as high as 8$\times 10^{4}$\,cm$^{-3}$ and 5$\times10^{24}$\,cm$^{-2}$, respectively, compared to outer CMZ values of $\sim$5-6$\times10^{3}$\,cm$^{-3}$ and $\sim$3-4$\times10^{22}$\,cm$^{-2}$. We will discuss comparisons of our measurements with those from other NGC\,253 studies in Section \ref{sec:others}.

The CRIR is also an order of magnitude larger in the central $\sim10^{\prime\prime}$ ($\sim4\times10^{-13}$\,s$^{-1}$) than in regions $\gtrsim15^{\prime\prime}$ away from the nucleus ($\sim 2 \times 10^{-14}$\,s$^{-1}$). However, the inferred CRIR map also indicates a ring of high-CRIR ($\gtrsim5\times10^{-13}$\,s$^{-1}$) regions exist around the edge of the CMZ. We do not believe these parameter estimates to be physically motivated and will discuss this issue in more detail in Section \ref{sec:high-zeta}. Comparison to other CRIR estimates, both observational and theoretical, will be addressed in Sections \ref{sec:meas_compare} and \ref{sec:theory_compare}.

\begin{deluxetable*}{cccccc}
\centering
\tablecolumns{6}
\tablecaption{Comparison of NGC\,253 volume and column density with the literature. \label{tab:dens_estimates}}
\tablehead{
 & \multicolumn{2}{c}{$r \lesssim 100$\,pc} & & \multicolumn{2}{c}{$r \gtrsim 100$\,pc} \\ 
 \cline{2-3} \cline{5-6} 
\colhead{ALCHEMI Study} & \colhead{$\log n_{\text{H}_2}$} & \colhead{$\log N_{\text{H}_2}$} & & \colhead{$\log n_{\text{H}_2}$} & \colhead{$\log N_{\text{H}_2}$} \\ [-4pt]
 &  (cm$^{-3}$) & (cm$^{-2}$) & &(cm$^{-3}$) & (cm$^{-2}$)
}
\startdata
\cite{Holdship2022} & 5.3 & 24.3 & & 4.5 & 22.6 \\
\cite{Behrens2022} & 5.0 & 24.0 & & $3.8-4.7$ & 23.0 \\
\cite{Tanaka2024} & $5.0-5.4$ & 23.7 & & $3.8 - 4.5$ & 23.0\\
This work & $4.7-5.0$ & $23.8-24.7$ & & $3.7-4.3$ & $22.5-23.5$\\
\enddata
\end{deluxetable*}

\vspace{-8mm}
Figure \ref{fig:allParam_hexmap} also shows a decrease in the beam-filling factor in NGC\,253's nucleus, which is consistent with the gas emission originating from smaller, higher-density regions and is supported by our estimation of higher volume and column densities in these regions. Finally, we see very little spatial correlation between the kinetic temperature and our other fitted parameters, with no consistent increase or decrease in temperature in the central versus outer CMZ regions. In general, we do not constrain the kinetic temperature well with our HCN and HNC measurements (see, for example, the kinetic temperature panels in Figures \ref{fig:hcnhnc_uclchem_comparison}, \ref{fig:allParam_hexmap}, and \ref{fig:errMaps}), which is consistent with the results from \cite{Behrens2022}. This finding is reinforced by the kinetic temperature uncertainty map in Figure \ref{fig:errMaps}. While the rest of our parameter maps show uncertainties are lowest in the center of the CMZ, where we have the highest S/N, the kinetic temperature uncertainty map demonstrates no such correlation.

\section{Discussion} \label{sec:disc}
\subsection{Column Density, Volume Density, Filling Factor, and Kinetic Temperature}
\label{sec:others}

In Section~\ref{sec:results_paramMaps} we found that there are significant gradients in the inferred values of hydrogen column density $N_{\text{H}_2}$ and volume density $n_{\text{H}_2}$ (Figures~\ref{fig:allParam_hexmap} and \ref{fig:errMaps}). Estimates for volume and column density from this work and other ALCHEMI studies are shown in Table \ref{tab:dens_estimates} for the inner ($r\lesssim100$\,pc) and outer ($r\gtrsim100$\,pc) CMZ. Note that with the exception of this work, all of the studies listed in Table \ref{tab:dens_estimates} estimated the volume and column densities on scales of $\sim28$\,pc, equivalent to the ALCHEMI beam, whereas we estimate these parameters on scales of 50\,pc. Therefore, our estimates of the volume and column density are largely consistent with the lower end of the ranges provided for these other studies. This slight discrepancy likely results from averaging these parameters over larger areas which may include more diffuse gas components, especially in the outer CMZ.

We also see a clear distribution across the CMZ in the dense gas beam-filling factor $\eta_{ff}$, with the lowest values found in the center. In the inner CMZ ($r\lesssim100$\,pc), $\eta_{ff} \lesssim 0.1$, while in the outer CMZ ($r \gtrsim 100$\,pc), $\eta_{ff} \gtrsim 0.3$. The gradients in $n_{\text{H}_2}$, $N_{\text{H}_2}$, and $\eta_{ff}$ are consistent with gas originating from smaller, higher-density and column density regions as one moves from the outer portions of the CMZ toward the nucleus of NGC\,253. Kinetic temperature, on the other hand, is poorly constrained by our models, exposing no systematic changes as a function of position in the CMZ.

The centrally-condensed nature of the NGC\,253 CMZ may reflect the process that catalyzed the starburst in this galaxy. Mergers are a common mechanism for generating starbursts in galaxies \citep{Armus1987,Renaud2022}.  As NGC\,253 shows no signs of having participated in a merger, the trigger for the burst of star formation in the NGC\,253 CMZ must be associated with its high concentration of dense gas.  Starburst galaxy evolution models can reproduce such a process.  \cite{Cenci2024MNRAS} have modeled starburst galaxy evolution using the \texttt{FIREbox} simulation \citep{Feldmann2023MNRAS}.  They find that global gravitational instabilities that drive central gas compaction are the main catalyst for starbursts in galaxies, independent of a galaxy's merger status.

\begin{figure*}
    \gridline{\leftfig{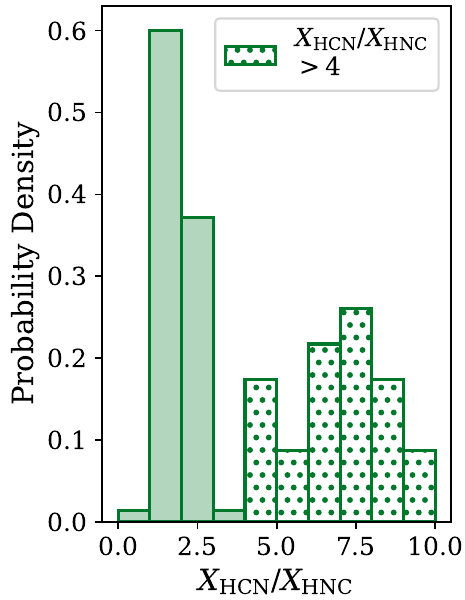}{0.315\textwidth}{(a)}
    \rightfig{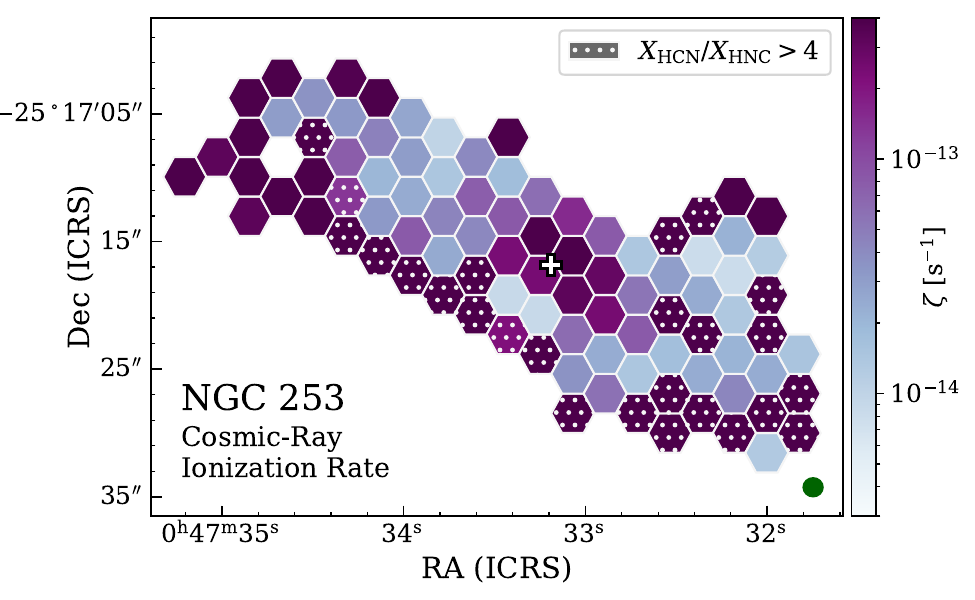}{0.68\textwidth}{(b)}}
    \caption{Probability density for HCN/HNC abundance ratio \textbf{(a)} and cosmic-ray ionization rate map \textbf{(b)} with dotted hatching in both plots corresponding to  $X_\text{HCN}/X_\text{HNC} > 4$.}
    \label{fig:abund_ratio}
\end{figure*}

\begin{figure*}   
    \centering
    \includegraphics[scale=0.7, trim = 3mm 3mm 0mm 6mm, clip=True]{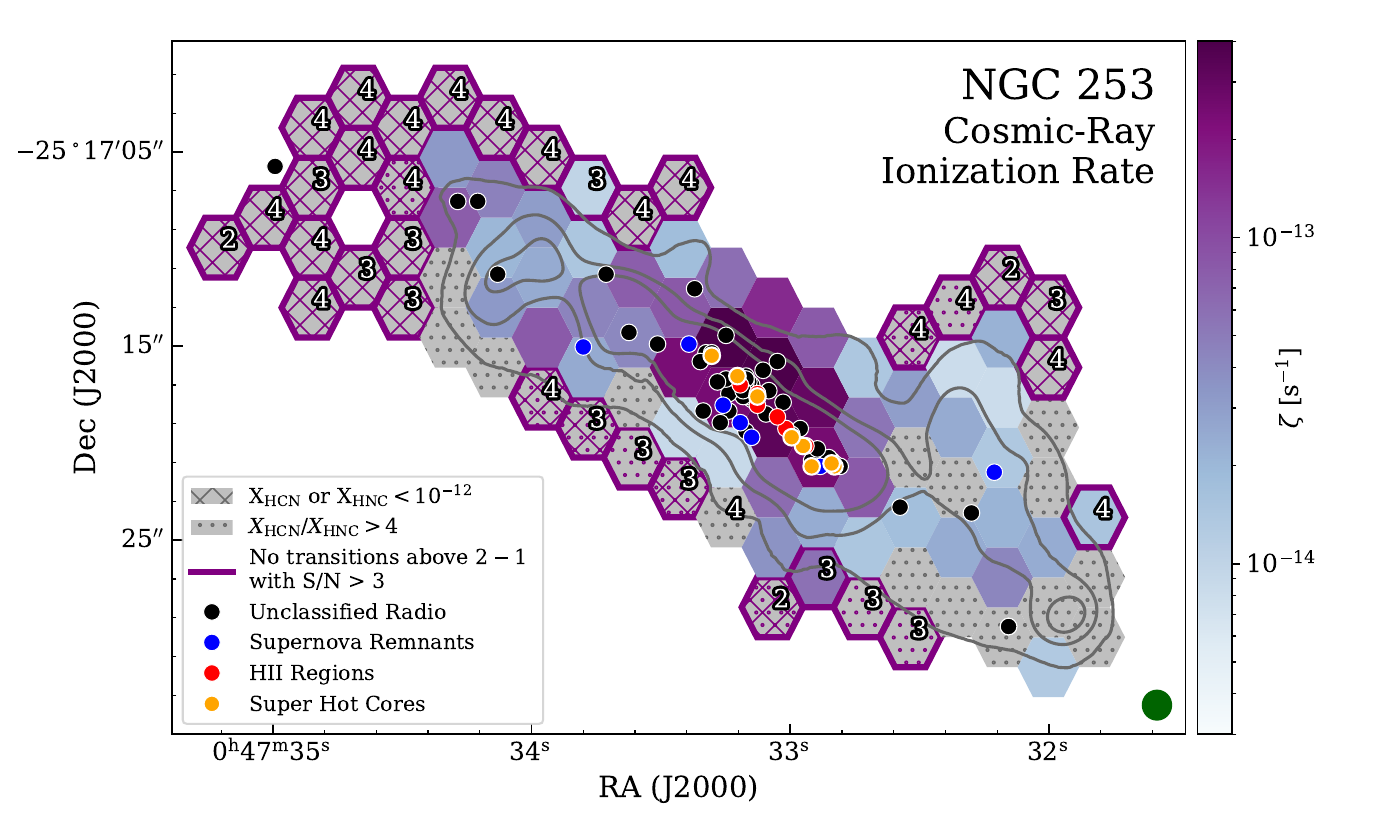}
    \caption{CRIR map estimated from our parameter inference algorithm. All grayed out, hatched regions are estimated by our algorithm to have $\zeta > 5\times 10^{-13}$\,s$^{-1}$ and one or more of the following issues. The `X' hatching indicates regions where the estimated HCN or HNC abundance was below 10$^{-12}$ with respect to hydrogen, and the dotted regions indicate locations where $X_{\text{HCN}}/X_{\text{HNC}} > 4$, with the two appearing together when both conditions are true. White and black numbers indicate the number of transitions that are detected with S/N $>$ 3 for a given region, and purple outlines indicate when there are no transitions above the $J=2-1$ that meet this threshold. Gray contours signify HCN $1-0$ emission at 1, 3, and 5 Jy\,km\,s$^{-1}$\,beam$^{-1}$. Colored dots indicate the locations of recent star formation via radio continuum sources \citep[supernova remnants and H\,II regions,][]{ua97} and super hot cores \citep{RV2020}. Note half of the unclassified sources are expected to be H\,II regions and half are thought to be supernova remnants. The green circle in the bottom right corner represents the ALCHEMI 1.\!\!$^{\prime\prime}$6 ($\sim28$\,pc) beam.}
    \label{fig:zeta_hatchMap}
\end{figure*}

\vspace{5mm}
\subsection{Apparent High-$\zeta$ Regions} \label{sec:high-zeta}

As noted above in Section \ref{sec:results_paramMaps} and in Figure \ref{fig:allParam_hexmap}, the CRIR maps feature regions primarily along the edge of the CMZ with CRIRs seemingly higher than those found in the nucleus ($\zeta\gtrsim 5\times10^{-13}$\,s$^{-1}$). \cite{Behrens2022} noted a positive correlation between the number of cosmic-ray sources (e.g. supernova remnants, see Figure \ref{fig:zeta_hatchMap}) and the CRIR in NGC\,253's CMZ\footnote{Note that of the `Unclassified Radio' sources labeled in Figure \ref{fig:zeta_hatchMap} \citep[identified in][]{ua97}, half are expected to be supernova remnants, and half are likely H\,II regions.}. As such, the lack of cosmic-ray sources around the edge of the CMZ begs questions about the legitimacy of these high inferred $\zeta$ values. We will explore two possible explanations for this behavior.

\subsubsection{High $X_\text{HCN}/X_\text{HNC}$}
\vspace{-2mm}
First, we will consider the HCN/HNC abundance ratio, $X_{\text{HCN}}/X_{\text{HNC}}$. \cite{Behrens2022} found a negative correlation between $X_{\text{HCN}}/X_{\text{HNC}}$ and the CRIR, with $X_{\text{HCN}}/X_{\text{HNC}} < 4$ for $\zeta > 10^{-13}$\,s$^{-1}$ and decreasing toward unity for even higher CRIRs. They explained that at high CRIRs, both HCN and HNC are expected to be formed and destroyed at similar rates, resulting in $X_{\text{HCN}}/X_{\text{HNC}}$ tending toward 1. However, as shown in Figure \ref{fig:abund_ratio}, we find $X_{\text{HCN}}/X_{\text{HNC}} > 4$ in many regions with high $\zeta$. Figure \ref{fig:abund_ratio}a shows a clear bimodal behavior in $X_{\text{HCN}}/X_{\text{HNC}}$, where all 24 regions with $X_{\text{HCN}}/X_{\text{HNC}} > 4$ also feature $\zeta > 5\times10^{-13}$\,s$^{-1}$. These regions are not located within the nucleus's innermost $5^{\prime\prime}$ and generally do not contain cosmic-ray-producing sources (see Figure \ref{fig:zeta_hatchMap}). 

Additionally, we find that most of the regions identified in Figure \ref{fig:abund_ratio} pair high CRIRs with very low kinetic temperatures such that $T_\text{K}$ is often pinned against the lower end of our prior distribution (see Figure \ref{fig:lowTcorners} for examples). Though we do not constrain $T_\text{K}$ well with our models, the high CRIRs that we estimate in these edge regions are not compatible with such low kinetic temperatures. Cosmic-ray ionization is expected to contribute significant energy to interstellar gas, raising the kinetic temperature to several hundreds of Kelvin \citep{bayet11,Behrens2022}, assuming the gas column density is not so high as to shield the inner layers from ionization \citep{Padovani2022A&A}. The unphysical combination of high $\zeta$ and very low $T_\text{K}$, in addition to anomalously high $X_{\text{HCN}}/X_{\text{HNC}}$ values, indicate that the chemistry included in our model is not well suited to the conditions in these regions.

As mentioned in Section \ref{sec:train_set}, we assume in our model that the gas we are probing is dense enough such that it is shielded from UV radiation and therefore dominated by cosmic-ray ionization. While this was true for the dense gas in the GMCs studied in \cite{Behrens2022}, it is likely not the case for the regions in more diffuse areas of the NGC\,253 CMZ. As a result, UV radiation is likely impacting the HCN and HNC chemistry. \cite{Santa-Maria2023} found that HCN was enhanced by far-UV radiation, increasing the HCN/HNC ratio as the incident radiation field increased. Thus, one explanation for the high CRIR values outside of the inner 5$^{\prime\prime}$ is that they are not shielded against UV radiation and are therefore not well-described by the chemistry in our model. 

Another possibility is that some of these regions are affected by shocks. \cite{meijerink11} discussed the influence of mechanical heating from shocks on the HCN/HNC ratio, showing that it is enhanced by shock activity. \cite{Tanaka2024} constrained the kinetic temperature in NGC\,253 and noted four high-temperature regions around the outskirts of the CMZ (approximately regions 34, 49, 51, and 71 from this work). They hypothesized that the gas in these regions is emitting from above the plane of the CMZ and could therefore be interacting with the large-scale outflow \citep{Bolatto2013}. In the southwest, region 86 (GMC 1) is another location potentially featuring shocks. \cite{Gorski2017} showed that this region is close to an expanding shell of dense gas, and \cite{Huang2023} also found evidence indicative of shocks here. Since the chemistry in shocked regions is likely not dominated by cosmic-ray ionization, the model presented here may struggle to  constrain the parameters in those areas.

\vspace{-3mm}
\subsubsection{Low $X_\text{HCN}$ and $X_\text{HNC}$}

After considering the regions marked by high $X_{\text{HCN}}/X_{\text{HNC}}$ that may involve chemistry not included in our model,
we see in Figure \ref{fig:abund_ratio} that there are still several regions, primarily in the northeast and northwest, that feature uncharacteristically high CRIRs. To address these regions, we then look beyond abundance ratios and consider the individual values of $X_{\text{HCN}}$ and $X_{\text{HNC}}$. As mentioned in Section \ref{sec:train_set}, \texttt{UCLCHEM} abundance estimates, upon which our neural network is based, are dominated by numerical errors when considering abundances below $10^{-12}$. As a result, we remove all parameter combinations from our training set that yield HCN or HNC abundances below this threshold. However, the neural network is able to characterize the functional relationships that exist between the parameters and molecular abundances during training. Since these functions are smooth around the abundance threshold we have imposed, it can still extrapolate and produce abundances below the $10^{-12}$ threshold if provided with the relevant parameter values by our nested sampling algorithm. 

We find that in many of the high-$\zeta$ regions along the outskirts of the CMZ, the most likely range of parameters yields $X_\text{HCN}$ and $X_\text{HNC}$ values below $10^{-12}$ (see the `X'-hatched regions in Figure \ref{fig:zeta_hatchMap}). Interestingly, despite not being trained on parameter combinations that result in such low abundances, the neural network does an excellent job at replicating the results that would be produced with \texttt{UCLCHEM}. Figure \ref{fig:corner1} in Appendix \ref{sec:addFigs} shows the posterior distributions achieved for region 1 using \texttt{UCLCHEM} and the neural network model, demonstrating nearly identical results in this low-abundance region. It is thus clear that our high-$\zeta$ result is not a product of the neural network, but of the chemical model's inefficacy at such low abundances. We see in Figure \ref{fig:zeta_hatchMap} that all low-abundance regions feature four or fewer combined HCN and HNC transitions that meet our S/N criteria (as compared to the full set of 8 that are available) and no higher energy transitions detected above $J=2-1$. The lack of HCN and HNC emission that spans the excitation ladder in these regions corroborate the low abundances estimated by \texttt{UCLCHEM} and the neural network. Since chemistry is largely irrelevant in this low-abundance regime, we determine that parameter results for any regions that correspond to such low abundances are unreliable.

\begin{figure*}
    \centering
    \includegraphics[scale=0.85, trim= 3mm 2mm 0mm 5mm, clip=True]{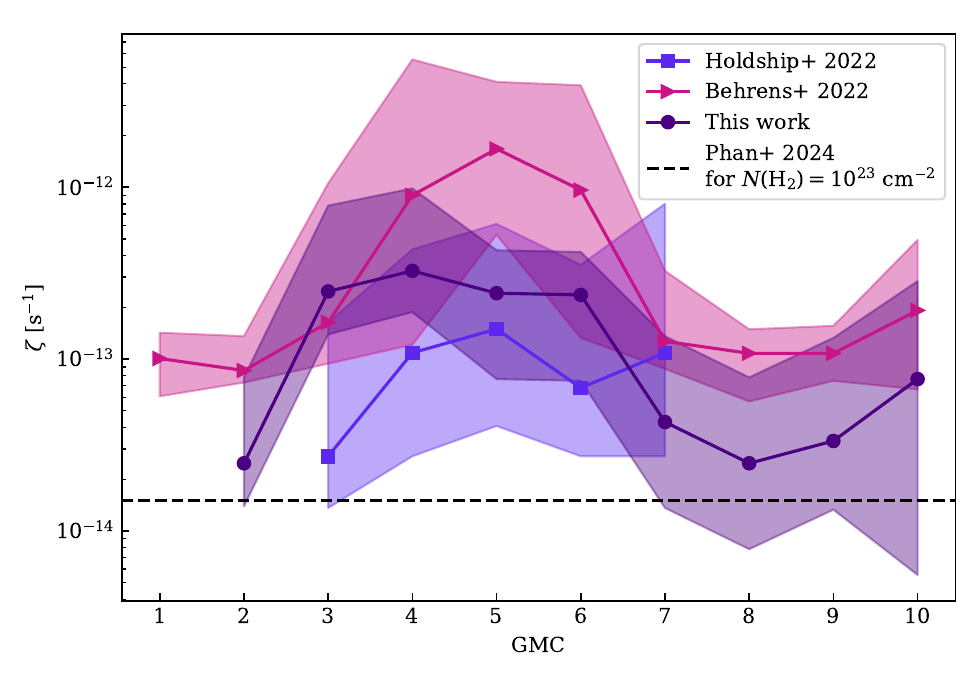}
    \caption{Comparison of our CRIR estimates with those from \cite{Holdship2022}, \cite{Behrens2022}, and \cite{Phan2024}. As proxies for GMCs 2--10, which were not studied in this work, we provide values for regions 75, 58, 53, 47, 36, 26, 20, and 16, which are the regions that overlap most significantly with the plotted GMCs. The widths of the shaded regions at each data point indicate the most likely 66\% of each $\zeta$ posterior distribution.}
    \label{fig:crir_meas}
\end{figure*}

\subsection{Comparison of Cosmic-Ray Ionization Rate to Other Measurements} \label{sec:meas_compare}

We compare our CRIR results to those predicted using similar methods in \cite{Holdship2022} and \cite{Behrens2022}. While \cite{Holdship2022} studied H$_3$O$^+$ and SO and \cite{Behrens2022} used HCN and HNC, both studies employed Bayesian nested sampling with a combination of chemical and radiative transfer models in order to infer the dense gas physical parameters in NGC\,253's CMZ. As mentioned in Section \ref{sec:intro}, both studies investigated a smaller number of GMC-sized regions (5 and 10, respectively) than in the study presented here. We compare our parameter estimations to those inferred in \cite{Holdship2022} and \cite{Behrens2022}, which are plotted as colored shaded regions in Figure \ref{fig:crir_meas}. Since GMCs 1--10 were not explicitly studied in this work, we associate each GMC with the hexagonal region with which it shares the most overlap (see Figure \ref{fig:num_ref}). Note that both of the aforementioned studies used smaller regions ($\sim28$\,pc circles, equal to the size of the ALCHEMI beam) than the 50\,pc hexagonal regions we study here. We are unable to plot a value for GMC 1, as region 86 (its closest proxy) was removed due to possible UV or shock contamination. Though \cite{Behrens2022} was able to obtain a reasonable CRIR estimate for GMC 1, which corresponded to a smaller region associated with denser gas, the larger area of our hexagonal regions results in the inclusion of more diffuse gas that is more likely to be influenced by UV radiation. Thus we are unable to obtain a reliable CRIR estimate for this region using our current chemical framework.

As shown in Figure \ref{fig:crir_meas}, this work predicts slightly lower CRIR values than those estimated previously with HCN and HNC in \cite{Behrens2022}, suggesting a CRIR range of $5\times10^{-15}-10^{-12}$\,s$^{-1}$, whereas \cite{Behrens2022} predicted a range of $6\times10^{-13}-5\times10^{-12}$\,s$^{-1}$. This lower range agrees better with the values reported in \cite{Holdship2022}, who estimated a CRIR range of $10^{-14}-10^{-12}$\,s$^{-1}$. The difference we see between our study's CRIR estimates and those estimated in \cite{Behrens2022} is likely a result of updates we implemented in our modeling procedure and not the effect of our neural network use. The model differences are as follows:

\vspace{-1mm}
\begin{enumerate}
    \item \textbf{Multi-phase modeling:} In this work, we use \texttt{UCLCHEM}'s two-phase model that includes the free-fall collapse of a cloud in order for the gas to start the second phase of the model in molecular form, rather than atomic or ionic. This first phase was not used in \cite{Behrens2022} but was indeed implemented in \cite{Holdship2022}. In cases where the CRIR was well constrained using only phase 2, the addition of phase 1 had little effect on our posterior distributions. In regions where our posterior distributions were less well constrained, introducing phase 1 into our modeling has a more significant effect on the shape of the posterior distributions but did not greatly alter the median values derived from each distribution. Figure \ref{fig:phase_comparison} demonstrates the differences in the posterior distributions with and without phase 1 modeling for sample regions 26 and 53.
    
    \item \textbf{New collisional excitation rates:} We implement more recent collisional excitation rates in our radiative transfer modeling with \texttt{SpectralRadex}. These rates are derived from calculations done by \cite{HernandezVera2017} that included collisions with ortho- and para-H$_2$, whereas \cite{Behrens2022} used older rates that were based on scaled calculations from collisions with helium \citep{Dumouchel2010}. These new rates did not result in significant changes in our posterior distributions for the majority of tested regions. 
    A detailed analysis of the new collisional rates and their impact on each of our \texttt{SpectralRadex} parameters can be found in Appendix \ref{sec:coll_rates}.

    \item \textbf{Beam-filling factor:} Recent work by \cite{Butterworth2024} suggests that HCN emission in NGC\,253 originates from sources significantly smaller than the $\sim28$\,pc ALCHEMI beam, indicating that a beam-filling factor is a necessary addition to an algorithm that models HCN, and likely also HNC. Thus, we also implement a beam-filling factor as one of our free parameters. The addition of this parameter did not greatly change the median CRIR values inferred for any of our test regions---small variations were noted, but any differences were within parameter uncertainties. However, including a beam-filling factor did slightly widen the posterior distributions derived for each parameter. This difference is demonstrated in Figure \ref{fig:bff_comparison}.
\end{enumerate}

For the most part, none of these changes individually have a large impact on the CRIR estimates in NGC\,253's CMZ. However, when applied together, we see a slight decrease (about half an order of magnitude) in the predicted CRIR estimates as compared to \cite{Behrens2022}. With the exceptions of GMCs 1 (discussed above) and 5, all CRIR uncertainties overlap with those from both \cite{Holdship2022} and \cite{Behrens2022}. It is important to remember, however, that the regions we study here are nearly twice as large in diameter and contain $\sim3.5$ times the area as those investigated in the aforementioned studies. Therefore, a decrease in the estimated CRIR could also be attributed to the gas parameters being averaged over a larger area, thus likely including both diffuse and dense components, whereas the smaller GMC structures, as defined in \cite{leroy15}, are centered on denser gas components.

\subsection{Comparison of Cosmic-Ray Ionization Rate to Theoretical Predictions}
\label{sec:theory_compare}

To confirm the validity of the high CRIRs derived from molecular emission in \cite{Behrens2022}, \cite{Holdship2022}, and this work, we consider CRIRs derived from theoretical calculations and fits to non-thermal radio, X-ray, and gamma-ray emission data from \cite{Phan2024}. This study considers a theoretical transport model for cosmic rays from which they derive cosmic-ray spectra (informed by observations of non-thermal emission) that are used to derive CRIRs as a function of molecular cloud H$_2$ column density in NGC\,253, M\,82, and Arp\,220. In developing the transport model, they consider cosmic-ray protons and electrons injected by supernova remnants, as well as secondary and tertiary electrons injected from proton-proton interactions and low-energy photon interactions with cosmic-ray-induced gamma rays, respectively. \cite{Phan2024} include four processes in their model to produce cosmic-ray-induced non-thermal emission: $\pi_0$ decay from proton-proton interactions, bremsstrahlung radiation, inverse Compton scattering, and synchrotron radiation. Using this model and parameters derived from fits to non-thermal emission from Fermi-LAT \citep{Abdo2010} and H.E.S.S. \citep{H.E.S.S.Collaboration2018}, they model cosmic-ray spectra and derive $\zeta$ as a function of H$_2$ column density, obtaining a value of $1.5\times10^{-14}$\,s$^{-1}$ for an H$_2$ column density of $10^{23}$\,cm$^{-2}$. However, the dependence of the CRIR on column density is weak, with the CRIR varying by less than 0.5 dex for $10^{22} < N(\text{H}_2) < 10^{25}$\,cm$^{-2}$.

This value is included as a reference in Figure \ref{fig:crir_meas} and is consistent with the lowest end of our modeled CRIR range. \cite{Phan2024} consider possible sources of discrepancy between their calculated values of $\zeta$ and those derived through molecular emission modeling from \cite{Holdship2022} and \cite{Behrens2022}. They note that the gamma-ray data, from which key parameters such as the supernova rate and ISM density are derived, have uncertainties that could alter the gamma-ray flux by a factor of 2, resulting in proportional variations of the CRIR. Additionally, \cite{Phan2024} note they are limited by the pointspread functions of the non-thermal emission data, which are on the order of a few arcminutes (one arcminute $\simeq 1020$\,pc in NGC\,253). Thus, any CRIR values derived from Fermi-LAT and H.\,E.\,S.\,S.~measurements must be treated as an average over the entire CMZ. As shown in Figures \ref{fig:allParam_hexmap} and \ref{fig:zeta_hatchMap}, the distributions of the gas parameters and locations of supernova remnants in NGC\,253 vary spatially, with a higher density of both molecular gas and supernova remnants in the center of the CMZ, where the CRIR is also enhanced. Thus we would expect any calculation of an average CRIR to be more consistent with values derived from molecular emission for the outer CMZ.

Finally, \cite{Phan2024} also consider the possibility of ionization contributions from sources that accelerate cosmic rays with an energy $E\lesssim280$\,MeV. At GeV and TeV energies, the cosmic-ray spectrum is dominated by $\pi_0$ decay, which drops off sharply for $E \lesssim 280$\,MeV, as $\pi_0$ cannot be produced at energies below this threshold. However, \cite{Phan2024} suggest there may be a population of sources that accelerate cosmic rays with MeV energies. These sources, which may include protostellar jets \citep{Padovani2015,Padovani2016,Gaches2018}, H\,II regions \citep{Padovani2019,Meng2019}, and stellar wind termination shocks \citep{Scherer2008}, are likely deeply embedded within the cloud and unable to penetrate into the more diffuse ISM due to the  high gas density. If so, these MeV sources could contribute to the CRIR but might not be observable to gamma-ray facilities.

\subsection{Cosmic Ray Enhancement in Star-forming Regions}

While the CRIR estimates presented by \cite{Phan2024} are consistent with the lowest values we infer in the outer CMZ, work by \cite{Semenov2021} could further explain the enhancement in CRIR values that we see in NGC\,253's nucleus. Many cosmic-ray transport models assume isotropic diffusion and therefore estimate CRIRs far from the sources that produce cosmic rays. However, as Figure \ref{fig:zeta_hatchMap} shows, many of the central regions in NGC\,253's CMZ contain sources that will produce and accelerate cosmic rays, such as supernova remnants and H\,II regions. Therefore it is important to consider how the CRIR and its effects on star formation would manifest in regions near these sources. \cite{Semenov2021} use simulations to test the effect of varying the cosmic-ray diffusion coefficient $\kappa_{\text{CR}}$ in star-forming regions. In environments with strong shocks from supernova explosions, the cosmic-ray current is much higher, which results in plasma instabilities and the excitation of nonresonant modes. The background magnetic field is therefore amplified, which can significantly suppress $\kappa_{\text{CR}}$. In the Milky Way, $\kappa_{\text{CR}}$ is assumed to be $\sim10^{28}$\,cm$^{2}$\,s$^{-1}$. However, studies of nearby supernova remnants and molecular clouds have shown that within $\sim50$\,pc of cosmic-ray injection sites, $\kappa_{\text{CR}}$ can be 10--100 times smaller \citep[see, for example,][]{Li2010,Ajello2012,Hanabata2014}, and theory predicts it could be suppressed by a factor of up to $10^6$ \citep{Blasi2007}. This suppression of cosmic-ray diffusion would result in a buildup of cosmic rays and increased ionization rates at injection sites, which has important implications for star formation and galaxy evolution.

\cite{Semenov2021} test this theory by running simulations (i) without cosmic-ray feedback, (ii) with cosmic-ray feedback and a constant diffusivity of $\kappa_{\text{CR}}=10^{28}$\,cm$^2$\,s$^{-1}$, and (iii) with cosmic-ray feedback where $\kappa_\text{CR}$ is suppressed in star-forming regions. Their simulations show that in models where $\kappa_\text{CR}$ is suppressed, the cosmic-ray pressure is significantly higher in the galaxy centers, and cosmic-ray pressure makes up a much larger fraction ($>75$\%) of the total pressure budget. They also find that in these same models, fewer dense, star-forming clumps are allowed to form and the overall SFR is reduced by a factor of $\sim$3--4. These effects are magnified in galaxies that were simulated to have a higher gas mass fraction $f_\text{g}$ of 40\% versus those with $f_\text{g}=20$\%. Additionally, \cite{Semenov2021} conclude that after the last supernova in the simulation explodes, cosmic rays are finally able to escape their injection sites and permeate the ISM, resulting in an extended vertical pressure gradient that can stabilize the disk against fragmentation and drive galactic winds. It is clear that cosmic rays play a crucial role in the regulation of star formation and galaxy evolution.

Though \cite{Semenov2021} show that cosmic-ray feedback has the power to disrupt, and even halt, star formation, NGC\,253 may be at too early a stage of its starburst phase to show symptoms of star formation suppressed by cosmic-ray feedback. Recent analyses have identified 14 super star clusters in NGC\,253 with a likely age range of 0.01 to 3\,Myr and an age gradient whereby the older SSCs are located in the center of the NGC\,253 CMZ \citep{leroy18,Mills2021ApJ,Levy2022ApJ,Butterworth2024}. This gradient may account for the greater number of supernova remnants and H\,II regions in the center of the NGC\,253 CMZ, as the younger SSCs farther from the center may not have existed long enough for massive stars to form and die. It is unclear, though, at exactly what point from the beginning of a starburst phase cosmic-ray feedback is expected to noticeably affect star formation. In the \cite{Semenov2021} simulations, supernova explosions occur between 3 and 43\,Myr after the start of the simulation, and supernova remnants are thought to have lifetimes $\sim10^5-10^6$\,yr \citep{Sarbadhicary2017,Bamba2022}. In non-interacting galaxies, such as NGC\,253, the starburst phase is thought to last $\sim70$\,Myr \citep{Cenci2024MNRAS}. However, \cite{Semenov2021} show that the effects of cosmic-ray feedback (e.g. decrease in SFR and number of dense gas clumps) are able to be seen 300--800\,Myr after the beginning of the simulation. Thus, the possible quenching effects of cosmic-ray feedback may not yet be visible in NGC\,253. Further investigations of starburst galaxies at later stages of evolution are necessary in order to better understand the full scope of cosmic-ray feedback's role in star formation and galaxy evolution.




\vspace{10mm}
\section{Conclusions} \label{sec:conc}

We present a neural network model that we have trained to replace the function of a chemical modeling code in a gas parameter inference algorithm. We use ALMA measurements of the first four rotational transitions of HCN and HNC to constrain our neural network and radiative transfer models in a Bayesian nested sampling framework in order to infer the gas conditions across the NGC\,253 CMZ at 50\,pc resolution. We find that the neural network model can very effectively reproduce the same molecular abundances and gas parameter values as would be found using a chemical modeling code but in $\sim10$\% of the time. Using this algorithm, we draw the following conclusions regarding the molecular gas conditions in NGC\,253:

\begin{enumerate}
    \item Our results show clear spatial gradients across the CMZ in the gas volume and column density as well as the beam-filling factor. Volume and column densities are higher by an order of magnitude in the center ($r\lesssim100$\,pc) of NGC\,253's CMZ ($n_{\text{H}_2}\sim5\times10^{5}$\,cm$^{-3}$, $N_{\text{H}_2}\sim10^{24}$\,cm$^{-2}$) than they are farther away from the nucleus. The beam-filling factor reaches its minimum ($\eta_{ff} < 0.1$) in NGC\,253's nucleus, which is consistent with the emission in this area originating from small, dense gas components.
    \item We estimate CRIR values of $\zeta \sim 4 \times 10^{-13}$\,s$^{-1}$ in the nucleus of the NGC\,253 CMZ and $\zeta \sim 2 \times 10^{-14}$\,s$^{-1}$ in its outer regions. These derived CRIRs are consistent with those derived for specific GMCs in the NGC\,253 CMZ by \cite{Holdship2022} and \cite{Behrens2022}.
    \item We are unable to derive CRIR estimates for regions with low HCN and HNC abundance as well as regions likely featuring UV-dominated or shock chemistry (X$_\text{HCN}$/X$_\text{HNC} \gtrsim 4$), which is not included in our models.  Chemical modeling of HCN and HNC in photodissociation regions and shock environments will be addressed in a future study of the sources of energy in the NGC\,253 CMZ. 
    \item The CRIR estimates in the outer CMZ are consistent within $\sim0.2$ dex with theoretical predictions of the CRIR \cite{Phan2024} derived from CMZ-averaged non-thermal emission.
    \item The high CRIR estimates in the inner CMZ are likely a result of their proximity to supernova remnants and other cosmic-ray-producing sources, as well as potential cosmic-ray diffusion suppression that prevents cosmic rays from escaping into the outer CMZ.
\end{enumerate}

This study demonstrates that neural networks can be an effective and efficient tool for replacing chemical models in larger parameter inference algorithms. This method has the potential to significantly decrease the time required for inferring physical parameters across sources in large samples or in individual images with large numbers of pixels. As discussed in Section \ref{sec:high-zeta}, this model is not appropriate for regions where the chemistry is very sensitive to small changes in temperature, i.e. photodissociation regions. Future possibilities include incorporating additional molecular tracers that trace the same gas component into a single neural network model in order to constrain the gas parameters even further. Additional future work will include utilizing the neural network's speed to investigate the limitations associated with using molecular emission to infer gas parameters, particularly focusing on how the number of constraining transitions affects the parameter inference outcomes.

\section*{Acknowledgements}
We thank Tommaso Grassi for his very helpful and constructive feedback in reviewing this article. We also thank Phoebe Behrens for her crucial support to the authors throughout this process. This work is part of a project that has received funding from the European Research Council (ERC) under the European Union’s Horizon 2020 research and innovation programme MOPPEX 833460. 
V.M.R. and L.C. acknowledge support from the grant PID2022-136814NB-I00 by the Spanish Ministry of Science, Innovation and Universities/State Agency of Research MICIU/AEI/10.13039/501100011033 and by ERDF, UE. V.M.R also ackowledges the grant RYC2020-029387-I funded by MICIU/AEI/10.13039/501100011033 and by ``ESF, Investing in your future", and from the Consejo Superior de Investigaciones Cient{\'i}ficas (CSIC) and the Centro de Astrobiolog{\'i}a (CAB) through the project 20225AT015 (Proyectos intramurales especiales del CSIC); and from the grant CNS2023-144464 funded by MICIU/AEI/10.13039/501100011033 and by ``European Union NextGenerationEU/PRTR”.
N.H. acknowledges support from JSPS KAKENHI Grant Number JP21K03634. PH is a member of and received financial support for this research from the International Max Planck Research School (IMPRS) for Astronomy and Astrophysics at the Universities of Bonn and Cologne. K.S. acknowledges the grant MOST 111-2112-M-001-039 from the Ministry of Science and Technology in Taiwan. 

This article makes use of the following ALMA data: ADS/JAO.ALMA\#2017.1.00161.L and ADS/JAO.ALMA\#2018.1.00162.S. ALMA is a partnership of ESO (representing its member states), NSF (USA) and NINS (Japan), together with NRC (Canada), MOST and ASIAA (Taiwan), and KASI (Republic of Korea), in cooperation with the Republic of Chile. The Joint ALMA Observatory is operated by ESO, AUI/NRAO and NAOJ.  The National Radio Astronomy Observatory is a facility of the National Science Foundation operated under cooperative agreement by Associated Universities, Inc. 


\facility{ALMA}

\software{CASA \citep{CASATeam2022PASP}, \texttt{Astropy} \citep{Astropy2013}, \texttt{MLFriends} \citep{ultranest14,ultranest19}, \texttt{UltraNest} \citep{ultranest21}, \texttt{SciPy} \citep{scipy2020}}


\vspace{5cm}
\newpage
\bibliography{HCN_HNC.bib}{}
\bibliographystyle{aasjournal}

\restartappendixnumbering\renewcommand{\theHfigure}{A\arabic{figure}}
\appendix 

\section{Neural Network Architecture} \label{sec:NN_appendix}

\begin{figure}
    \centering
    \includegraphics[scale=0.6]{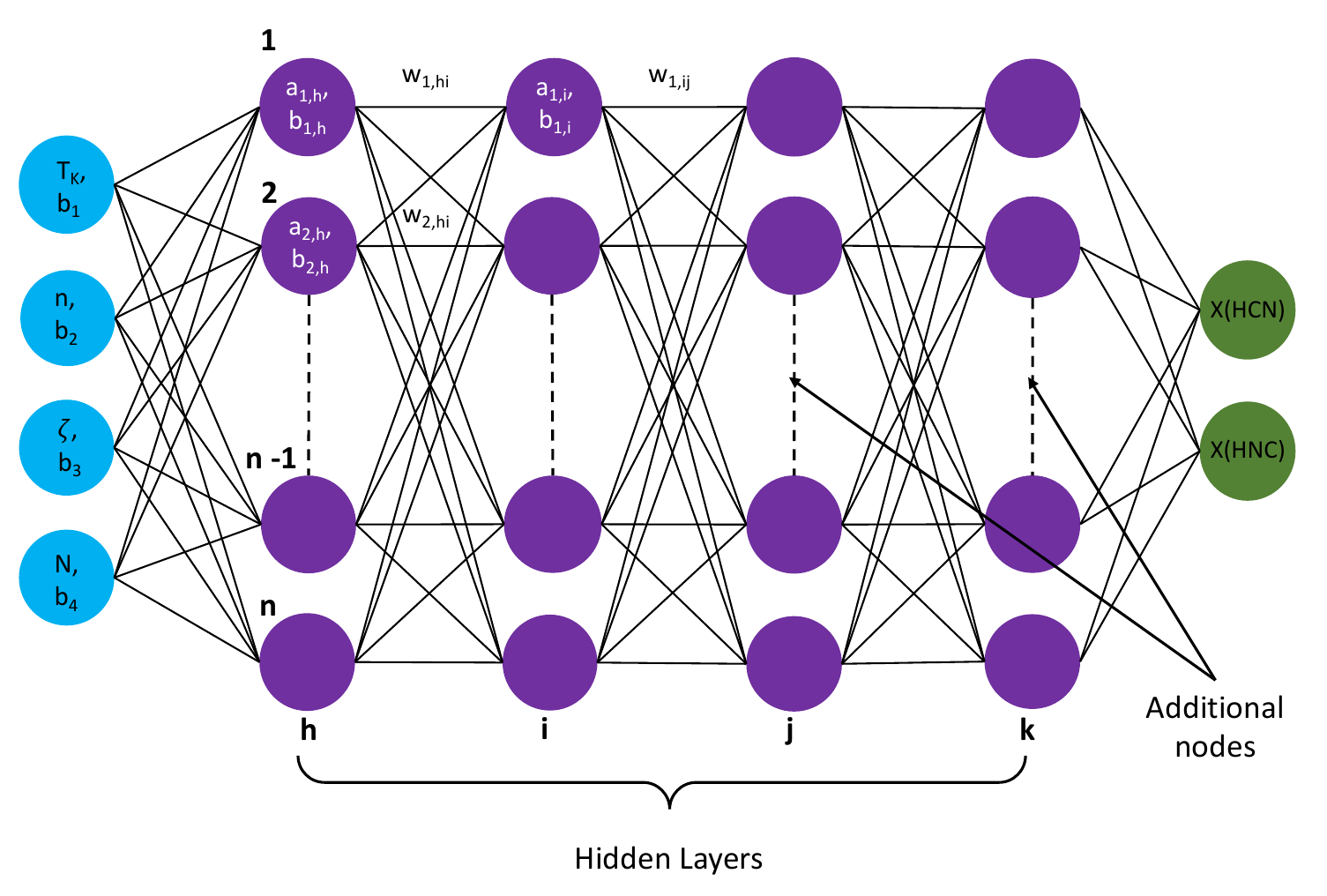}
    \caption{Schematic of our neural network architecture. Blue circles represent the nodes in the input layer of the neural network which contain our input parameters, purple circles represent the nodes that make up the neural network's inner hidden layers, and the green circles represent the output nodes in the final layer, which contain the abundances for HCN and HNC, respectively.}
    \label{fig:NN_arch}
\end{figure}

Neural networks consist of multiple layers of nodes, where each node is connected to all nodes in the layers immediately preceding and following it (see Figure \ref{fig:NN_arch}). The number of layers and nodes needed is typically unique to the problem at hand and is found primarily through trial and error---in our case, we found a neural network with 4 layers each containing 2000 nodes produced results that aligned best with chemical model calculations. Every node has a unique value and bias, and each connection between nodes in different layers is given a weight (see Equation \ref{eq:nn}). Altogether, these quantities are used to calculate the values of nodes in the following layer. The values of the nodes in the first layer are the values of our input parameters (kinetic temperature $T_K$, H$_2$ volume density $n$, CRIR $\zeta$, and H$_2$ column density $N$), and each node is assigned an arbitrary starting bias and each node connection an arbitrary starting weight. During the neural network training process, the node values, biases, and weights in each layer are successively manipulated to produce molecular abundances, which are stored in the final layer of the neural network. To calculate the value $a_{1,j}$ of a node in row 1 and layer $j$, we use the following prescription:
\begin{equation} \label{eq:nn}
    a_{1,j} = \Phi\left(\sum_{n=1}^{2000} \space a_{n,i} w_{n,ij} + b_{1,j}\right) 
\end{equation} 
where $a_{n,i}$ is the value of a node in the previous layer $i$ and some row $n$, $w_{n,ij}$ is the weight associated with the unique connection between that node and our target node $a_{1,j}$, $b_{1,j}$ is the bias of our target node in layer $j$, and $\Phi$ is an activation function through which the sum of the combinations for each node in the previous layer are passed. Activation functions, such as the rectified linear unit (ReLU) that we employ, allow neural networks to consider non-linear solutions. 

\begin{figure}
\centering
    \includegraphics[trim = 4mm 2mm 15mm 13mm,scale=0.68,clip=True]{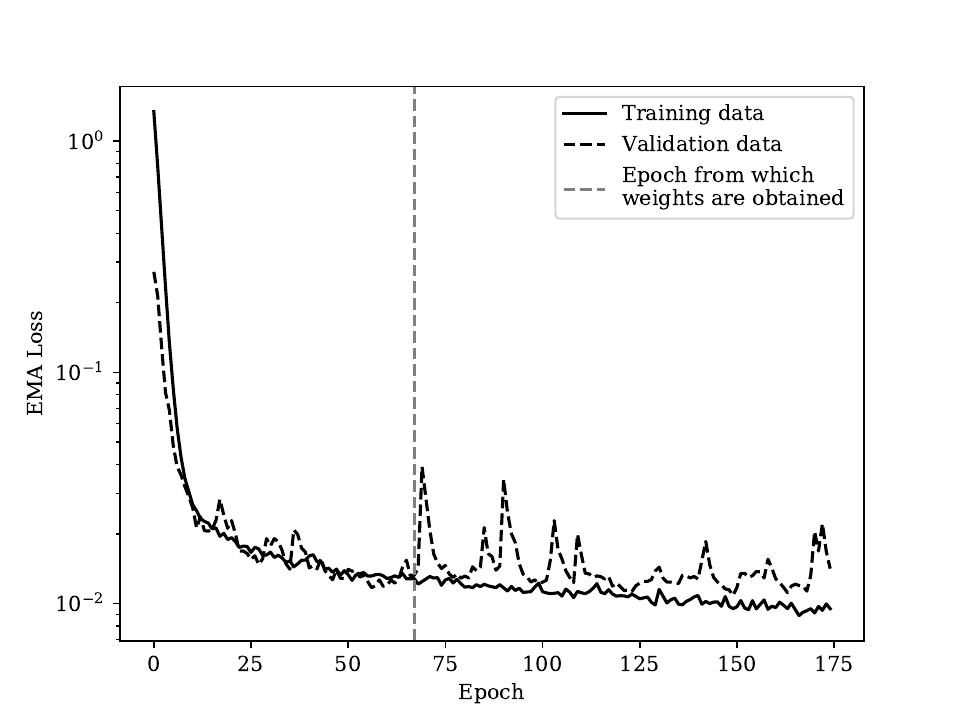}
    \caption{Exponential moving average of the mean squared error loss as a function of training epoch calculated for both the training and validation datasets. Here the neural network was trained for many epochs past our early stopping criteria (described in Section \ref{sec:architecture}) to demonstrate the divergence between the training and validation set losses. The dashed gray line indicates the epoch from which the final weights would have been obtained, as the validation loss does not improve for more than 20 epochs after this point.}
    \label{fig:loss}
\end{figure}

As discussed in Section \ref{sec:architecture}, we assess the difference between the \texttt{UCLCHEM} and neural network abundances using a mean squared error loss function, where the loss is calculated on the training set throughout each epoch and calculated on a subset of the validation set just once at the end of each epoch. We calculate the validation loss using mini batches with a batch size of 64. The exponential moving average of the loss associated with our neural network training is shown in Figure \ref{fig:loss}.

\section{Additional Figures} \label{sec:addFigs}

\begin{figure}[h]
    \gridline{\leftfig{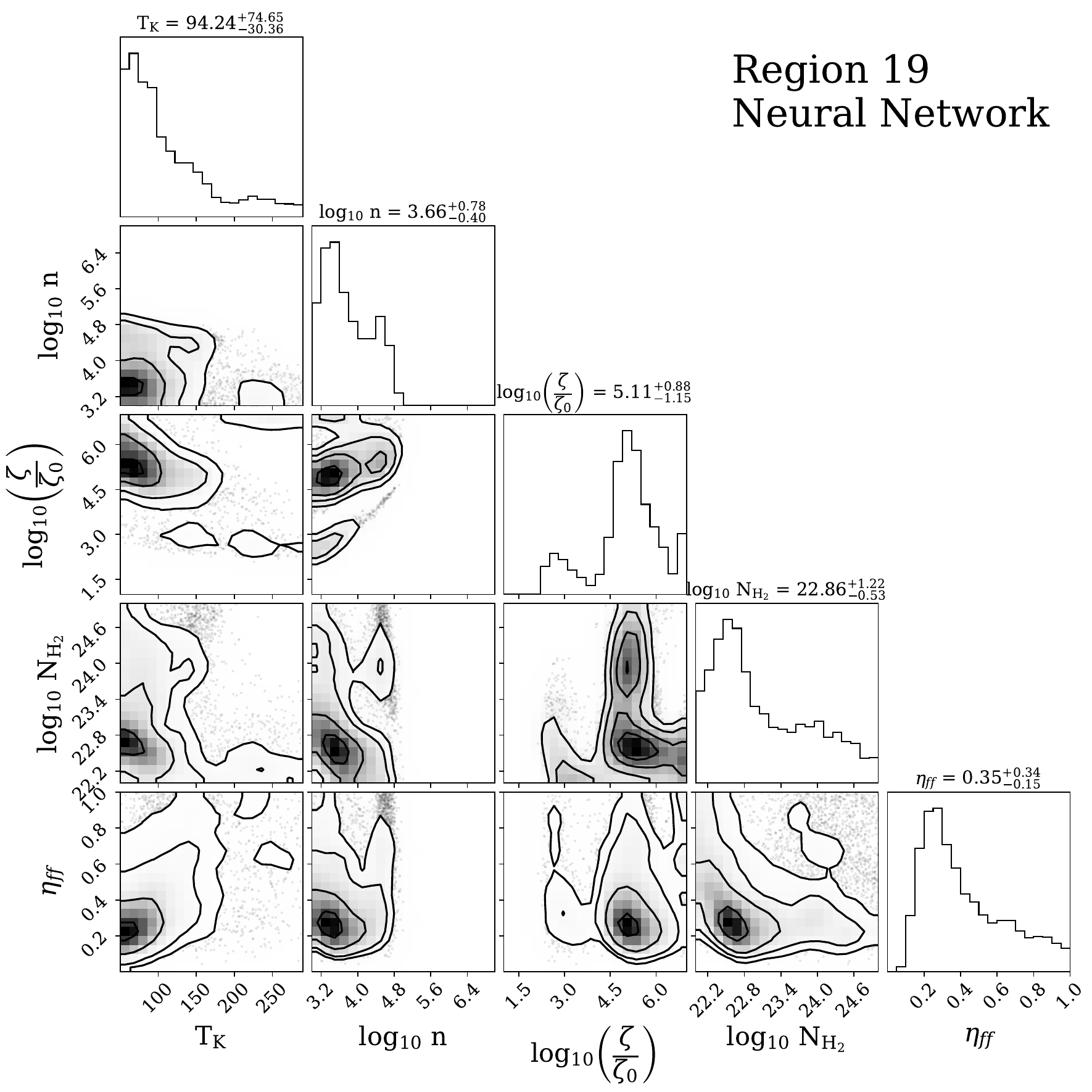}{0.5\textwidth}{(a)} \rightfig{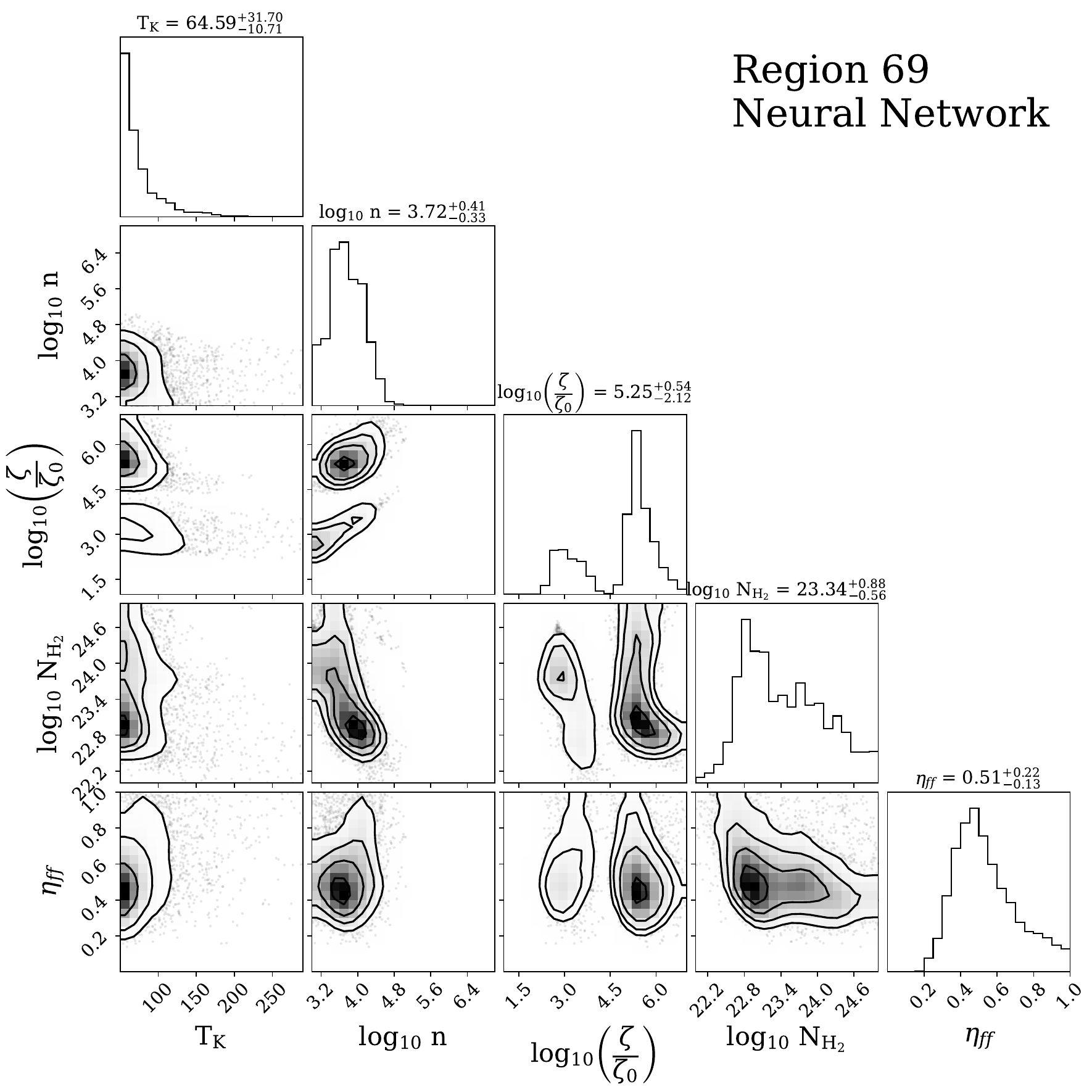}{0.5\textwidth}{(b)}}
    \gridline{\leftfig{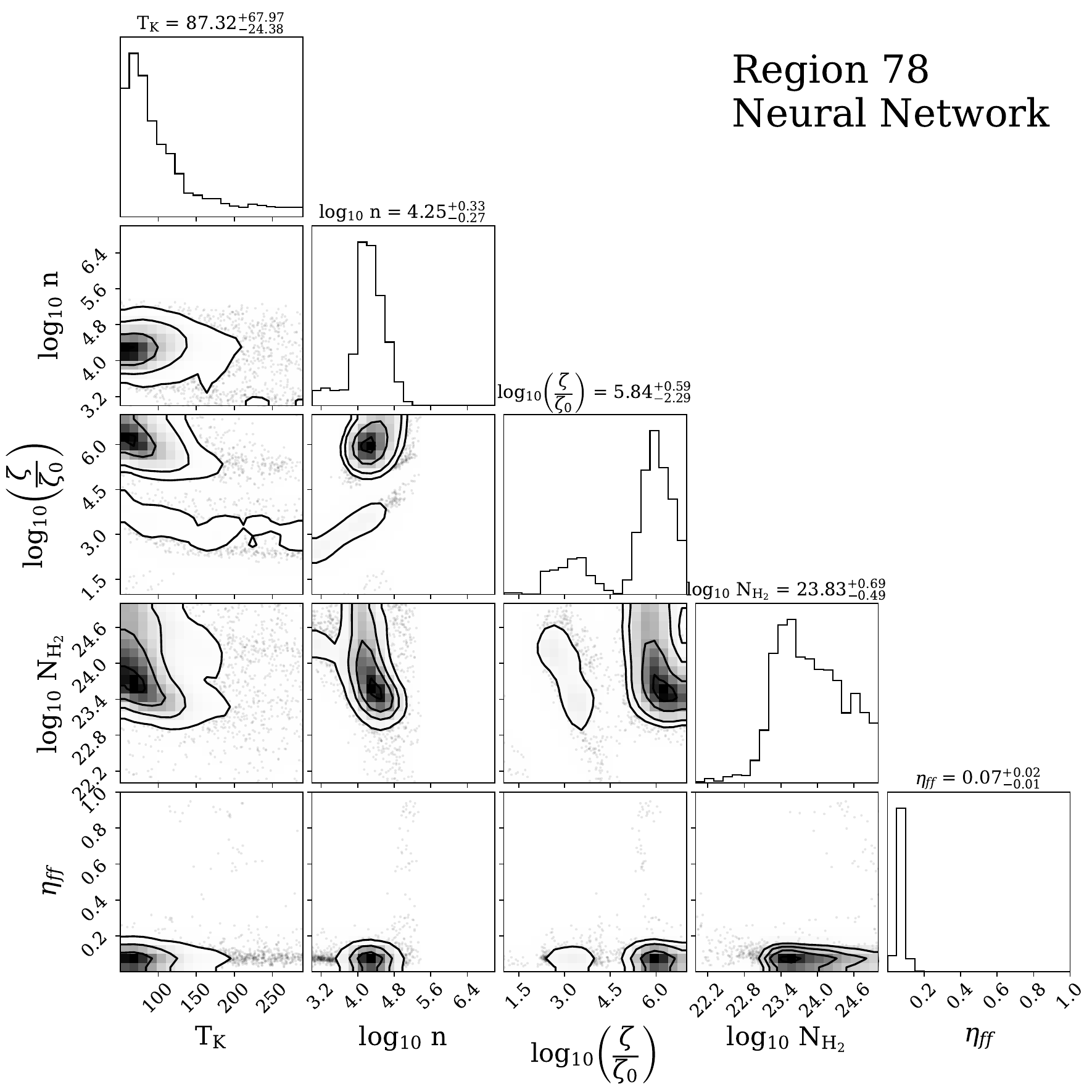}{0.5\textwidth}{(c)} \rightfig{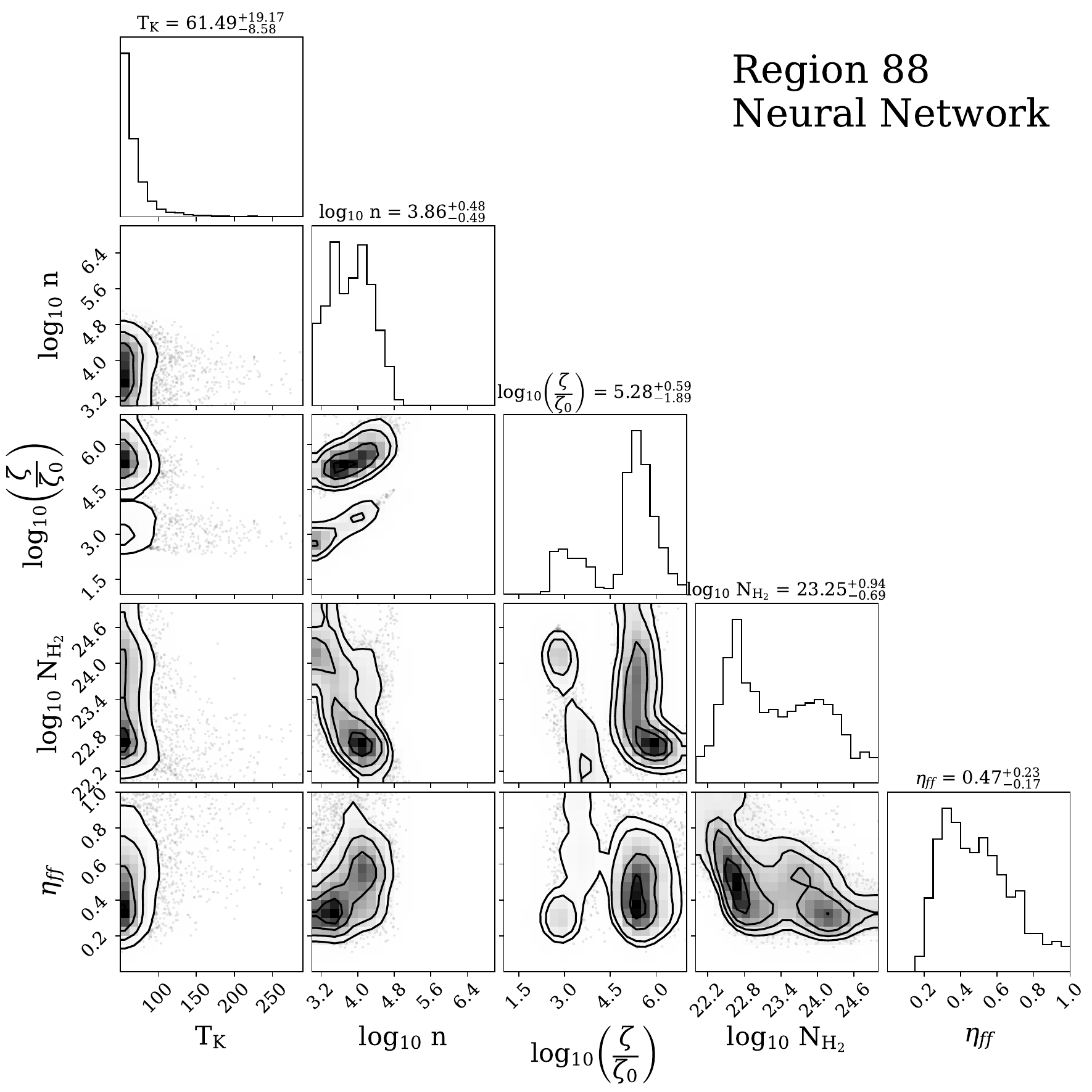}{0.5\textwidth}{(d)}}
    \caption{Posterior distributions for regions 19 (a), 69 (b), 78 (c), and 88 (d), which all feature high CRIRs paired with low kinetic temperatures.}
    \label{fig:lowTcorners}
\end{figure}

\begin{figure}[h]
    \gridline{\leftfig{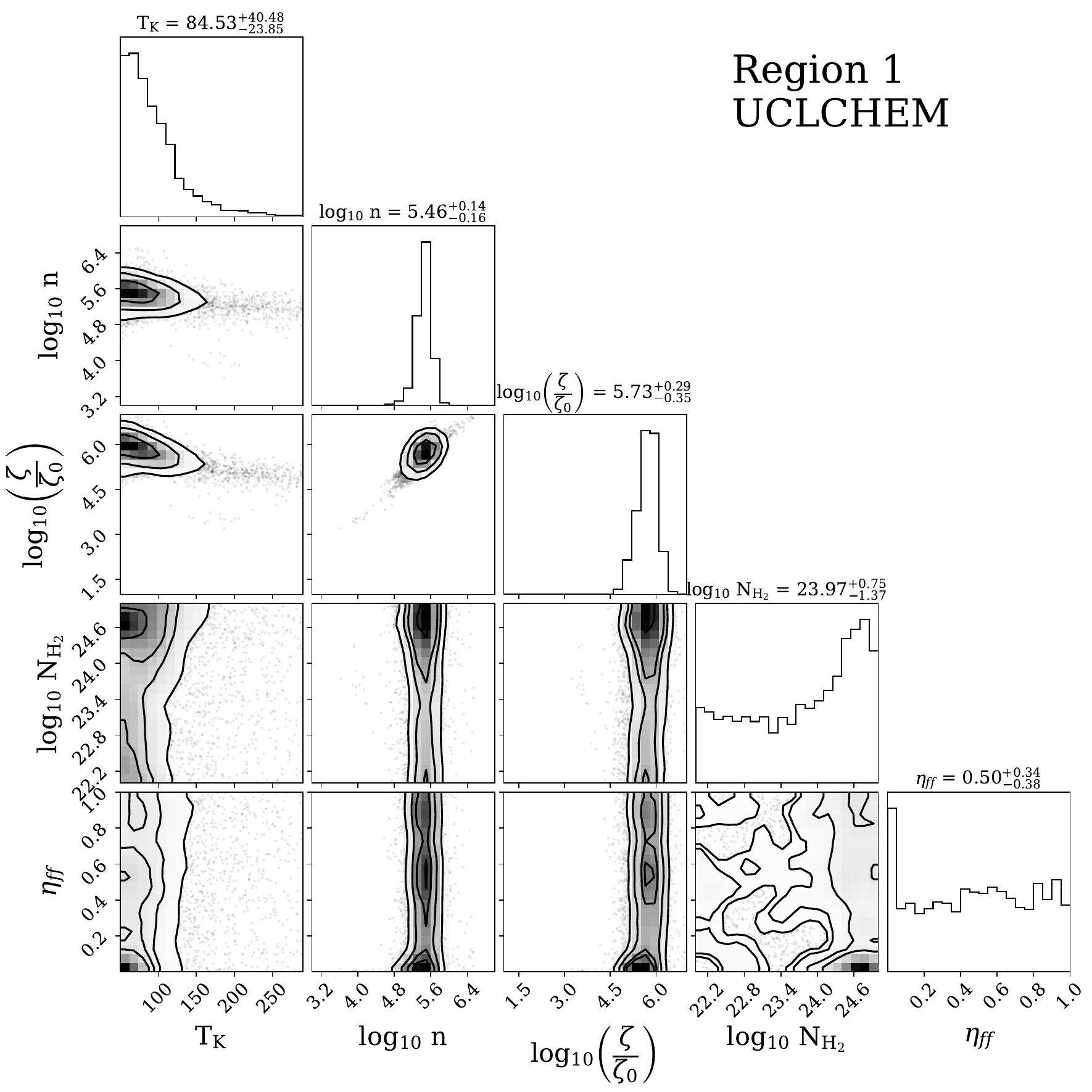}{0.5\textwidth}{(a)} \rightfig{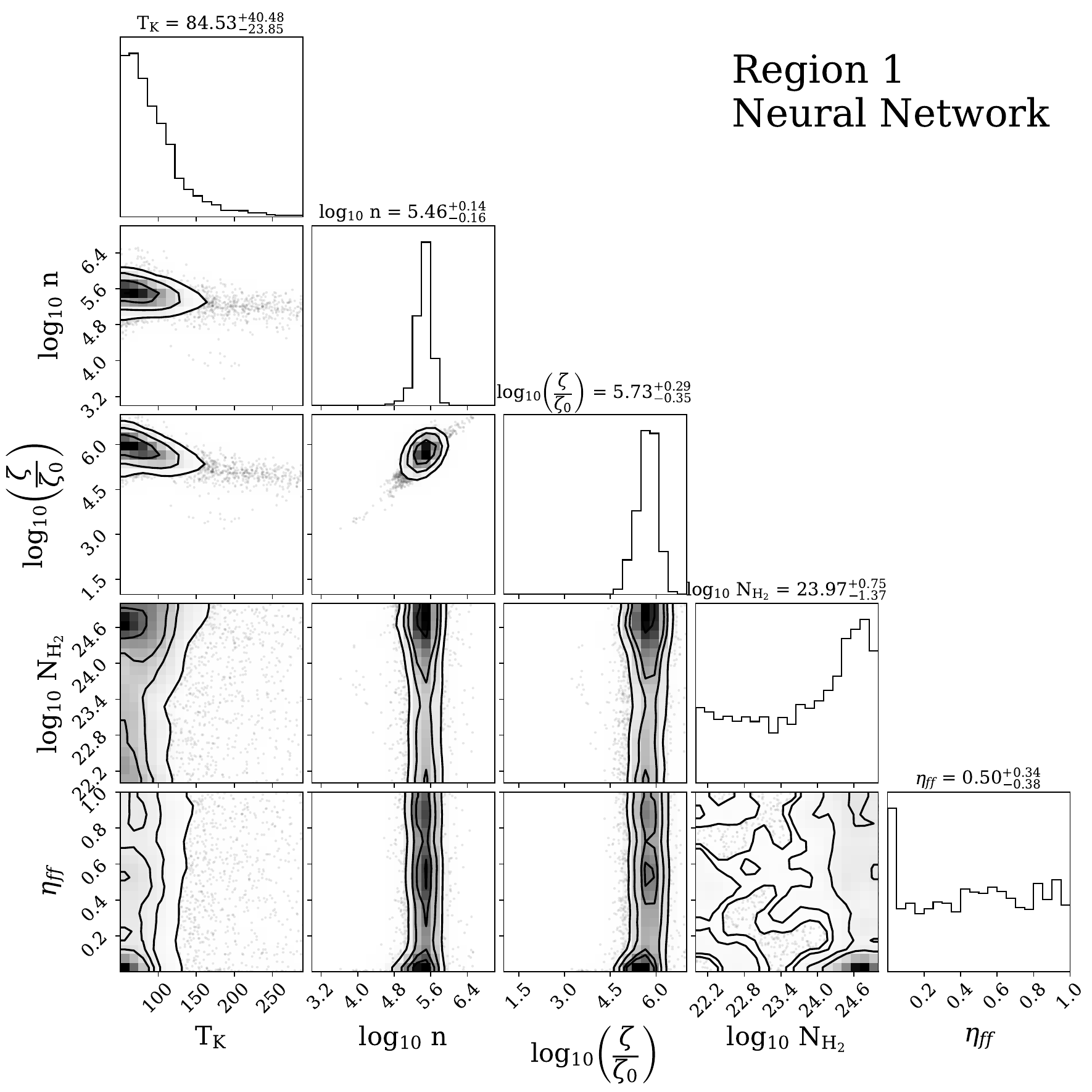}{0.5\textwidth}{(b)}}
    \caption{Posterior distributions for region 1 using \texttt{UCLCHEM} (a) versus the neural network (b).}
    \label{fig:corner1}
\end{figure}

\begin{figure}[h]
    \gridline{\leftfig{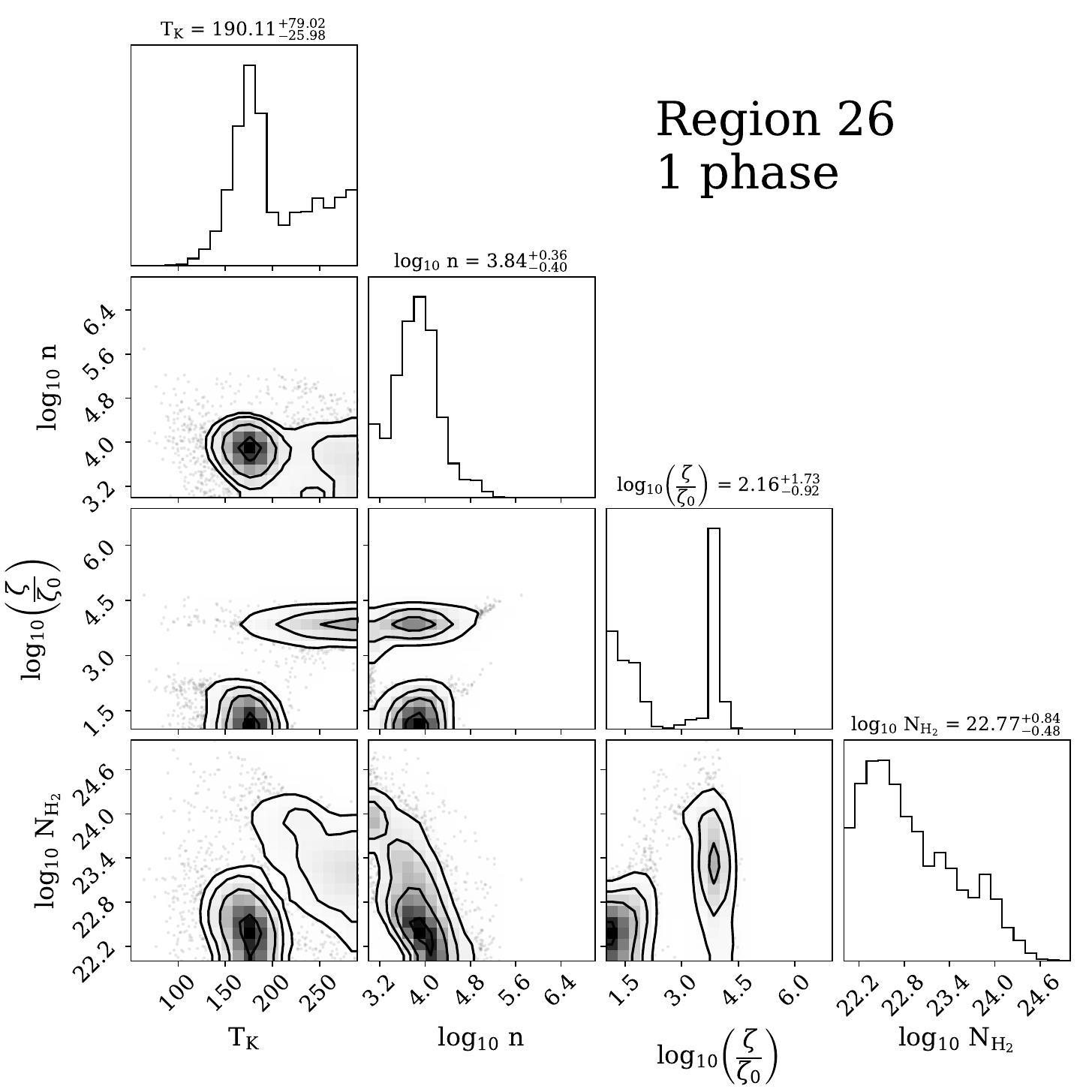}{0.5\textwidth}{(a)} \rightfig{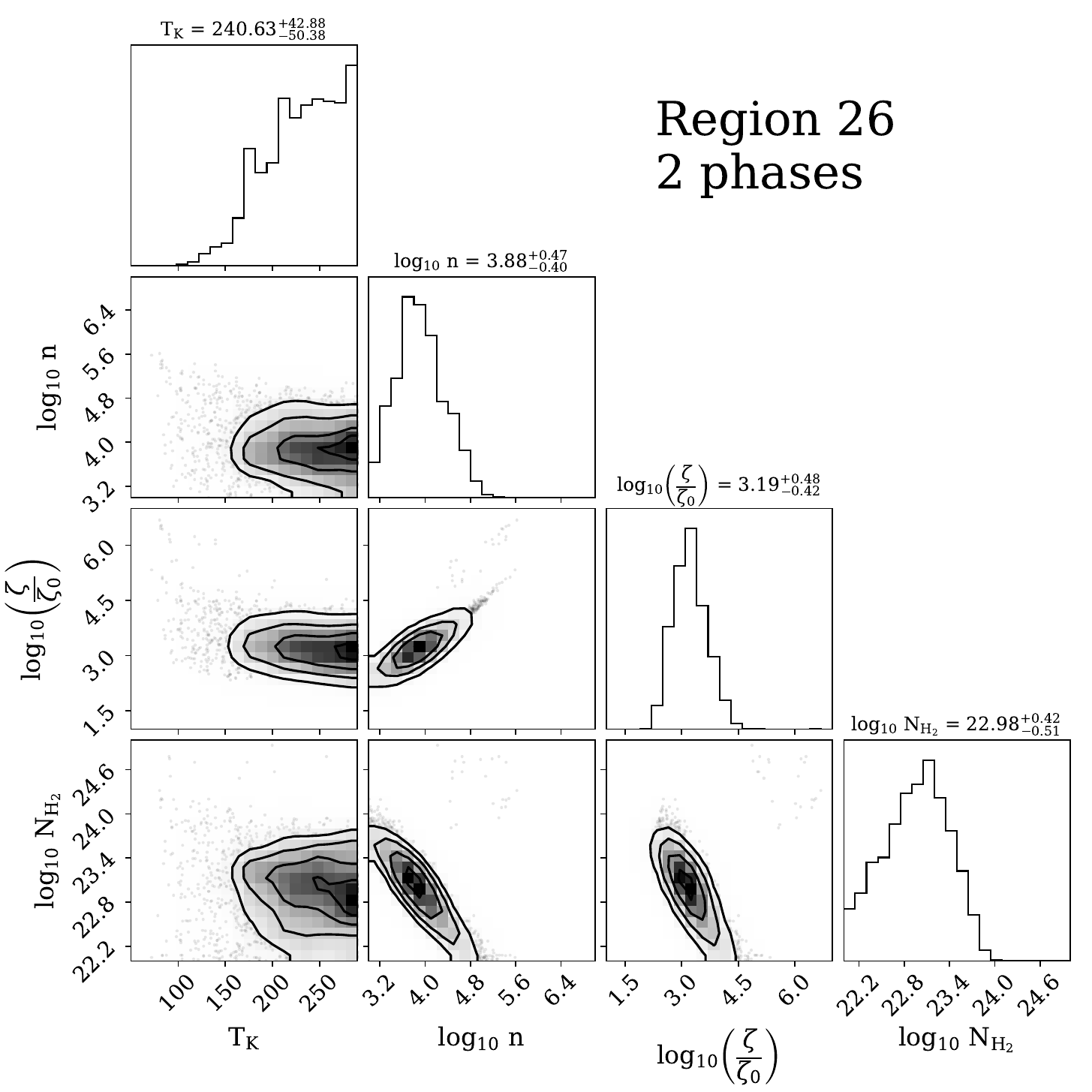}{0.5\textwidth}{(b)}}
    \gridline{\leftfig{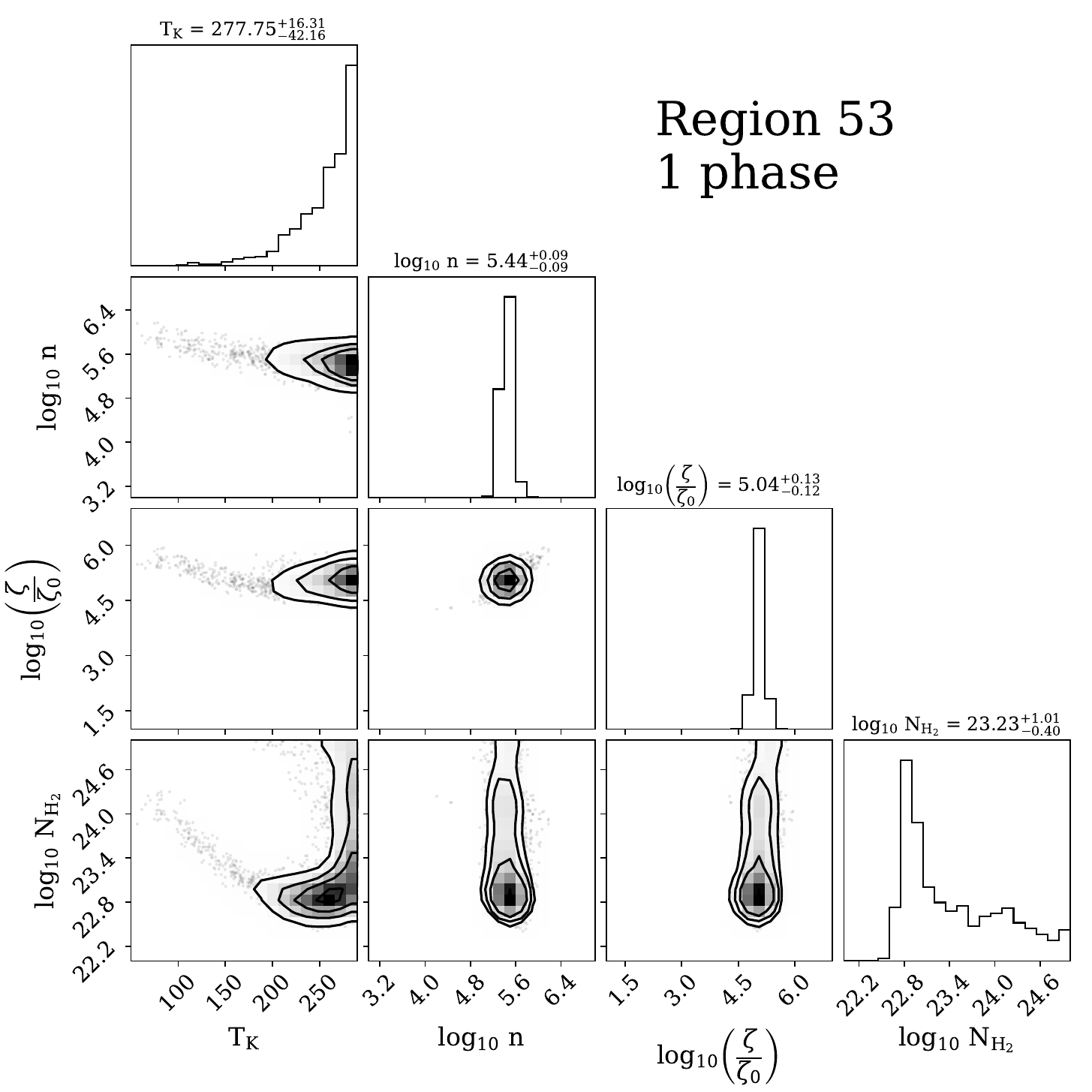}{0.5\textwidth}{(c)}\rightfig{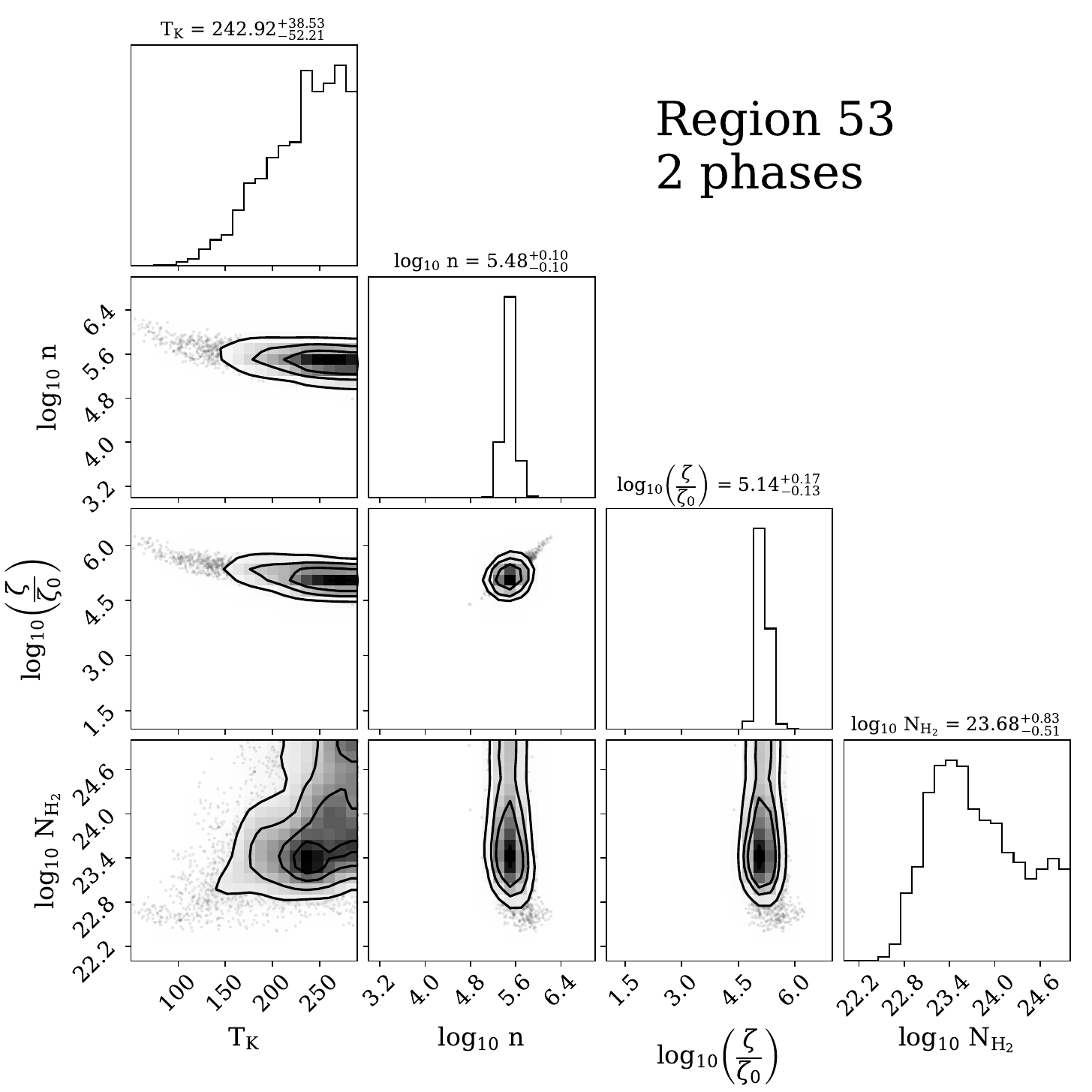}{0.5\textwidth}{(d)}}
    \caption{Posterior distributions for regions 26 (top) and 53 (bottom) using only one phase of chemical modeling (left column) versus two phases (right column).}
    \label{fig:phase_comparison}
\end{figure}

\begin{figure}[h]
    \gridline{\leftfig{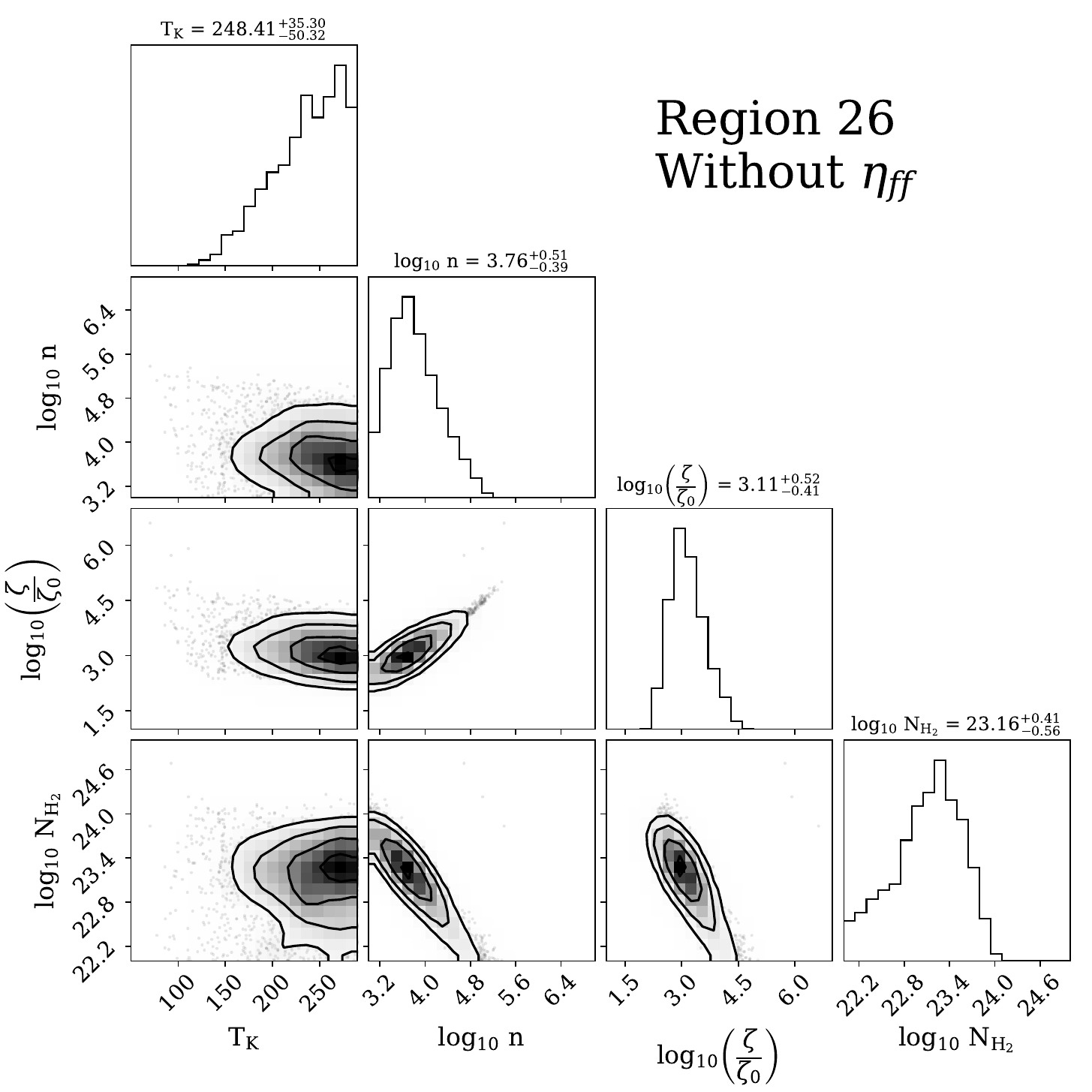}{0.5\textwidth}{(a)}\rightfig{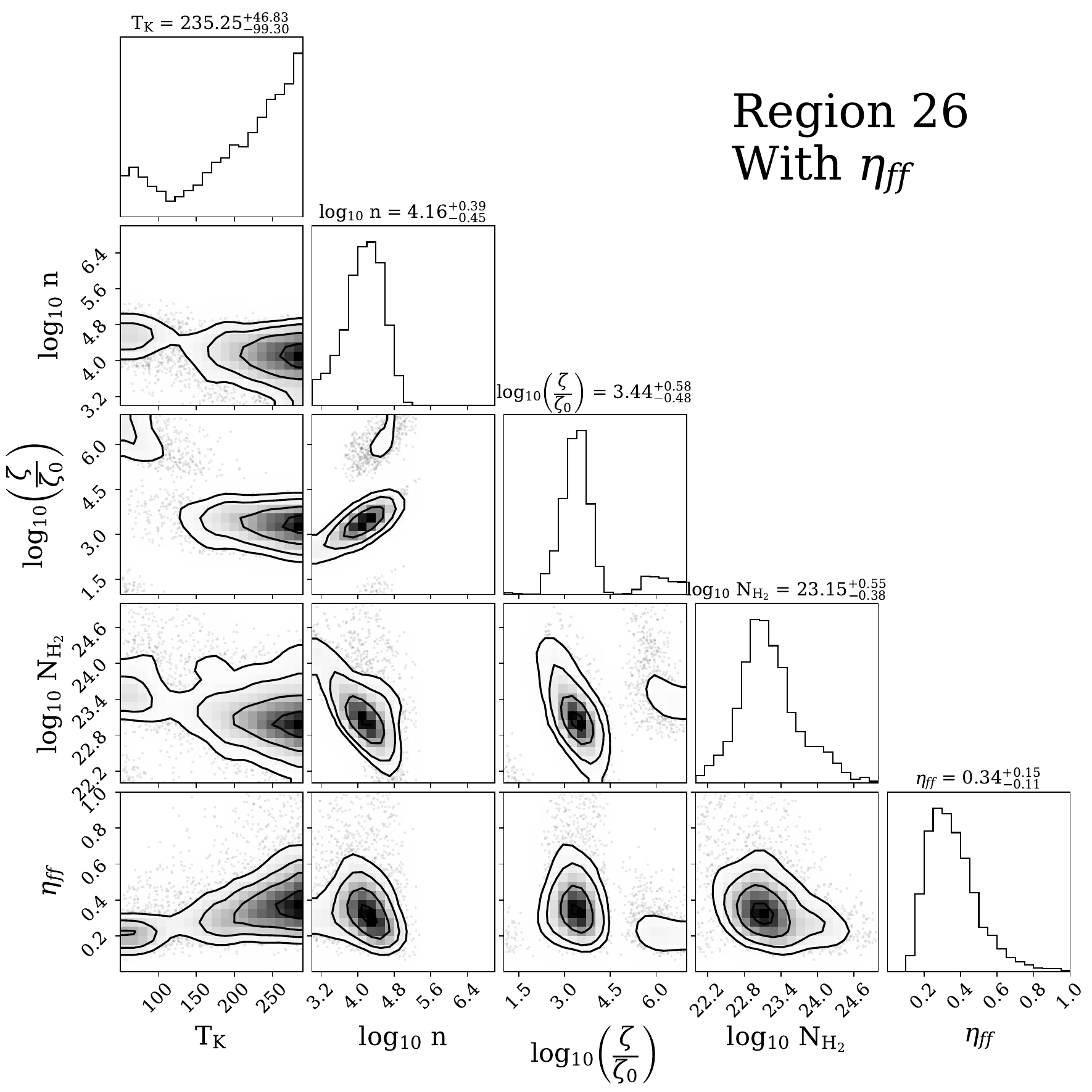}{0.5\textwidth}{(b)}}
    \gridline{\leftfig{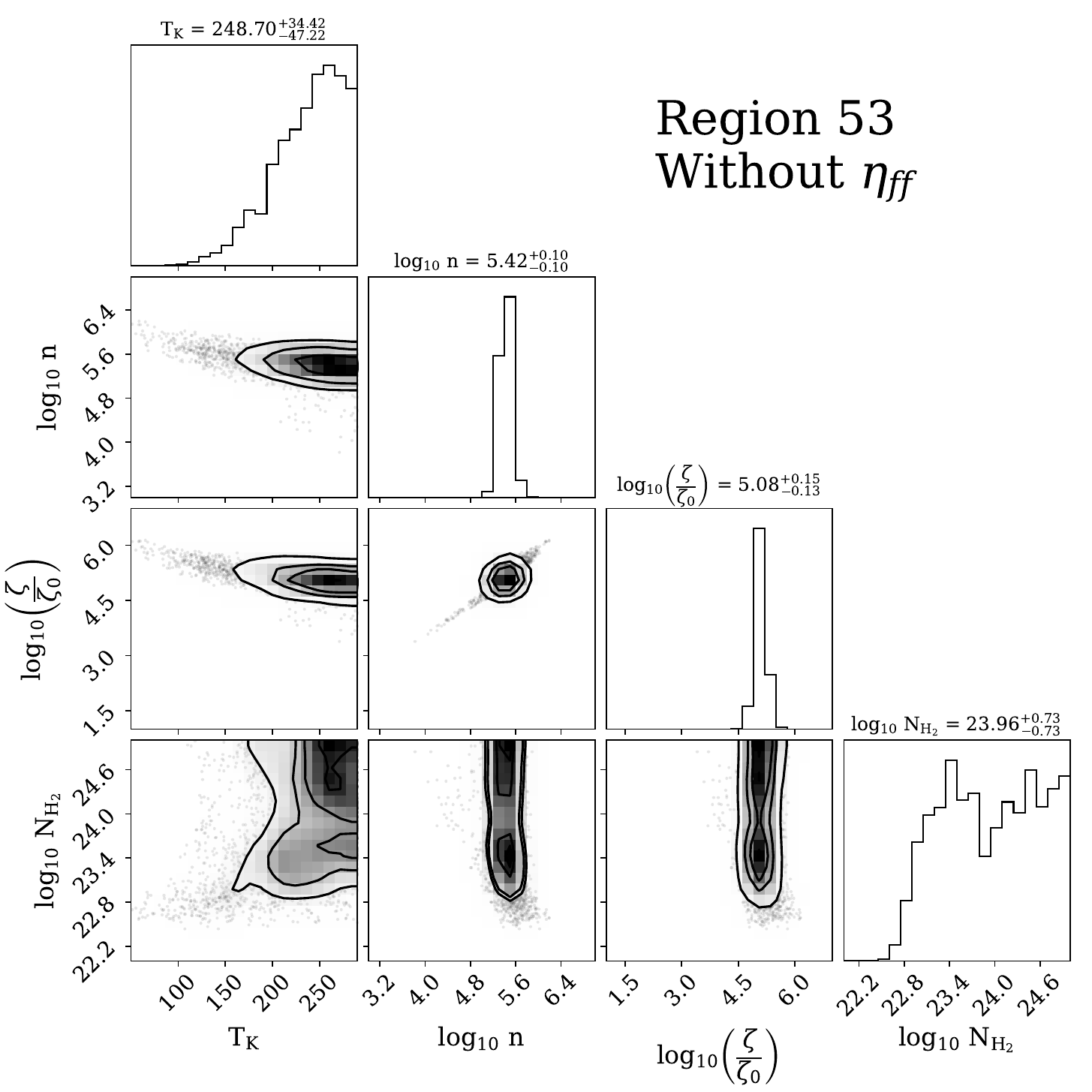}{0.5\textwidth}{(c)}\rightfig{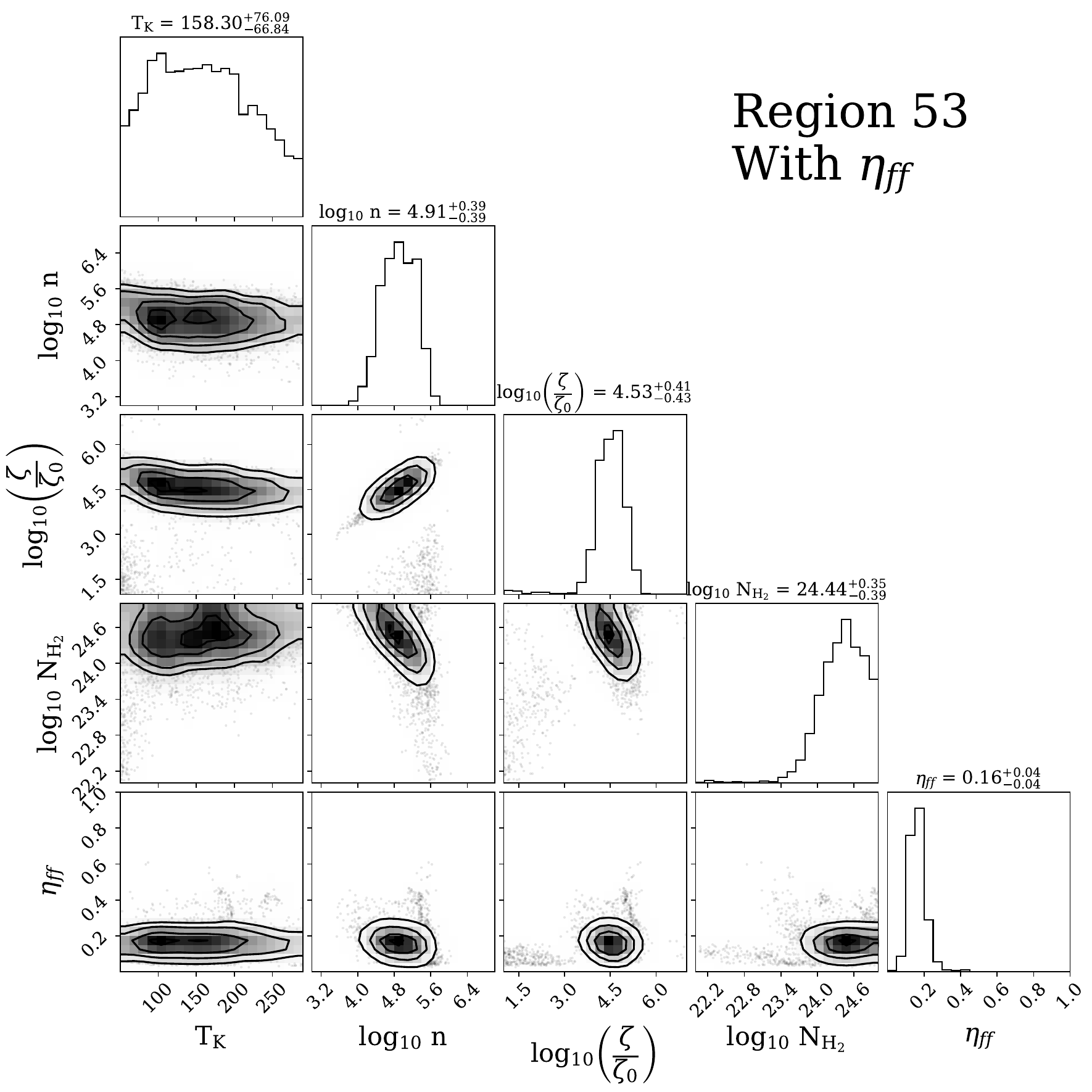}{0.5\textwidth}{(d)}}
    \caption{Posterior distributions for regions 26 (top) and 53 (bottom) before (left) and after (right) including a beam-filling factor as a free parameter in our modeling algorithm.}
    \label{fig:bff_comparison}
\end{figure}

This appendix includes a number of figures provided in support of the discussion of Section~\ref{sec:high-zeta} (Figures~\ref{fig:lowTcorners}, \ref{fig:corner1}, \ref{fig:phase_comparison}, and \ref{fig:bff_comparison}).

\section{New Collisional Rates Analysis} \label{sec:coll_rates}

In this appendix we compare published HCN and HNC collisional excitation rates used in radiative transfer and statistical equilibrium calculations to predict HCN and HNC abundances.  The intent of this comparison is to highlight differences between excitation rates used for each isomer that might impact derived physical and chemical properties.  In the following we highlight the salient features of the published HCN and HNC excitation rates (Table~\ref{tab:colrates}).  We also provide a direct comparison to the two HCN and HNC excitation rate sets predominant in the astrophysical literature.

\begin{deluxetable}{lcccc}
\centering
\tablecolumns{5}
\tablewidth{0pt}
\tablecaption{Published HCN and HNC Excitation Rates\label{tab:colrates}}
\tablehead{
\colhead{Reference} &
\colhead{Isomer(s)} & \colhead{Col.~Partner} & \colhead{No.~Energy Levels} & \colhead{T$_{\rm K}$ Range (K)}
}
\startdata
\cite{Dumouchel2010} & 
HCN/HNC & He & 26 & 5--500 \\
\cite{Dumouchel2011PCCP} & HNC & para-/ortho-H$_2$ & 16 & 5--100 \\
\cite{Vera2014JChPh} & HCN & para-/ortho-H$_2$ & 13 & 5--100 \\
\cite{HernandezVera2017} & HCN/HNC & para-/ortho-H$_2$ & 26 & 10--500 \\
\enddata
\end{deluxetable}

The He excitation rates published by \cite{Dumouchel2010} were considered the default HCN and HNC excitation rates before subsequent excitation rates involving H$_2$ as the collision partner were available.  The \cite{Dumouchel2010} rates are available in the Leiden Atomic and Molecular Database\footnote{https://home.strw.leidenuniv.nl/~moldata/} \citep[LAMDA; ][]{Schoier2005A&A} where a scaling factor of 1.363 is used to correct these He-based collisional excitation rates by the difference in mass between He and H$_2$. This approximation thus makes these rates appropriate to use for modeling the interstellar medium where ground-state (j=0) para-H$_2$ is a common collision partner. \cite{Dumouchel2010} noted the following propensity rules for collisional de-excitation:
\begin{itemize}
    \item HCN--He favor even $\Delta{J}$ transitions.
    \item HNC--He favor odd $\Delta{J}$transitions.
    \item Propensity favoring even/odd transitions in HCN and HNC decreases with increasing temperature.
\end{itemize}

\cite{Vera2014JChPh}  and \cite{Dumouchel2011PCCP} published some of the first excitation rates for HCN and HNC, respectively, which used para- (j=0 and 2) and ortho- (j=1) H$_2$ as the collision partner.  
Both studies noted that the rate coefficients for collisions with different H$_2$ rotational states are very different.  Collisions with para-H$_2$ (j=0) are significantly smaller than those for collisions with ortho-H$_2$ (j=1) and with excited para-H$_2$ (j=2).
For the HNC excitation rates, \cite{Dumouchel2011PCCP} noted that propensity rules in favor of $\Delta{J}=1$  rates are larger than those involving $\Delta{J}=2$ for all H$_2$ rotational states.  For the HCN excitation rates, \cite{Vera2014JChPh} noted that propensity rules in favor of: (1) Even $\Delta{J}$ transitions were found for HCN in collisions with para-H$_2$ (j=0) and (2) Odd $\Delta{J}$ transitions were found for HCN in collisions with excited para- and ortho-H$_2$ (j$\geq 1$).

The newest and most complete sets of collisional excitation rates for HCN and HNC involving collisions with para- and ortho-H$_2$ have been published by \cite{HernandezVera2017}. These studies extended the \cite{Dumouchel2011PCCP} HCN and \cite{Vera2014JChPh} HNC excitation rates 
to the first 26 rotational energy levels and to kinetic temperatures of 500\,K for both molecules.  \cite{HernandezVera2017} compared their HCN and HNC para- and ortho-H$_2$ excitation rates with the He-scaled rates published by \cite{Dumouchel2010}, noting differences\footnote{In this comparison we have applied the standard scaling factor of 1.363 to convert He excitation rates to para-H$_2$, which \cite{HernandezVera2017} did not appear to apply.} of $\sim2$ for HCN--H$_2$(J=0)/HCN--He. \cite{HernandezVera2017} also noted that collisions for HCN and HNC involving ortho-H$_2$ (j=1) are $\sim 2-3$ times larger than those involving para-H$_2$ (j=0).  Propensity rules derived by \cite{HernandezVera2017} for HCN and HNC collisions with para- and ortho-H$_2$ are the same as those derived for collisions with He.

All of these studies have noted that collisional excitation of HCN is very different than that due to HNC, which should have a strong influence on the calculated intensities of these two isomers. We should also note the collisional excitation rates for the C, N, and H isotopologues of HCN and HNC published by \cite{Navarro-Almaida2023A&A}.

\begin{figure}[h]
    \gridline{\leftfig{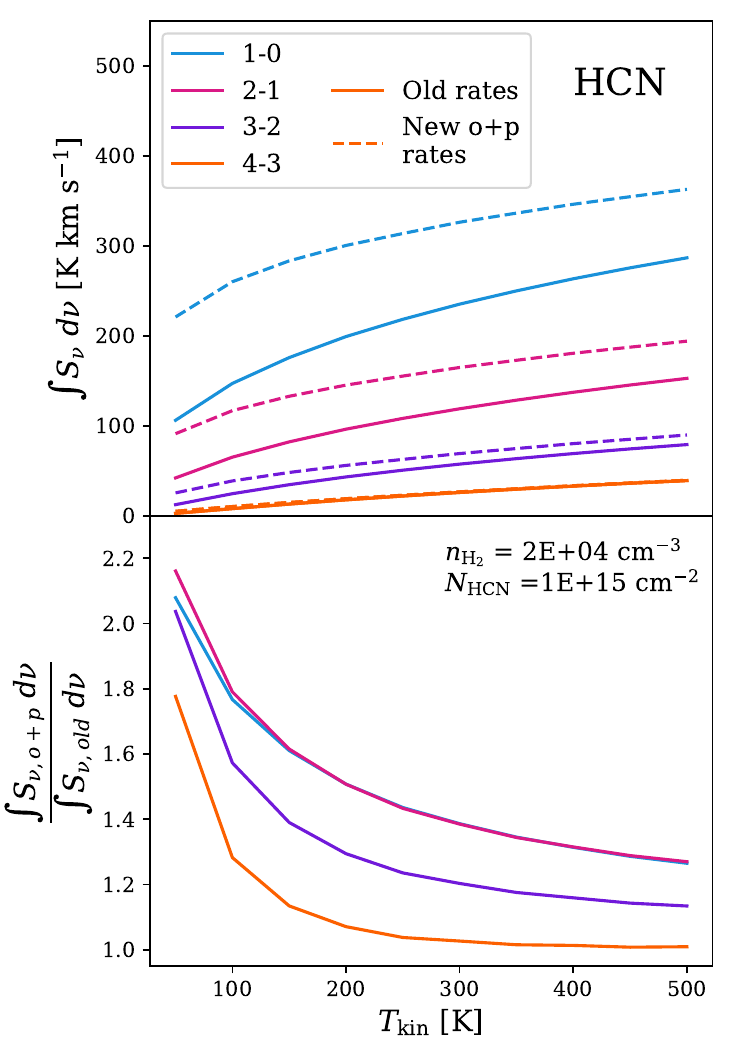}{0.4\textwidth}{(a)} \rightfig{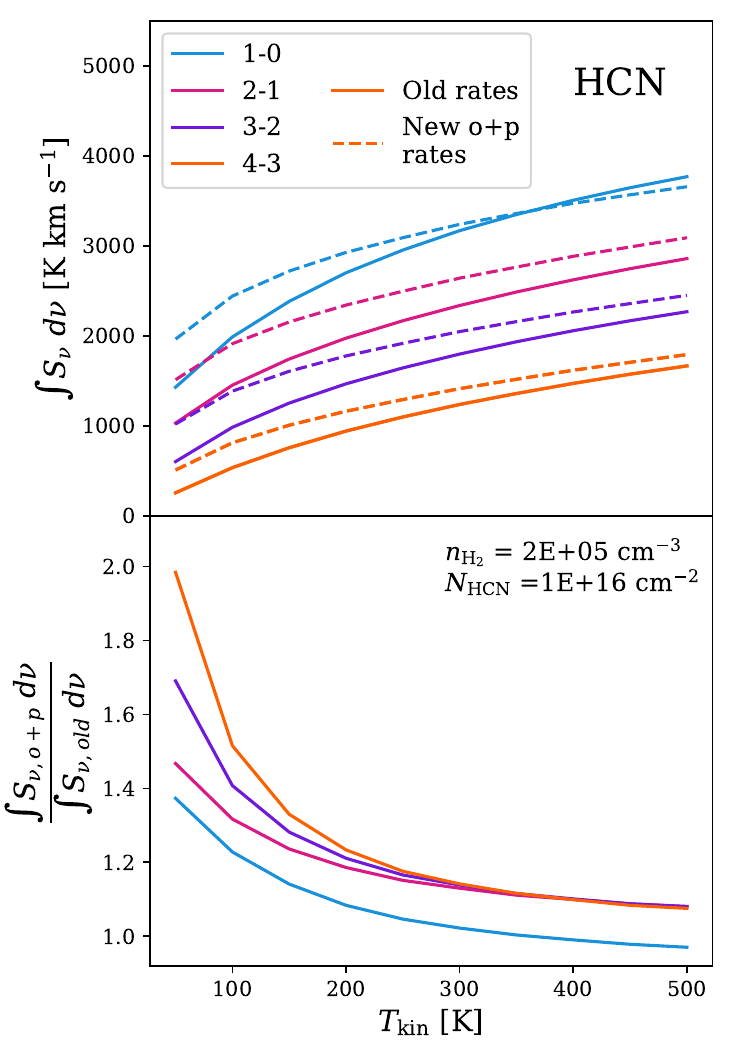}{0.4\textwidth}{(b)}}
    \gridline{\leftfig{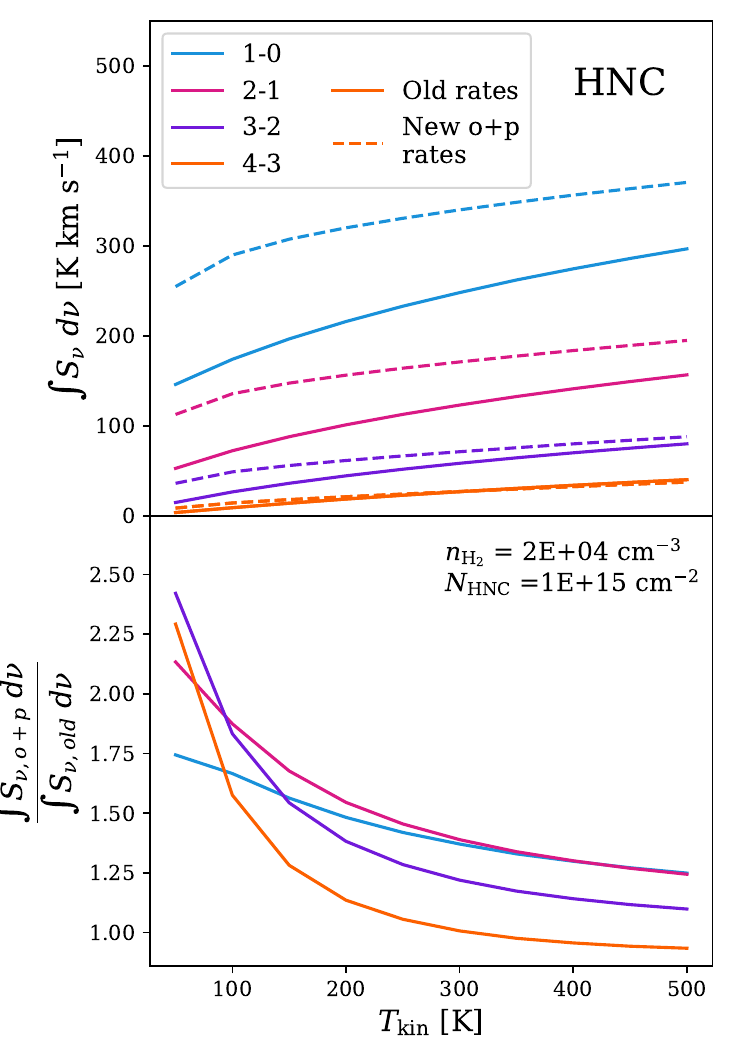}{0.4\textwidth}{(c)} \rightfig{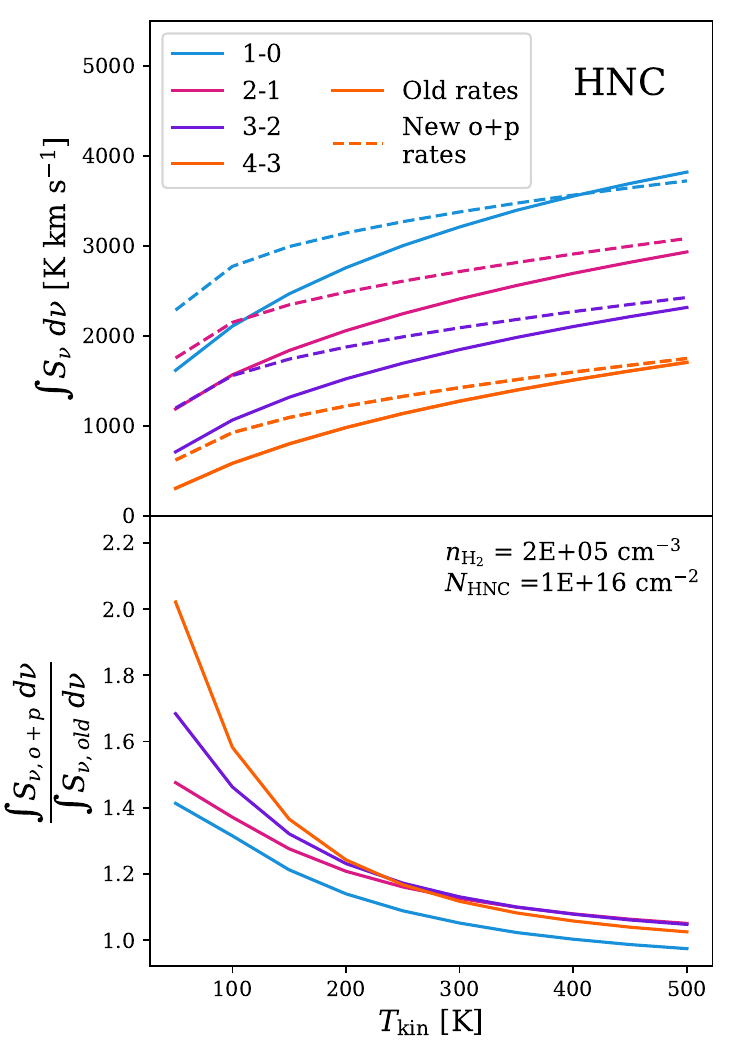}{0.4\textwidth}{(d)}}
    \caption{HCN (\textbf{a} and \textbf{b}) and HNC (\textbf{c} and \textbf{d}) integrated intensities (top panel in each pair) and integrated intensity ratios (new rates divided by old rates, bottom panels in each pair) as a function of kinetic temperature. Volume and column density are held constant at the listed values for each comparison.}
    \label{fig:collRates_tkin}
\end{figure}

\begin{figure}
    \gridline{\leftfig{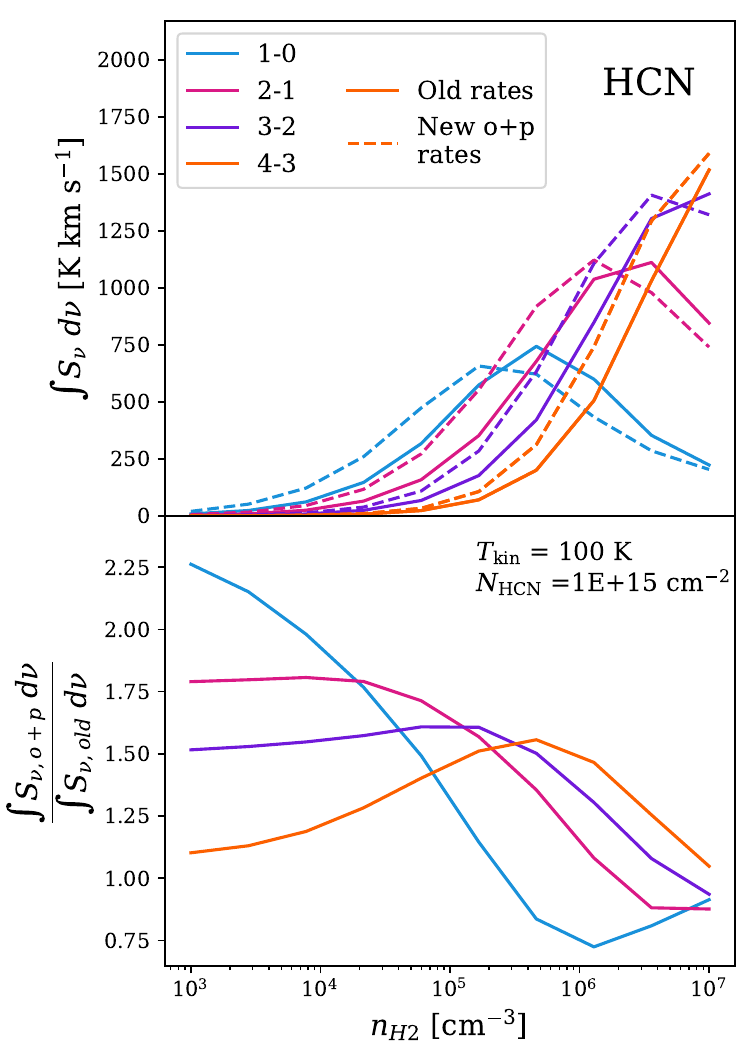}{0.4\textwidth}{(a)} \rightfig{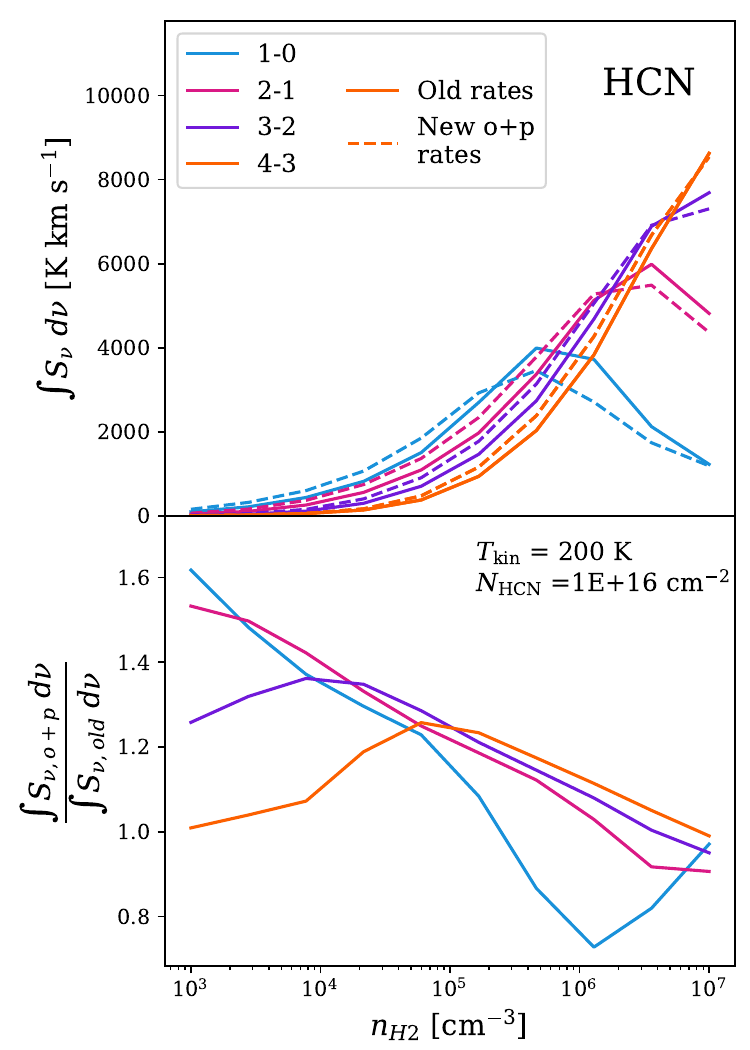}{0.4\textwidth}{(b)}}
    \gridline{\leftfig{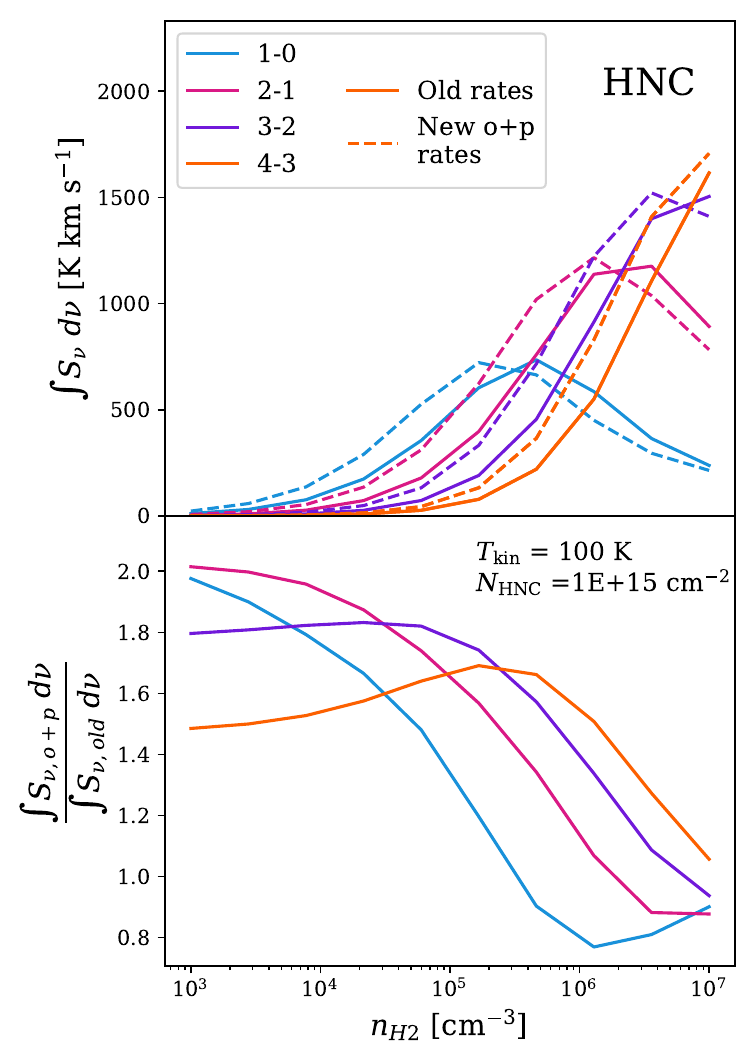}{0.4\textwidth}{(c)} \rightfig{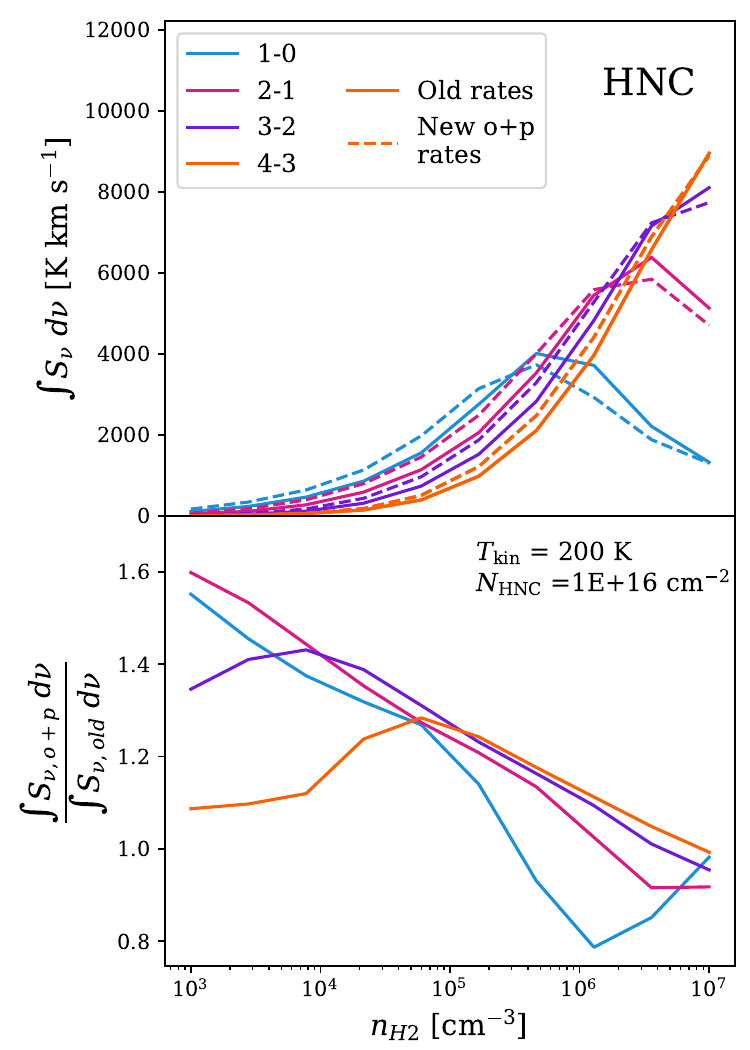}{0.4\textwidth}{(d)}}
    \caption{Same as in Figure \ref{fig:collRates_tkin} but plotted as a function of volume density, where kinetic temperature and column density are held constant.}
    \label{fig:collRates_n}
\end{figure}

\begin{figure}[h]
    \gridline{\leftfig{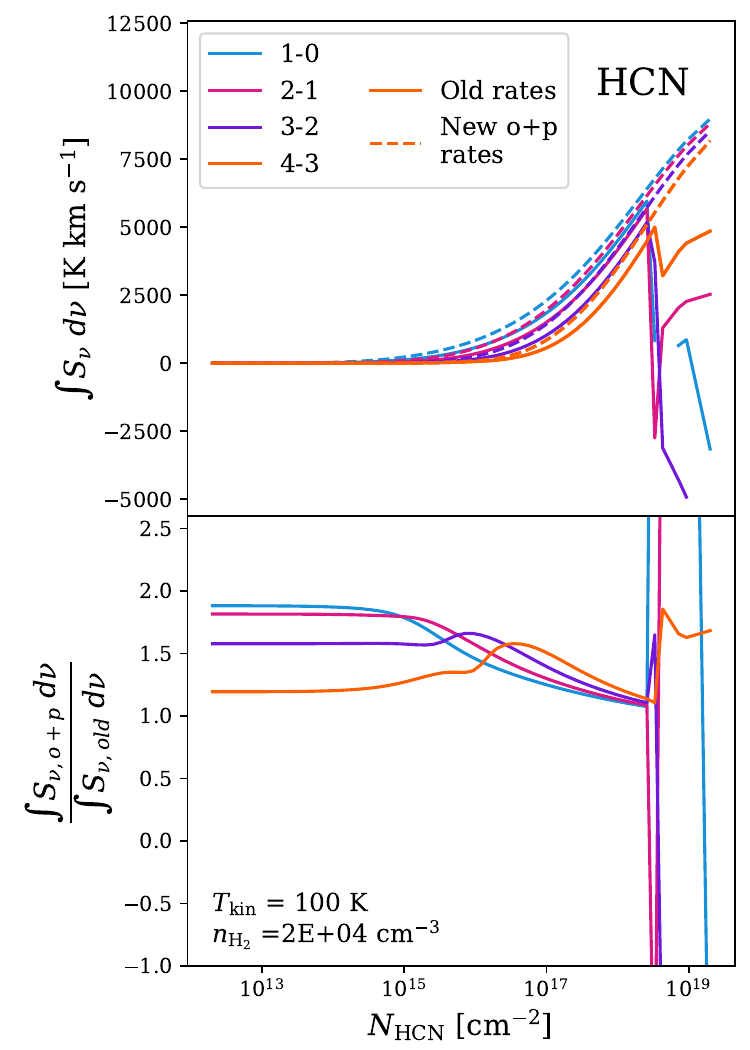}{0.4\textwidth}{(a)} \rightfig{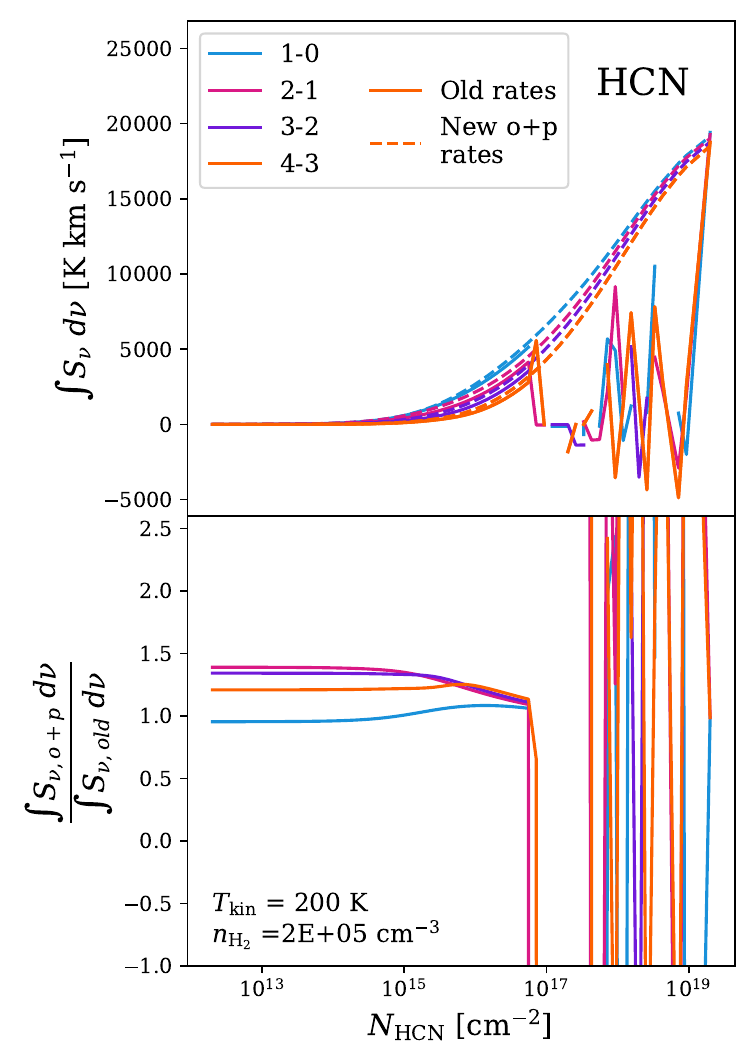}{0.4\textwidth}{(b)}}
    \gridline{\leftfig{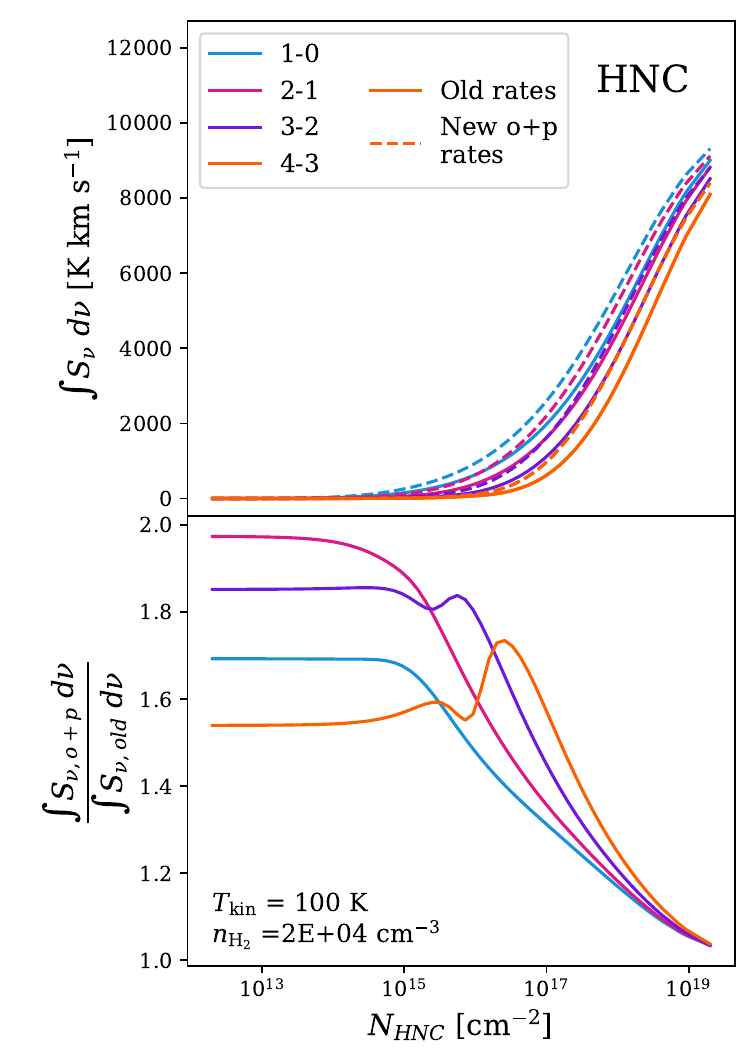}{0.4\textwidth}{(c)} \rightfig{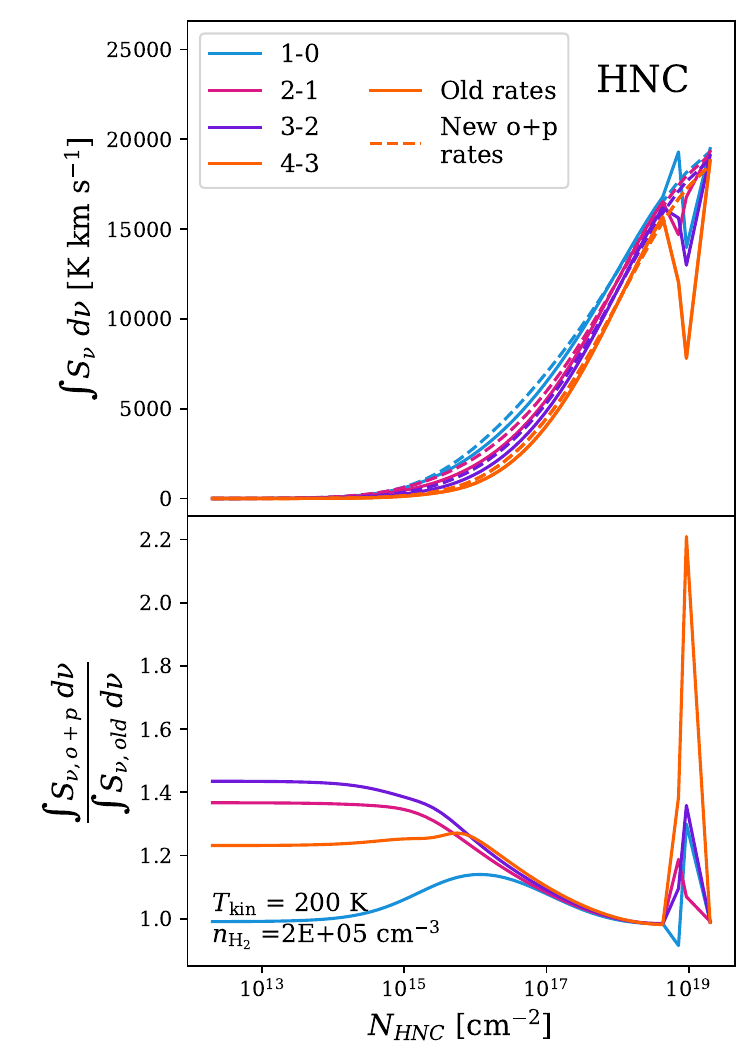}{0.4\textwidth}{(d)}}
    \caption{Same as in Figures \ref{fig:collRates_tkin} and \ref{fig:collRates_n} but plotted as a function of column density, where kinetic temperature and volume density are held constant.}
    \label{fig:collRates_N}
\end{figure}

To provide comparison of the commonly-used He excitation rates \citep{Dumouchel2010} to the H$_2$-based excitation rates \citep{HernandezVera2017} used in the present analysis, we provide in the following a series of comparison diagrams meant to illuminate the differences between these two sets of collisional excitation rates. These modeled intensities were calculated using both the ``old" (He-based) and ``new" (ortho+para-H$_2$-based) rates for the kinetic temperature $T_\text{K}$, H$_2$ volume density $n_{\text{H}_2}$, and H$_2$ column density $N_{\text{H}_2}$ ranges provided in Table \ref{tab:priors}. Note that the CRIR was not involved in these calculations, and no beam-filling factor was applied. We combine $N_{\text{H}_2}$ with $X_\text{HCN}$ and $X_\text{HNC}$ values in the range of $10^{-12} \leq X_\text{mol} \leq 10^{-6}$ to obtain molecular column densities $N_\text{HCN}$ and $N_\text{HNC}$ using $N_\text{mol} = 2 \times N_{\text{H}_2} \times X_\text{mol}$. This molecular column density calculation yields values in the range of $2\times 10^{12} \leq N_\text{mol} \leq 2\times10^{19}$\,cm$^{-2}$.    
Figures~\ref{fig:collRates_tkin}, \ref{fig:collRates_n}, and  \ref{fig:collRates_N} show a series of comparisons as a function of $T_\text{K}$, $n_{\text{H}_2}$, and $N_\text{mol}$, respectively, where the other two parameters in each figure are held constant at ``low" (panel (a) in each figure) and ``high" (panel (b) in each figure) values. A summary of these comparisons concludes that:
\begin{itemize}
    \item Use of the new ortho+para-H$_2$ excitation rates result in higher (by up to a factor of 2) predicted HCN and HNC integrated transition intensities.
    \item Differences between predicted line intensities derived from the old (He-scaled) versus new (ortho+para-H$_2$) excitation rates are highest ($\sim$ factor of 2) at low temperatures, volume densities, and column densities.
    \item Differences between line intensities for each $J$ transition diminish at higher volume and column densities for both sets of collisional excitation rates (see Figure~\ref{fig:collRates_tkin}).
    \item At very high HCN and HNC column densities ($\gtrsim 10^{18}$\,cm$^{-2}$), the predicted HCN and HNC transition intensities and opacities become so large that the \texttt{SpectralRadex} models do a poor job of predicting integrated intensities. Note that these HCN and HNC column densities are unrealistically large; more than a factor of 1000 larger than measured in the ISM.
\end{itemize}

\begin{figure}
    \gridline{\leftfig{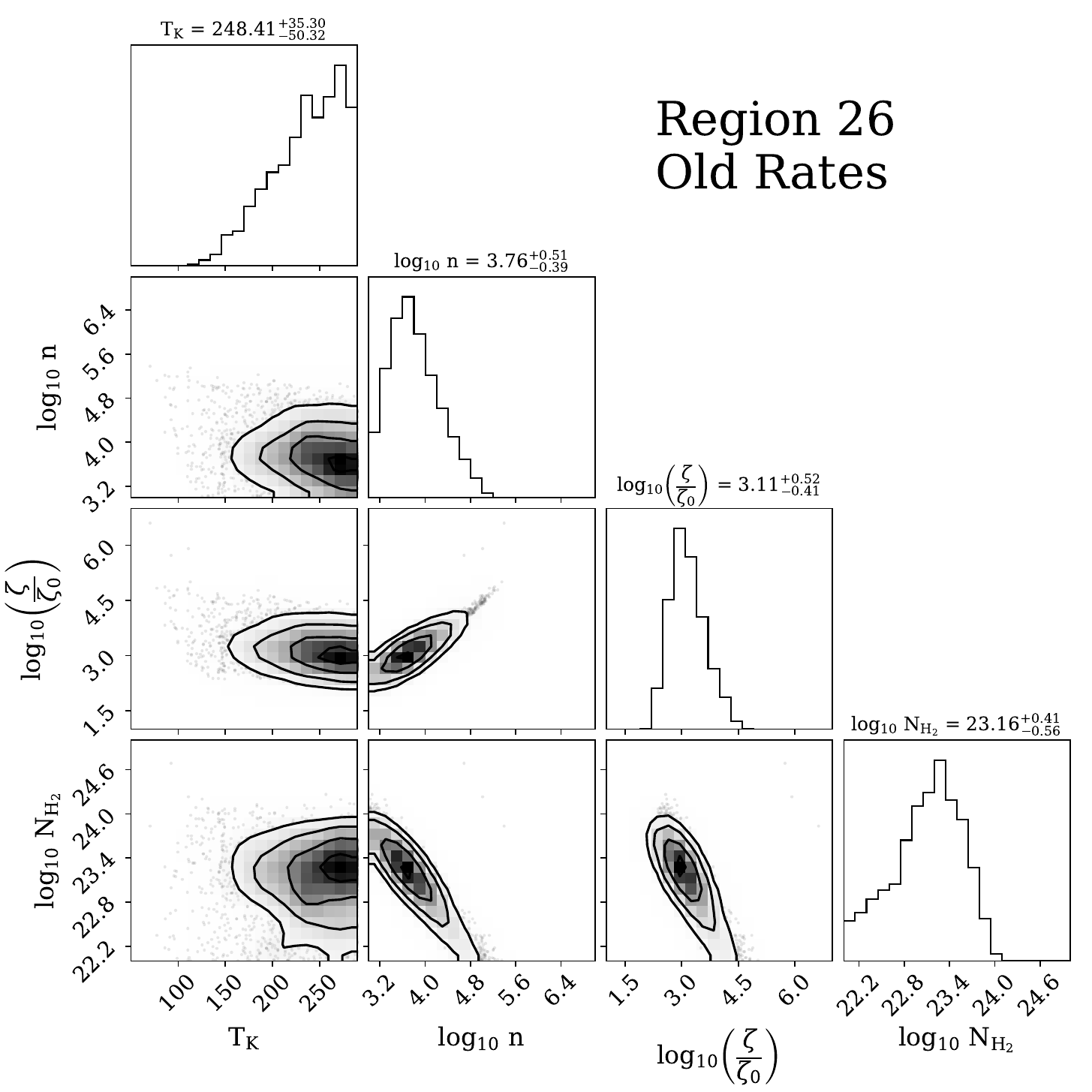}{0.5\textwidth}{(a)} \rightfig{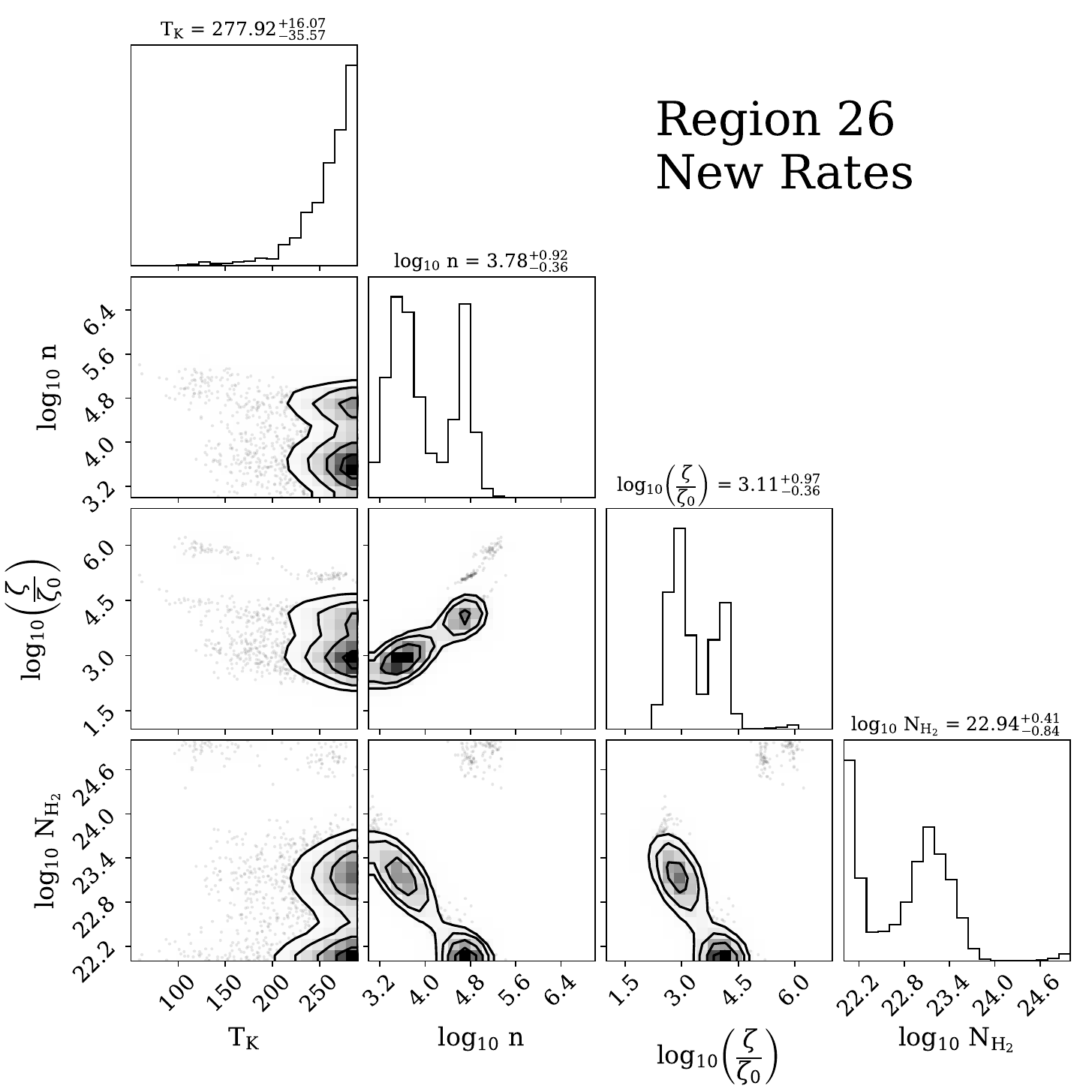}{0.5\textwidth}{(b)}}
    \gridline{\leftfig{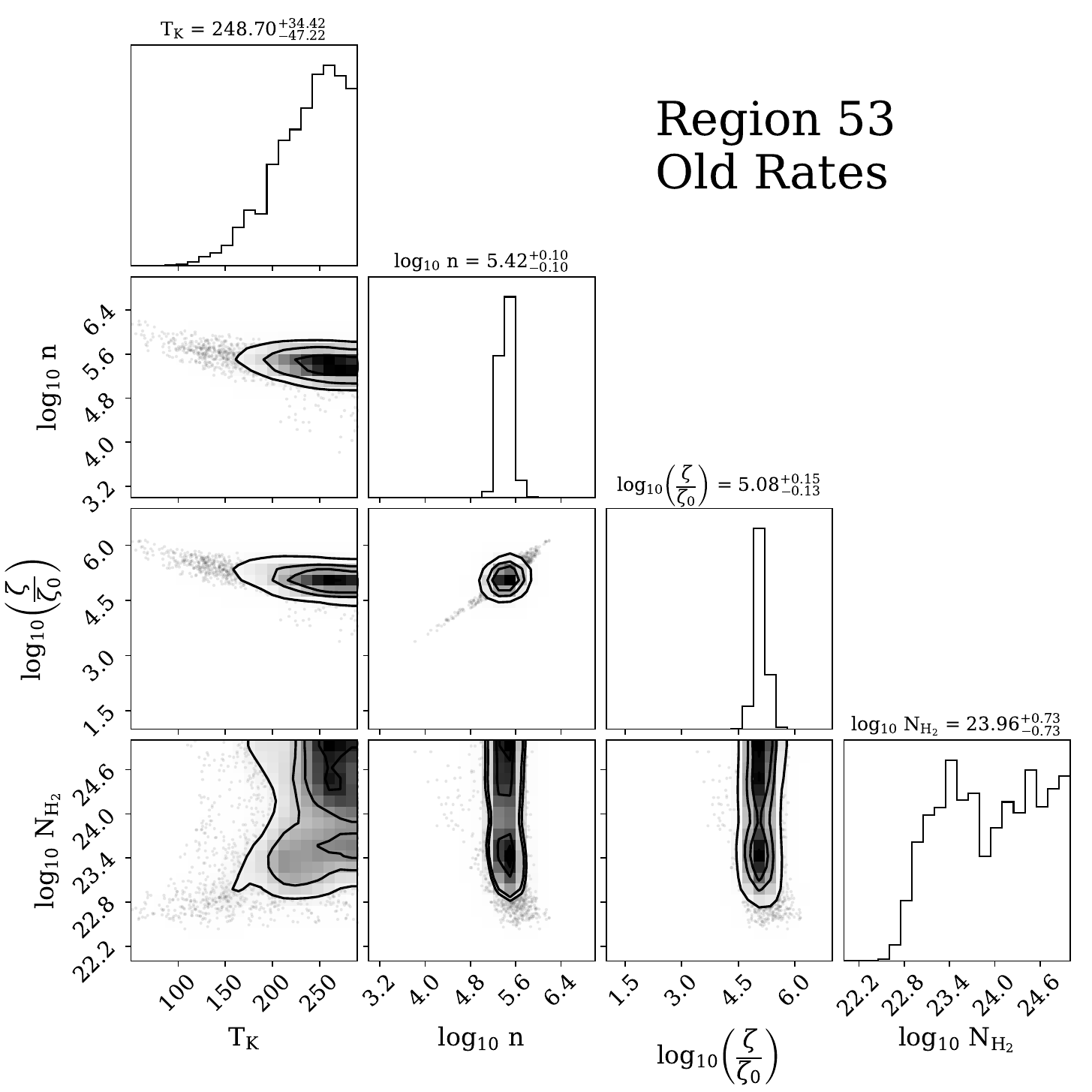}{0.5\textwidth}{(c)} \rightfig{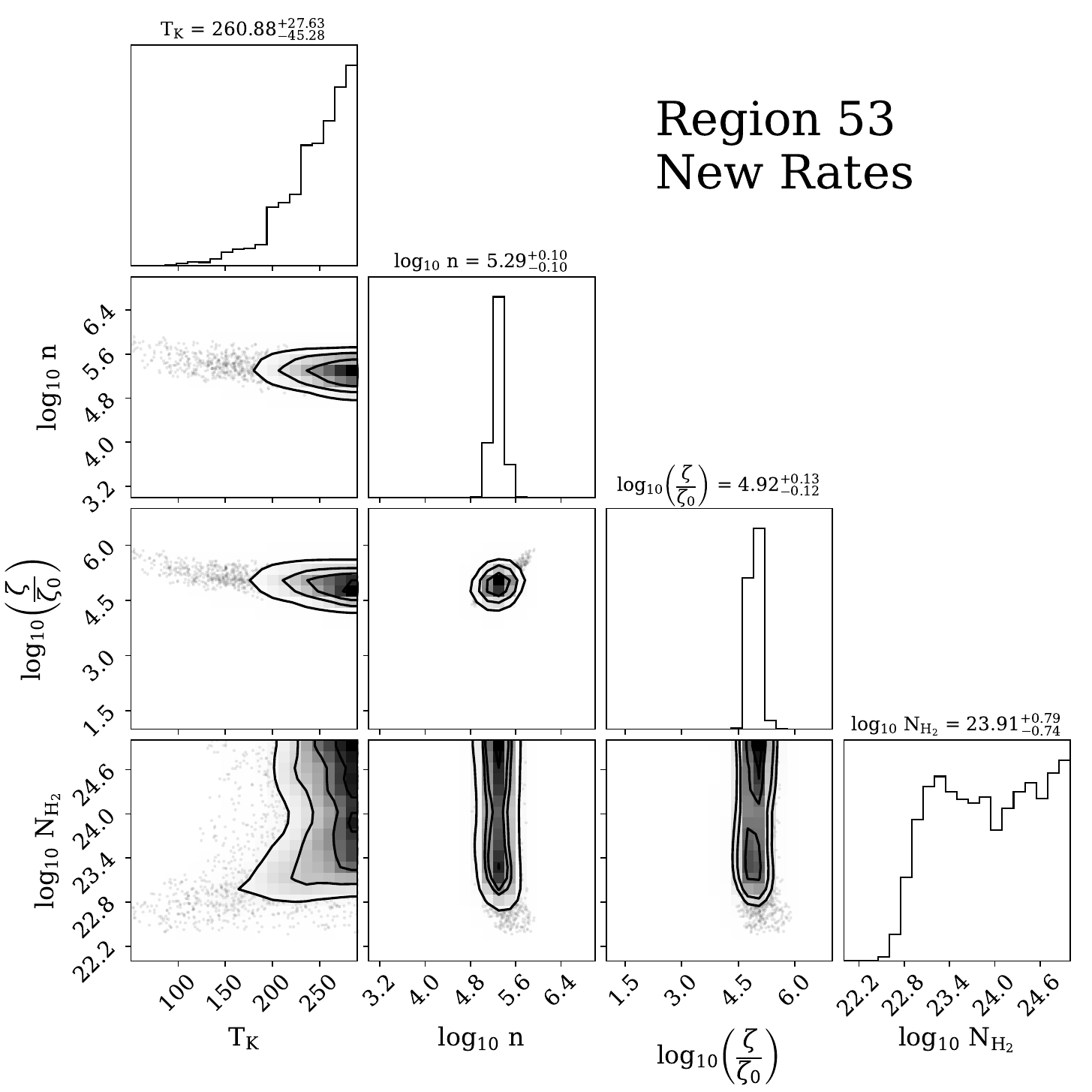}{0.5\textwidth}{(d)}}
    \caption{\texttt{UltraNest} inference results derived from HCN and HNC data for regions 26 (top) and 53 (bottom) using old (left) versus new (right) HCN and HNC collisional excitation rates.}
    \label{fig:HCNHNCreg26+53Rates}
\end{figure}

We also test the differences between the \cite{Dumouchel2010} and \cite{HernandezVera2017} rates in the context of our neural network + Bayesian nested sampling algorithm. The \texttt{UltraNest} inference results using the old and new rates are shown in Figure~\ref{fig:HCNHNCreg26+53Rates} for regions 26 and 53. Here we include CRIR as a free parameter using the prior distribution provided in Table \ref{tab:priors}. Note that no beam-filling fator is applied in these tests. Overall, only minor differences result from the use of the new versus old excitation rates. Though panels (a) and (b) in Figure \ref{fig:HCNHNCreg26+53Rates} for region 26 show a bimodality in the posterior distributions for H$_2$ volume density, CRIR, and H$_2$ column density, the median predicted values for these parameters are nearly identical using the new and old rates. The minimal differences between parameter values derived from the two sets of rates likely also result from higher predicted kinetic temperatures in both regions and higher H$_2$ volume and column density estimates in region 53. As noted above, the differences in intensities derived from the two sets of collisional excitation rates diminish at larger values of $T_\text{K}$, $N_{\text{H}_2}$, and $N{\text{H}_2}$.




\end{CJK*}
\end{document}